\documentclass[a4paper,11pt]{article}
\usepackage{jheppub} 
\usepackage[table,xcdraw,dvipsnames]{xcolor}
\usepackage{multirow}
\usepackage{graphicx} 
\usepackage{subfig}
\usepackage{tikz}

\usepackage{cleveref}
\crefname{table}{Table}{Tables}
\crefname{equation}{Eq.}{Eqs.}
\crefname{appendix}{App.}{Apps.}
\crefname{section}{Sec.}{Secs.}
\crefname{figure}{Fig.}{Figs.}

\makeatletter
\newcommand{\ostar}{\mathbin{\mathpalette\make@circled{*}}}
\newcommand{\make@circled}[2]{%
  \ooalign{$\m@th#1\smallbigcirc{#1}$\cr\hidewidth$\m@th#1#2$\hidewidth\cr}%
}
\newcommand{\smallbigcirc}[1]{%
  \vcenter{\hbox{\scalebox{0.77778}{$\m@th#1\bigcirc$}}}%
}
\makeatother

\title{\boldmath Astrophysical Uncertainties in Sub-GeV Dark Matter Detection via Single Phonon Excitations}

\author[a]{Xu-Xiang~Li,}
\author[a]{Navaneetha Valsan,}
\author[a]{and Zhengkang~Zhang}

\affiliation[a]{Department of Physics \& Astronomy, University of Utah, Salt Lake City, UT 84112, USA}

\emailAdd{xuxiang.li@utah.edu}
\emailAdd{navaneetha.valsan@utah.edu}
\emailAdd{z.k.zhang@utah.edu}

\abstract{
We present the first systematic study of how local dark matter velocity distribution uncertainties propagate into direct detection rates for dark matter--single phonon scattering. We consider three benchmark halo models---Standard Halo Model, Tsallis and empirical---and vary the astrophysical parameters within observationally motivated ranges. To compare halo models on equal footing, we introduce an rms-matching prescription that holds the mean dark matter kinetic energy fixed across models. With this prescription, differences between halo models prove subdominant to parameter variations within each model, so that astrophysical uncertainties can be effectively captured by varying parameters within the Standard Halo Model alone. We find $\mathcal{O}(1\%)$ to $\mathcal{O}(100\%)$ fractional deviations in the predicted rates across the dark matter mass range of interest. For the daily modulation signal, astrophysical parameter variations rescale the amplitude but leave the phase robust. These results provide timely input for reliably interpreting upcoming phonon-based direct detection experiments targeting sub-GeV dark matter.
}

\begin{document}
\maketitle
\flushbottom

\section{Introduction}

The nature of dark matter remains one of the most profound open questions in modern physics. Direct detection experiments, which search for signals from dark matter particles scattering off laboratory targets, play an important role in probing its particle nature. The dominant paradigm for such searches has long been the weakly interacting massive particle (WIMP), whose thermal relic abundance naturally favors masses above the GeV scale. Nuclear recoil experiments designed to detect WIMPs have achieved remarkable sensitivity over the past few decades, progressively closing much of the available parameter space for dark matter heavier than $\mathcal{O}(\mathrm{GeV})$~\cite{LUX:2016ggv,XENON:2018voc,PandaX:2024qfu,LZ:2024zvo,XENON:2025vwd}. On the other hand, a wide class of theoretically motivated scenarios beyond the WIMP paradigm---including freeze-in production~\cite{Hall:2009bx, Bernal:2017kxu}, hidden sector dark matter~\cite{Strassler:2006im, Arkani-Hamed:2008kxc, Cheung:2009qd, Morrissey:2009ur}, asymmetric dark matter~\cite{Kaplan:2009ag, Cohen:2010kn, Petraki:2013wwa, Zurek:2013wia}, and strongly self-interacting models~\cite{Hochberg:2014dra, Hochberg:2014kqa}---naturally accommodate sub-GeV dark matter. 
Searching for such candidates with conventional nuclear recoil detectors is, however, fundamentally limited: the kinematic mismatch between a light dark matter particle and a heavy target nucleus means that only a tiny fraction of the dark matter kinetic energy can be transferred in a collision, pushing recoil signals well below detector thresholds. 

A broad program of novel detection strategies has therefore emerged, exploiting low-energy excitations in condensed matter and atomic targets. These include electronic excitations in semiconductors and noble liquids~\cite{Settimo:2020cbq, SENSEI:2020dpa, SuperCDMS:2020ymb}, the Migdal effect~\cite{Ibe:2017yqa, EDELWEISS:2022ktt, SuperCDMS:2023sql}, excitations of phonon and roton modes in superfluid helium~\cite{Schutz:2016tid, Knapen:2016cue, Hertel:2018aal}, quasiparticle production in superconducting targets~\cite{Hochberg:2015pha, Hochberg:2019cyy, Das:2022srn}, electronic excitations in three-dimensional Dirac and topological semimetals with tunable meV-scale band gaps~\cite{Hochberg:2017wce, Geilhufe:2019ndy}, magnon excitations in magnetically ordered materials~\cite{Trickle:2019ovy, Trickle:2020oki}, excitations of phonon polaritons in polar materials~\cite{Knapen:2017ekk, Mitridate:2020kly}, and electronic excitations in two-dimensional materials such as graphene~\cite{Hochberg:2016ntt}.

Among these emerging channels, single phonon excitation in crystalline targets stands out for several reasons. First, the characteristic energies of phonon modes in crystalline targets span the $\mathcal{O}(\mathrm{meV})$ range~\cite{Trickle:2019nya, Griffin:2019mvc}, enabling sensitivity to dark matter masses as low as a few keV. Second, crystal targets often exhibit strong lattice anisotropy, meaning the phonon response depends not only on the magnitude of the momentum transfer but also on its direction relative to the crystal axes. As the Earth rotates, the orientation of a terrestrial detector relative to the incoming dark matter wind changes over a sidereal day, inducing a periodic daily modulation in the scattering rate~\cite{Griffin:2018bjn, Coskuner:2021qxo}. This directional signature provides a powerful handle for discriminating a dark matter signal from isotropic backgrounds. This combination of low-mass coverage and directional sensitivity makes single phonon excitation a uniquely powerful probe of sub-GeV dark matter. Meanwhile, detector technologies based on transition-edge sensors and microwave kinetic inductance devices are rapidly maturing, with demonstrated energy resolutions now approaching the thresholds necessary for single phonon detection~\cite{Fink:2020noh, Ramanathan:2024frr}, making this an especially timely channel to characterize thoroughly.

The scattering rate in any direct detection experiment depends not only on the underlying particle physics model and the properties of the detector material, but also on the local velocity distribution function (VDF) of dark matter in the Galactic halo~\cite{McCabe:2010zh, Pato:2012fw}. The VDF, however, is not precisely known: both its functional form and the astrophysical parameters that characterize it are subject to considerable observational uncertainties. Moreover, high-resolution $N$-body simulations consistently find that $\Lambda$CDM halos deviate from the simple Maxwell--Boltzmann form assumed in the Standard Halo Model (SHM)~\cite{Vogelsberger:2008qb, Ling:2009eh, Kuhlen:2009vh}, with residual deviations persisting in the high-velocity tail even when baryonic physics is included~\cite{Bozorgnia:2016ogo, Bozorgnia:2017brl, Kelso:2016qqj, Sloane:2016kyi}. The impact of these uncertainties on direct detection observables has been studied for nuclear recoil~\cite{Vergados:2007nc, Kuhlen:2009vh, McCabe:2010zh, Green:2017odb} and electron scattering~\cite{Hryczuk:2020trm, Radick:2020qip,Maity:2020wic,Li:2022acp}. It is the purpose of this work to extend such studies to the single phonon channel, quantifying how VDF uncertainties propagate into both the projected reach and the daily modulation signal, which is essential for ongoing and future phonon-based searches to be reliably interpreted.

Concretely, in this work we consider three benchmark halo models---the SHM~\cite{Drukier:1986tm}, the Tsallis distribution~\cite{Hansen:2004dg, Hansen:2005yj}, and an empirical model motivated by cosmological simulations~\cite{Mao:2012hf, Mao:2013nda}---and examine the effects of varying the key astrophysical parameters within observationally motivated ranges. We focus on spin-independent interactions and study four benchmark crystalline targets selected for their complementary physical properties and their established suitability for light dark matter detection via phonon excitations~\cite{Knapen:2017ekk, Griffin:2018bjn, Coskuner:2021qxo}. GaAs, a polar semiconductor with a simple crystal structure, features gapped optical phonons around 30--40~meV and strong Born effective charges that enable efficient coupling to dark photons~\cite{Knapen:2017ekk}. $\text{Al}_2\text{O}_3$ (sapphire) combines a large range of phonon energies with an anisotropic lattice structure that enhances directional sensitivity and daily modulation effects~\cite{Griffin:2018bjn}. $\text{SiO}_2$ ($\alpha$-quartz) has been identified as a promising light-element polar crystal with strong projected reach across several benchmark interaction models~\cite{Coskuner:2021qxo}. Finally, $\text{CaWO}_4$, already deployed in cryogenic experiments such as CRESST, exhibits multiple optical phonon branches and favorable modulation prospects due to its anisotropy~\cite{Coskuner:2021qxo}. Together, these targets span a representative range of phonon energies, lattice anisotropies, and Born effective charges, making them well-suited benchmarks for exploring the impact of astrophysical uncertainties on sub-GeV dark matter searches. For each target, we compute the projected dark matter--phonon cross section reach and the daily modulation signal, and quantify the relative importance of each astrophysical parameter. 

Such a halo uncertainty study also raises an important methodological question: how should predictions from different VDF models be compared? The standard practice in the literature has been to identify each model's characteristic velocity scale $v_0$ directly with the local circular velocity $v_\mathrm{c}$~\cite{Green:2017odb, Radick:2020qip,Maity:2020wic,Li:2022acp}. However, this approach conflates differences in the functional form of the distribution with differences in its overall energy scale, making it difficult to isolate the effect of functional shape alone. We argue instead that a more physically meaningful comparison is obtained by requiring the root-mean-square (rms) velocity---equivalently, the mean dark matter kinetic energy---to be equal across models. This rms-matching prescription ensures that any differences in predicted rates can be attributed to the shape of the velocity distribution rather than to a mismatch in its overall energy scale, placing the comparison on equal dynamical footing.

The paper is organized as follows. In~\cref{sec:formalism}, we review the dark matter--single phonon scattering formalism. In~\cref{sec:VDF_astro}, we introduce the VDF models and astrophysical parameters considered in this work, and define our rms-matching prescription. Our main results on projected reach and daily modulation are presented in~\cref{sec:randd}, and we conclude in~\cref{sec:conclusion}. Supplementary tables and additional results using alternative parameter ranges are provided in the appendices.

\section{Dark matter--single phonon scattering formalism}
\label{sec:formalism}

The rate for spin-independent scattering of a dark matter particle $\chi$ with mass $m_\chi$ on a crystalline target can be written as a product of astrophysical, particle physics, and material response factors~\cite{Trickle:2019nya}:
\begin{equation}
     R(t) = \frac{1}{\rho_\mathrm{T}} \frac{\rho_\chi}{m_\chi}
     \frac{\pi \overline{\sigma}_\psi}{\mu^2_{\chi\psi}}
     \int \mathrm{d}^3 \boldsymbol{v} \, f_\chi(\boldsymbol{v},t)
     \int \frac{\mathrm{d}^3\boldsymbol{q}}{(2\pi)^3} \, \mathcal{F}^2_{\mathrm{med}}(q) \, S(\boldsymbol{q},\omega_{\boldsymbol{q}}) \,,
     \label{eq:rate}
\end{equation}
where $\rho_\mathrm{T}$ is the target mass density, $\boldsymbol{v}$ and $\boldsymbol{q}$ are the incoming dark matter velocity and momentum transfer, and energy-momentum conservation fixes the energy deposition to
\begin{equation}
     \omega_{\boldsymbol{q}} = \boldsymbol{q}\cdot\boldsymbol{v}-\frac{q^2}{2m_\chi} \,.
\end{equation}
The three ingredients entering~\cref{eq:rate} are:

\paragraph{Astrophysical input.}
The local dark matter density $\rho_\chi$ and velocity distribution $f_\chi(\boldsymbol{v},t)$ together set the dark matter flux impinging on the detector. The density serves as an overall normalization: since $R \propto \rho_\chi$, a different value of $\rho_\chi$ rescales the projected reach without affecting its shape. We adopt $\rho_\chi = 0.4\,\mathrm{GeV}/\mathrm{cm}^3$ as a commonly used benchmark. The velocity distribution, on the other hand, enters the rate in a nontrivial, kinematically entangled way, and its modeling is the central subject of this work; we defer a detailed discussion to~\cref{sec:VDF_astro}.

\paragraph{Particle physics input.}
The reference cross section
\begin{equation}
     \overline{\sigma}_\psi = \frac{\mu^2_{\chi\psi}}{\pi} \, \overline{\bigl|\mathcal{M}_{\chi\psi}(q=q_0)\bigr|^2}
\end{equation}
parametrizes the overall interaction strength, with $\psi = n$ or $e$ standing for nucleon or electron. $\mu_{\chi\psi}$ is the dark matter--$\psi$ reduced mass, and $q_0$ is a reference momentum transfer that is usually chosen to be $q_0 = m_\chi v_0$ for $\psi = n$ and $q_0 = \alpha m_e$ for $\psi = e$. The mediator form factor encodes the $q$~dependence of the propagator:
\begin{equation}
\mathcal{F}_{\mathrm{med}}(q) =
\begin{cases}
    1  & \text{(heavy mediator),} \\
    (q_0/q)^2  & \text{(light mediator).}
\end{cases}
\end{equation}

\paragraph{Material response.}
The dynamic structure factor $S(\boldsymbol{q},\omega)$ encodes the target's response to a momentum and energy deposition $(\boldsymbol{q},\omega)$. For single phonon excitations at zero temperature, the target transitions from the vacuum with no phonons to a one-phonon state $|\nu,\boldsymbol{k}\rangle$, labeled by branch index $\nu$ and momentum $\boldsymbol{k}$ within the first Brillouin zone (1BZ). Lattice momentum conservation requires $\boldsymbol{q} = \boldsymbol{k} + \boldsymbol{G}$ for a reciprocal lattice vector $\boldsymbol{G}$, while energy conservation requires $\omega = \omega_{\nu,\boldsymbol{k}}$. The resulting dynamic structure factor is~\cite{Trickle:2019nya}
\begin{equation}
     S(\boldsymbol{q},\omega) = \frac{\pi}{\Omega}\sum_{\nu}\frac{1}{\omega_{\nu,\boldsymbol{k}}} \,
     \Biggl|\sum_{j}\frac{e^{-W_j(\boldsymbol{q})}}{\sqrt{m_j}} \, e^{i\boldsymbol{G}\cdot x_j^0} \, (\boldsymbol{Y}_j\cdot\boldsymbol{\epsilon}_{\nu,\boldsymbol{k},j}^\ast)\Biggr|^2
     \delta(\omega-\omega_{\nu,\boldsymbol{k}}) \,,
\end{equation}
where $\Omega$ is the primitive cell volume, $j$ runs over the ions in the primitive cell, and $m_j$, $\boldsymbol{x}_j^0$, $\omega_{\nu,\boldsymbol{k}}$, $\boldsymbol{\epsilon}_{\nu,\boldsymbol{k},j}$ are the ion masses, equilibrium positions, phonon energies, and phonon eigenvectors, respectively. The Debye--Waller factor $W_j(\boldsymbol{q}) = \frac{\Omega}{4m_j}\sum_\nu\int_\mathrm{1BZ}\frac{\mathrm{d}^3k}{(2\pi)^3}\frac{|\boldsymbol{q}\cdot\boldsymbol{\epsilon}_{\nu,\boldsymbol{k},j}|^2}{\omega_{\nu,\boldsymbol{k}}}$ suppresses the response at large momentum transfer due to thermal ion displacements.

The dark matter--ion coupling vectors $\boldsymbol{Y}_j$ depend on the interaction model. For the two benchmark scenarios we consider---dark photon and hadrophilic scalar mediators---they are given by~\cite{Trickle:2019nya}
\begin{equation}
\boldsymbol{Y}_j =
\begin{cases}
    -\dfrac{\boldsymbol{q}\cdot\boldsymbol{Z}_j^\star}{\hat{\boldsymbol{q}}\cdot\varepsilon_\infty\cdot\hat{\boldsymbol{q}}} & \text{(dark photon mediator),} \\[8pt]
    \boldsymbol{q}\,A_j\,F_{N_j}(q)  & \text{(hadrophilic scalar mediator),}
\end{cases}
\end{equation}
where $\boldsymbol{Z}_j^\star$ is the Born effective charge tensor, $\varepsilon_\infty$ is the high-frequency dielectric tensor, $A_j$ is the mass number, and $F_{N_j}(q)$ is the nuclear form factor, which is well approximated by unity at the momentum transfers relevant for this work. These mediator benchmarks are complementary in their phonon coupling patterns: the dark photon has opposite-sign couplings to oppositely charged ions in polar crystals and primarily excites gapped optical modes, while the hadrophilic scalar has same-sign couplings to all ions and primarily drives gapless acoustic modes in the low $q$ limit. The mediator mass further differentiates the kinematics: a heavy mediator emphasizes large momentum transfers $q$, whereas a light mediator shifts the weight to smaller $q$. By combining these benchmark models, we cover a representative range of coupling patterns and kinematic sensitivities.

\subsection{Kinematic function and the role of the velocity distribution}

To make the dependence of the rate on the VDF explicit, it is useful to perform the velocity integral in~\cref{eq:rate} for fixed $\boldsymbol{q}$ and $\omega$. Define the kinematic function~\cite{Trickle:2019nya}
\begin{equation}
    g(\boldsymbol{q},\omega)\equiv \int \mathrm{d}^3v \, f_\chi(\boldsymbol{v},t) \, 2\pi \, \delta(\omega-\omega_{\boldsymbol{q}}) \,,
    \label{eq:kinematic_term}
\end{equation}
which encapsulates all astrophysical dependence at a given point in $(\boldsymbol{q},\omega)$ space, and serves as a weight function for the material response. The rate becomes
\begin{equation}
     R(t) = \frac{1}{m_{\mathrm{cell}}} \frac{\rho_\chi}{m_{\chi}}
     \frac{\pi\overline{\sigma}_\psi}{2\mu^2_{\chi\psi}}
     \int \frac{\mathrm{d}^3q}{(2\pi)^3} \, \mathcal{F}^2_{\mathrm{med}}(q) \sum_{\nu}g(\boldsymbol{q},\omega_{\nu,\boldsymbol{k}}) \, \frac{1}{\omega_{\nu,\boldsymbol{k}}}\Biggl|\sum_{j}\frac{e^{-W_j(\boldsymbol{q})}}{\sqrt{m_j}} \, e^{i\boldsymbol{G}\cdot x_j^0} \,(\boldsymbol{Y}_j\cdot\boldsymbol{\epsilon}_{\nu,\boldsymbol{k},j}^\ast)\Biggr|^2 \,,
     \label{eq:new_rate}
\end{equation}
where $m_\mathrm{cell} = \rho_\mathrm{T}\Omega$ is the mass of a single crystal cell.

The factorized form of~\cref{eq:new_rate} clarifies why the impact of VDF uncertainties is dark matter model- and material-dependent: the kinematic function $g(\boldsymbol{q},\omega)$ acts as a filter in $\boldsymbol{q}$ space, and the region it selects depends on the VDF, while the integrand it weights depends on the mediator type and phonon spectrum. Different VDF choices therefore lead to different rates, but the magnitude of the effect varies with the interaction model and target.

The VDF in the detector frame, $f_\chi(\boldsymbol{v},t)$, is related to the Galactic-frame VDF, $f_\mathrm{gal}(\boldsymbol{v})$, by
\begin{equation}
    f_\chi(\boldsymbol{v},t) = f_{\mathrm{gal}}(\boldsymbol{v}+\boldsymbol{v}_\mathrm{e}(t)) \,,
    \label{eq:fchi_fgal}
\end{equation}
where $\boldsymbol{v}_\mathrm{e}(t)$ is the Earth's velocity in the Galactic frame. For an isotropic Galactic-frame VDF, $f_\mathrm{gal}(v)$, the delta function in~\cref{eq:kinematic_term} can be used to reduce the velocity integral to one dimension:
\begin{equation}
    g(\boldsymbol{q},\omega) = \frac{(2\pi)^2}{Kq}\int_{v\_}^{v_{\mathrm{esc}}}v \,\mathrm{d}v\, f_{\mathrm{gal}}(v) \,,
    \label{eq:gterm}
\end{equation}
where $v_\mathrm{esc}$ is the Galactic escape velocity and $K$ is a normalization constant,
\begin{equation}
    K = 4\pi\int_{0}^{v_{\mathrm{esc}}}v^2 \,\mathrm{d}v\, f_{\mathrm{gal}}(v) \,.
\end{equation}
The lower limit of the integral in~\cref{eq:gterm} is given by
\begin{equation}
    v\_ = \mathrm{min}\biggl\{\frac{1}{q}\biggl|\boldsymbol{q}\cdot\boldsymbol{v}_\mathrm{e}+\frac{q^2}{2m_{\chi}}+\omega\biggr|,\;v_{\mathrm{esc}}\biggr\} \,.
    \label{eq:v_minus}
\end{equation}
Importantly, $v\_$ sets the minimum dark matter speed needed to produce a phonon of energy $\omega$ at momentum transfer $\boldsymbol{q}$. Since the kinematic function $g(\boldsymbol{q},\omega)$ integrates the VDF above the minimum speed $v\_$, it is sensitive to the behavior of $f_\mathrm{gal}(v)$ near and above this cutoff. The shape of $f_\mathrm{gal}(v)$ is controlled by both the assumed functional form and astrophysical parameters such as $v_\mathrm{e}$ and $v_\mathrm{esc}$, so uncertainties in these inputs directly propagate into the rate through $g$, effectively reweighting the region of $\boldsymbol{q}$ space that contributes to the signal. Near the tail of the VDF, where it falls steeply with velocity, even small variations can lead to large fractional changes in $g$ and hence in the predicted rate. This sensitivity is more pronounced for smaller $m_\chi$ and larger $\omega$, which push $v\_$ closer to the high-velocity tail of the distribution, where different VDF models diverge most significantly as we will see in~\cref{sec:VDF_astro}. Therefore, accurately modeling the VDF and its uncertainties is crucial for reliable predictions of single phonon excitation rates and interpreting potential signals.

\subsection{Daily modulation}

The time dependence of the rate arises from that of the dark matter velocity distribution in the detector frame, $f_\chi(\boldsymbol{v},t)$, which in turn inherits its time dependence from the Earth's velocity $\boldsymbol{v}_\mathrm{e}(t)$ in the Galactic frame via \cref{eq:fchi_fgal}. Variations in $|\boldsymbol{v}_\mathrm{e}|$ over the year produce the well-known annual modulation~\cite{Drukier:1986tm,Freese:1987wu}. In this work, however, we focus on the daily modulation that arises from variations in the direction of $\boldsymbol{v}_\mathrm{e}$. As the Earth rotates, the orientation of a terrestrial crystal detector with respect to the dark matter wind changes over a sidereal day. For anisotropic targets whose response to an energy-momentum transfer depends on the direction of $\boldsymbol{q}$ relative to the crystal axes, this induces a periodic modulation of the rate that is absent in isotropic detectors.

Daily modulation is especially interesting because it provides a directional handle for discriminating dark matter signals from isotropic backgrounds~\cite{Griffin:2018bjn,Coskuner:2021qxo}. The amplitude and phase of the modulation are shaped by the time dependence of the kinematic function $g(\boldsymbol{q},\omega)$, which comes from the $\boldsymbol{q}\cdot\boldsymbol{v}_\mathrm{e}(t)$ term in the expression for $v\_$ in~\cref{eq:v_minus}. As a result, the velocity integral defining $g(\boldsymbol{q},\omega)$ is periodically modulated, enhancing or suppressing contributions from different regions of $\boldsymbol{q}$ space depending on their alignment with the dark matter wind. Uncertainties in the VDF therefore affect not only the total rate but also the daily modulation pattern, motivating a joint study of both observables.

To isolate the daily modulation signal, we fix $|\boldsymbol{v}_\mathrm{e}|$ and retain only the time-varying direction of $\boldsymbol{v}_\mathrm{e}(t)$ arising from the Earth's rotation. Following the benchmark detector orientation of Refs.~\cite{Griffin:2018bjn, Coskuner:2021qxo}, which is independent of geographic location, $\boldsymbol{v}_\mathrm{e}(t)$ takes the form
\begin{equation}
    \boldsymbol{v}_\mathrm{e}(t)= v_\mathrm{e}
    \begin{pmatrix}
        \sin{\theta_\mathrm{e}}  \sin{\phi(t)}\\
        \sin{\theta_\mathrm{e}} \cos{\theta_\mathrm{e}} (\cos{\phi(t)}-1)\\
        \cos^2{\theta_\mathrm{e}} + \sin^2{\theta_\mathrm{e}} \cos{\phi(t)}
    \end{pmatrix} \,,
\end{equation}
where $\theta_\mathrm{e} = 42^\circ$ is the angle between the Earth's rotation axis and the dark matter wind direction, and $\phi(t) = 2\pi t/(24\,\mathrm{hr})$. A comprehensive analysis exploring different detector locations and crystal orientations is left to future work.

\section{Velocity distribution and astrophysical inputs}
\label{sec:VDF_astro}

The dark matter scattering rate in~\cref{eq:rate} depends on the velocity distribution $f_\chi$ through both its functional form and the astrophysical parameters entering it. Previous studies of VDF uncertainties in direct detection~\cite{McCabe:2010zh,Green:2017odb,Radick:2020qip,Maity:2020wic,Li:2022acp,Herrera:2024zrk,Herrera:2026lqi} have shown that these uncertainties can significantly affect projected sensitivities for nuclear recoils and electronic excitations, particularly for lighter dark matter masses where the kinematics probe the high-velocity tail of the distribution. In this section, we describe the three benchmark VDF models used in this work---the SHM, the Tsallis distribution, and an empirical model---following the benchmark set introduced in Ref.~\cite{Radick:2020qip}. We then discuss the astrophysical velocity parameters, their observational constraints, and the prescription we adopt for comparing different VDF models.

\subsection{VDF models}

\paragraph{Standard Halo Model.}
The SHM~\cite{Drukier:1986tm} assumes dark matter particles reside in an isothermal sphere, yielding an isotropic Maxwell--Boltzmann (MB) velocity distribution truncated at the Galactic escape velocity $v_\mathrm{esc}$:
\begin{equation}
f_\mathrm{gal,SHM}(v) = \frac{1}{N_0} e^{-v^2 / v_0^2} \, \Theta(v_{\mathrm{esc}} - v) \,,
\end{equation}
where $N_0$ is a normalization constant given by
\begin{equation}
N_0 = \pi^{3/2} v_0^3 \left[ \operatorname{erf}\left( \frac{v_{\mathrm{esc}}}{v_0} \right) - \frac{2}{\sqrt{\pi}} \frac{v_{\mathrm{esc}}}{v_0} e^{-(v_{\mathrm{esc}} / v_0)^2} \right] \,.
\end{equation}
Despite its analytical convenience, the SHM has well-documented limitations. Dark matter halos formed within the $\Lambda$CDM framework are not isothermal, and cosmological simulations consistently find that the local velocity distribution deviates from Maxwellian form~\cite{Vogelsberger:2008qb,Ling:2009eh,Kuhlen:2009vh}. The SHM imposes an ad hoc sharp truncation at $v_\mathrm{esc}$, whereas simulations predict a smoother fall-off in the high-velocity tail. We note that the inclusion of baryonic physics in hydrodynamic simulations partially ameliorates these deviations, bringing the VDF closer to Maxwellian form compared with dark matter-only runs~\cite{Bozorgnia:2016ogo,Bozorgnia:2017brl,Kelso:2016qqj,Sloane:2016kyi}; nevertheless, residual departures persist, motivating the study of non-Maxwellian alternatives. 

\paragraph{Tsallis distribution.}
For a self-gravitating system of collisionless particles interacting through long-range gravity, the assumptions underlying Boltzmann--Gibbs statistics---thermal equilibrium, short-range interactions, and extensive entropy---are not expected to hold. Non-extensive statistical mechanics~\cite{Tsallis:1987eu} provides a generalized framework that accommodates such systems. The resulting velocity distribution for dark matter halos was derived in Refs.~\cite{Hansen:2004dg,Hansen:2005yj} by applying a factorization approximation to the Tsallis entropy, and its application to direct detection phenomenology was developed in Refs.~\cite{Vergados:2007nc,Ling:2009eh}. The Galactic-frame velocity distribution takes the form
\begin{equation}
f_\mathrm{gal,Tsa}(v) \propto
\begin{cases}
    \Bigl[1-(1-q)\dfrac{v^2}{v_0^2}\Bigr]^{1/(1-q)} & v<v_{\mathrm{esc}}  \\[4pt]
    0 & v\geq{v_{\mathrm{esc}}}
\end{cases}
\end{equation}
where $q$ is a shape parameter that quantifies deviations from classical Boltzmann--Gibbs statistics; the MB distribution is recovered in the limit $q \to 1$. For $q < 1$, the distribution possesses a built-in cutoff at $v_\mathrm{esc}^2 = v_0^2/(1-q)$, producing a smooth, continuous fall-off near the tail---in contrast to the SHM's sharp truncation---that is consistent with the behavior seen in $N$-body simulations~\cite{Hansen:2004dg,Vogelsberger:2008qb,Ling:2009eh}. For $q > 1$, the escape velocity must be imposed as a separate parameter. In this work, we focus on the $q < 1$ regime and vary $v_0$ and $v_\mathrm{esc}$ as independent parameters, which determine $q$ through $q = 1 - v_0^2/v_\mathrm{esc}^2$.

\paragraph{Empirical distribution.}%
An effective two-parameter model was introduced in Refs.~\cite{Mao:2012hf,Mao:2013nda} to describe the velocity distribution of dark matter halos in cosmological simulations. The Galactic-frame distribution is
\begin{equation}
f_\mathrm{gal,Emp}(v) \propto
\begin{cases}
    e^{-v/v_0}(v_{\mathrm{esc}}^2-v^2)^p & v<v_{\mathrm{esc}}  \\[4pt]
    0 & v\geq{v_{\mathrm{esc}}}
\end{cases}
\label{eq:EMP_VDF}
\end{equation}
where the exponential term captures the non-Gaussian core of the distribution, while the power-law factor $(v_\mathrm{esc}^2 - v^2)^p$ enforces a smooth cutoff at the escape velocity. The parameter $p$ primarily reflects halo-to-halo and directional variation, with the 90\% scatter across simulated halos spanning the range $p \in [1,3]$~\cite{Mao:2013nda,Radick:2020qip}. We adopt $p = 1.5$ as our fiducial value, following Refs.~\cite{Radick:2020qip,Li:2022acp}; this corresponds to the best-fit value found for the dark matter-only ErisDark simulation by Ref.~\cite{Kuhlen:2013tra}, while the baryonic Eris simulation yields $p = 2.7$.

\subsection{Astrophysical parameters}

The VDF models discussed above are specified by a set of astrophysical velocity parameters: $v_0$, $v_\mathrm{esc}$, and, when boosted to the detector frame, $v_\mathrm{e}$. In the SHM, $v_0$ is usually set to the local circular velocity $v_\mathrm{c}$, which is the velocity of a test particle on a circular orbit at the Sun's Galactocentric radius. We will discuss prescriptions for setting $v_0$ in the Tsallis and empirical models given $v_\mathrm{c}$ in \cref{sec:prescriptions}.

In this subsection, we discuss the observational uncertainties in $v_\mathrm{c}$, $v_\mathrm{esc}$, and $v_\mathrm{e}$, which propagate into the predicted scattering rate. To account for the range of values found in the literature, we consider two sets of velocity parameters: a broader (``conservative'') set and a narrower (``aggressive'') set, similar to Ref.~\cite{Radick:2020qip}. In the conservative set, we choose widely adopted parameters as the benchmark and remain agnostic about the time of year at which the signal occurs. The aggressive choice incorporates more recently determined parameters. These are summarized in~\cref{tab:Halo_Parameters} and discussed in the following paragraphs. Unless stated otherwise, results in the main text use the conservative ranges, with the aggressive choices explored in \cref{App_C}. 

\begin{table}[t]
\centering
\begin{tabular}{l|c|r}
\hline
Halo Parameter & Conservative & Aggressive\\
\hline
$v_\mathrm{c}~[\mathrm{km}\,\mathrm{s}^{-1}]$ & $220_{-20}^{+60}$ & $238\pm1.5$\\
$v_{\mathrm{esc}}~[\mathrm{km}\,\mathrm{s}^{-1}]$  & $544_{-94}^{+56}$ & $528_{-25}^{+24}$\\
$v_\mathrm{e}~[\mathrm{km}\,\mathrm{s}^{-1}]$ & $232\pm15$ & $250.6\pm1.4$\\
\hline
\end{tabular}
\caption{Uncertainty ranges of halo parameters used in this study.
\label{tab:Halo_Parameters}}
\end{table}

\paragraph{Circular velocity.}
The legacy benchmark $v_\mathrm{c} = 220~\mathrm{km\,s^{-1}}$ traces to the IAU recommendation of Ref.~\cite{Kerr:1986cei}, which consolidated kinematic measurements spanning a broad interval (${\sim}\,200$--$280~\mathrm{km\,s^{-1}}$). This range is commonly parameterized in the direct detection literature as $v_\mathrm{c} = 220_{-20}^{+60}~\mathrm{km\,s^{-1}}$~\cite{Radick:2020qip}, which we adopt as the conservative choice. A tighter determination of $v_\mathrm{c}$ can be obtained from the apparent proper motion of Sagittarius~$A^*$, the supermassive black hole at the Galactic center. The proper motion measurement of Ref.~\cite{Reid_2004} yields the Sun's total angular velocity in the Galactic plane as $\Omega_\odot = 30.24 \pm 0.12~\mathrm{km\,s^{-1}\,kpc^{-1}}$~\cite{Baxter:2021pqo}. Combining this with the GRAVITY measurement of the Galactocentric distance, $R_0 = 8275 \pm 9\,(\mathrm{stat.}) \pm 33\,(\mathrm{syst.})~\mathrm{pc}$~\cite{Abuter:2021yys}, gives the Sun's tangential speed relative to the Galactic center, $v_\mathrm{tan} = \Omega_\odot \times R_0 = 250.2 \pm 1.4~\mathrm{km\,s^{-1}}$. This tangential speed includes both the circular motion and the tangential component of the Sun's peculiar velocity $\boldsymbol{v}_{\ostar}$. Using the peculiar velocity measurement of Ref.~\cite{Sch_nrich_2010}, $\boldsymbol{v}_{\ostar} = (U, V, W)_{\ostar} = (11.1_{-0.75}^{+0.69},\, 12.24_{-0.47}^{+0.47},\, 7.25_{-0.36}^{+0.37})~\mathrm{km\,s^{-1}}$, one obtains $v_\mathrm{c} = v_\mathrm{tan} - V_{\ostar} = 238.0 \pm 1.5~\mathrm{km\,s^{-1}}$~\cite{Baxter:2021pqo}, which we adopt as the aggressive choice.

\paragraph{Escape velocity.}
The value of $v_\mathrm{esc}$ is inferred from the high-velocity tail of the stellar halo distribution, typically modeled with a power-law ansatz truncated at $v_\mathrm{esc}$~\cite{Smith:2006ym}. The RAVE survey analysis of Ref.~\cite{Smith:2006ym} reported a median likelihood estimate of $v_\mathrm{esc} = 544~\mathrm{km\,s^{-1}}$ with a 90\% confidence interval of $498$--$608~\mathrm{km\,s^{-1}}$ (i.e., $544_{-46}^{+64}~\mathrm{km\,s^{-1}}$). A more recent determination using Gaia DR2 data with self-consistent Milky Way mass modeling found $v_\mathrm{esc} = 528_{-25}^{+24}~\mathrm{km\,s^{-1}}$~\cite{Deason_2019}; accounting for kinematic substructure in the high-velocity tail can further refine such estimates~\cite{Necib:2021yhq,Necib:2021vxr}. For our conservative benchmark, we adopt $v_\mathrm{esc} = 544_{-94}^{+56}~\mathrm{km\,s^{-1}}$ (corresponding to a range of 450--600~$\mathrm{km\,s^{-1}}$), following the envelope of Ref.~\cite{Radick:2020qip} which encompasses Ref.~\cite{Smith:2006ym} and subsequent measurements. This is deliberately broader than any single measurement's confidence interval, reflecting remaining systematic uncertainties in the stellar velocity anisotropy, the Galactic potential, and the selection of tracer populations. The aggressive choice, $v_\mathrm{esc} = 528_{-25}^{+24}~\mathrm{km\,s^{-1}}$, adopts the Gaia-era determination of Ref.~\cite{Deason_2019}.

\paragraph{Earth's velocity.}
The Earth's velocity in the Galactic frame receives contributions from three sources, $\boldsymbol{v}_\mathrm{e} = \boldsymbol{v}_\mathrm{c} + \boldsymbol{v}_{\ostar} + \boldsymbol{v}_\oplus$, where $\boldsymbol{v}_\oplus$ is the Earth's orbital velocity around the Sun. For the conservative estimate, we take the benchmark $v_\mathrm{e} = 232 \pm 15~\mathrm{km\,s^{-1}}$ from Ref.~\cite{Radick:2020qip}, which accounts for the variation over the course of the year. For the aggressive estimate, we compute $v_\mathrm{e}$ from the same inputs used to derive the aggressive $v_\mathrm{c}$ while neglecting the Earth's orbital velocity, finding $v_\mathrm{e} = \sqrt{v_\mathrm{tan}^2 + U_{\ostar}^2+ W_{\ostar}^2} = 250.6 \pm 1.4~\mathrm{km\,s^{-1}}$~\cite{Baxter:2021pqo}.

\vspace{8pt}
The three velocity parameters $v_\mathrm{c}$, $v_\mathrm{esc}$, and $v_\mathrm{e}$ are not independent: determinations of $v_\mathrm{e}$ and $v_\mathrm{c}$ are correlated, and $v_\mathrm{esc}$ and $v_\mathrm{c}$ are both tied to the Galactic potential. Nevertheless, in this work we vary each parameter independently within its allowed range to map out the full envelope of VDF uncertainty. This procedure is conservative by construction---it spans a larger region of parameter space than a joint variation accounting for possible correlations would allow. Recent work by Folsom et al.~\cite{Folsom:2025lly}, using the TNG50 cosmological simulation to extract correlated halo parameters, finds that the effective astrophysical uncertainty in direct detection rates is smaller than what independent variation suggests. Our independent variation approach therefore provides an upper bound on the astrophysical uncertainty.

\subsection{$v_0$ prescriptions and VDF comparison}
\label{sec:prescriptions}

The parameter $v_0$ sets the overall scale of the dark matter velocity in all three halo models. In the SHM, $v_0$ is the most probable speed of the truncated Maxwell--Boltzmann distribution, and is commonly identified with the local circular velocity $v_\mathrm{c}$. For the Tsallis and empirical models, previous analyses~\cite{Green:2017odb,Radick:2020qip,Maity:2020wic,Li:2022acp} have also identified $v_0$ with $v_\mathrm{c}$; we refer to this as the {\it standard prescription}.

As an alternative, we advocate an {\it rms-matching prescription} that places different models on a more equal dynamical footing. For a given set of astrophysical parameters $(v_\mathrm{c}, v_\mathrm{esc})$ defining the SHM, we determine $v_{0,\mathrm{Tsa}}$ and $v_{0,\mathrm{Emp}}$ for the Tsallis and empirical distributions by requiring that the rms velocity of each model match that of the SHM:
\begin{equation}
\langle v^2 \rangle_\mathrm{SHM}(v_0 = v_\mathrm{c}, v_\mathrm{esc}) =\langle v^2 \rangle_\mathrm{Tsa}(v_{0,\mathrm{Tsa}}, v_\mathrm{esc}) = \langle v^2 \rangle_\mathrm{Emp}(v_{0,\mathrm{Emp}}, v_\mathrm{esc}) \,.
\label{eq:rms_matching}
\end{equation}
This ensures that the average kinetic energy of the dark matter population at the solar position is the same across all models, so that any difference in predicted rates can be attributed to the shape of the distribution rather than to a mismatch in its overall energy scale. This approach is operationally similar to the dispersion-matching procedure used in Ref.~\cite{McCabe:2010zh}.

\begin{figure}[t]
\centering
\includegraphics[width=0.75\linewidth]{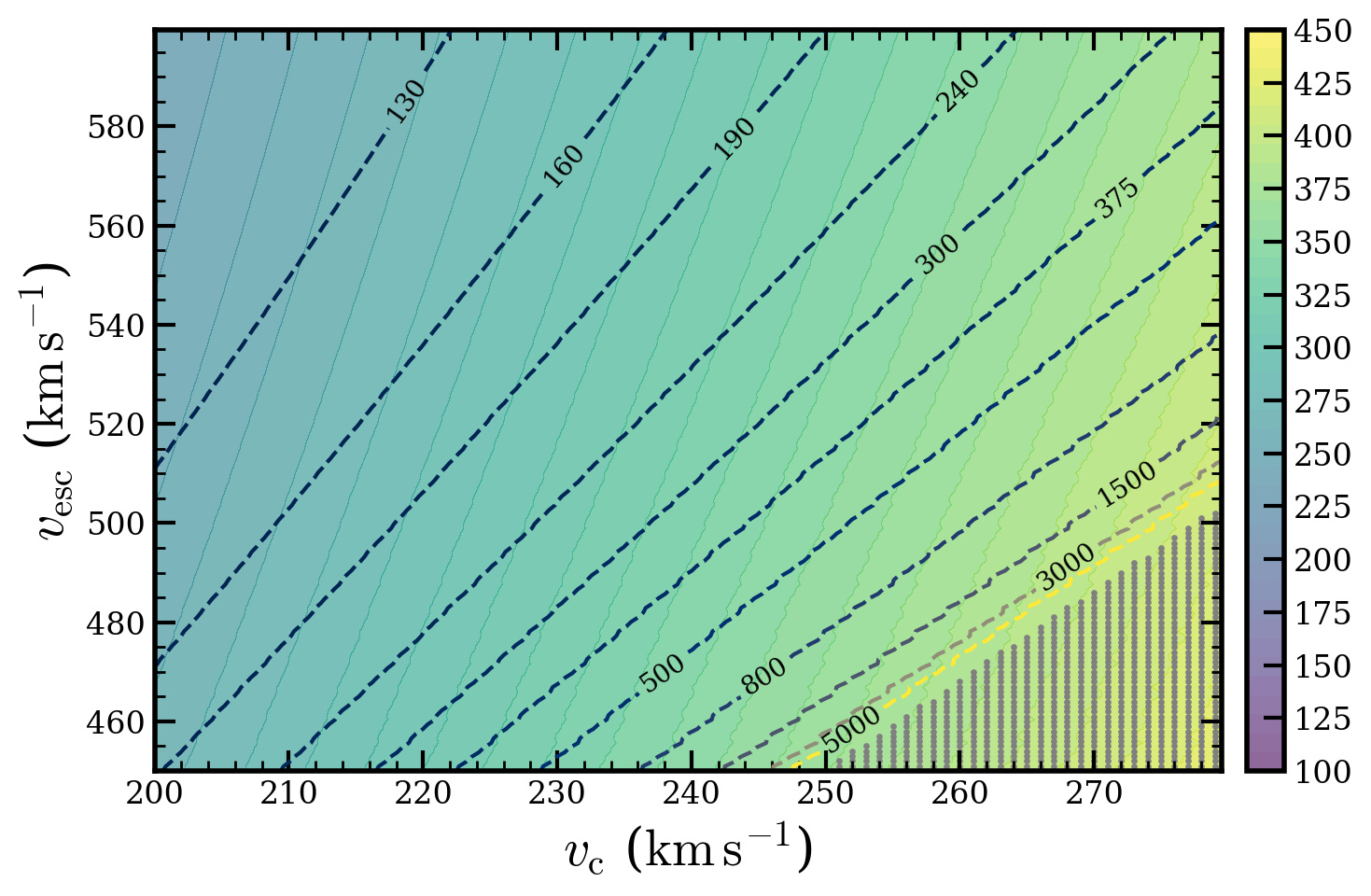}
\caption{Contour plot of $v_0$ in the Tsallis and empirical models obtained using the rms-matching prescription \cref{eq:rms_matching} as a function of the local circular velocity $v_\mathrm{c}$ and escape velocity $v_{\mathrm{esc}}$. Color-filled contours correspond to the Tsallis $v_{0,\mathrm{Tsa}}$, while dashed contours show the empirical distribution $v_{0,\mathrm{Emp}}$. The hatched region in the bottom-right corner indicates parameter combinations for which the rms-matching procedure does not yield a finite solution for $v_{0,\mathrm{Emp}}$, in which case we set $v_{0,\mathrm{Emp}}\rightarrow\infty$ in our analysis.
\label{fig:rms_scaling}}
\end{figure}

\cref{fig:rms_scaling} shows contours of $v_{0,\mathrm{Tsa}}$ and $v_{0,\mathrm{Emp}}$ obtained from the rms-matching condition \cref{eq:rms_matching}. 
For the Tsallis distribution, we find solutions across the full range of conservative halo parameters. For the empirical model, certain combinations of $(v_\mathrm{c}, v_\mathrm{esc})$ admit no finite solution for $v_{0,\mathrm{Emp}}$. This can be traced to the functional form of~\cref{eq:EMP_VDF}: the exponential core and power-law cutoff together suppress the high-velocity tail more strongly than the MB form, so matching the SHM's rms velocity requires progressively larger $v_{0,\mathrm{Emp}}$. In the low-$v_\mathrm{esc}$, high-$v_\mathrm{c}$ region (hatched bottom-right corner in \cref{fig:rms_scaling}), the rms-matching condition \cref{eq:rms_matching} cannot be satisfied at any finite $v_{0,\mathrm{Emp}}$. In such cases, we adopt the limit $v_{0,\mathrm{Emp}} \to \infty$ (so $f \propto (v_\mathrm{esc}^2 - v^2)^p$) in our analysis. 

\begin{figure}[ht]
\centering
\subfloat[]{%
    \includegraphics[width=0.455\textwidth]{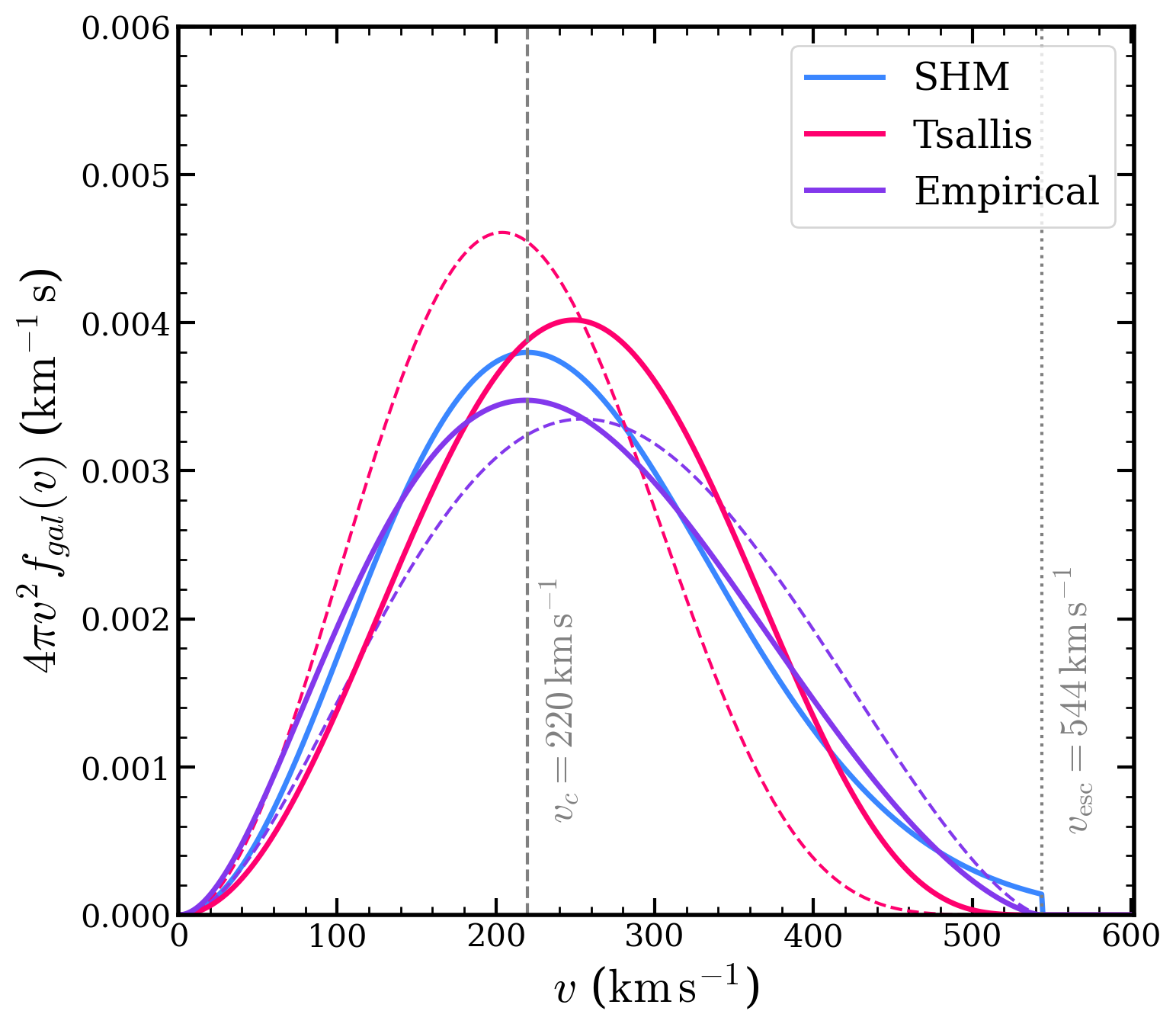}
}
\subfloat[]{%
    \includegraphics[width=0.26\textwidth]{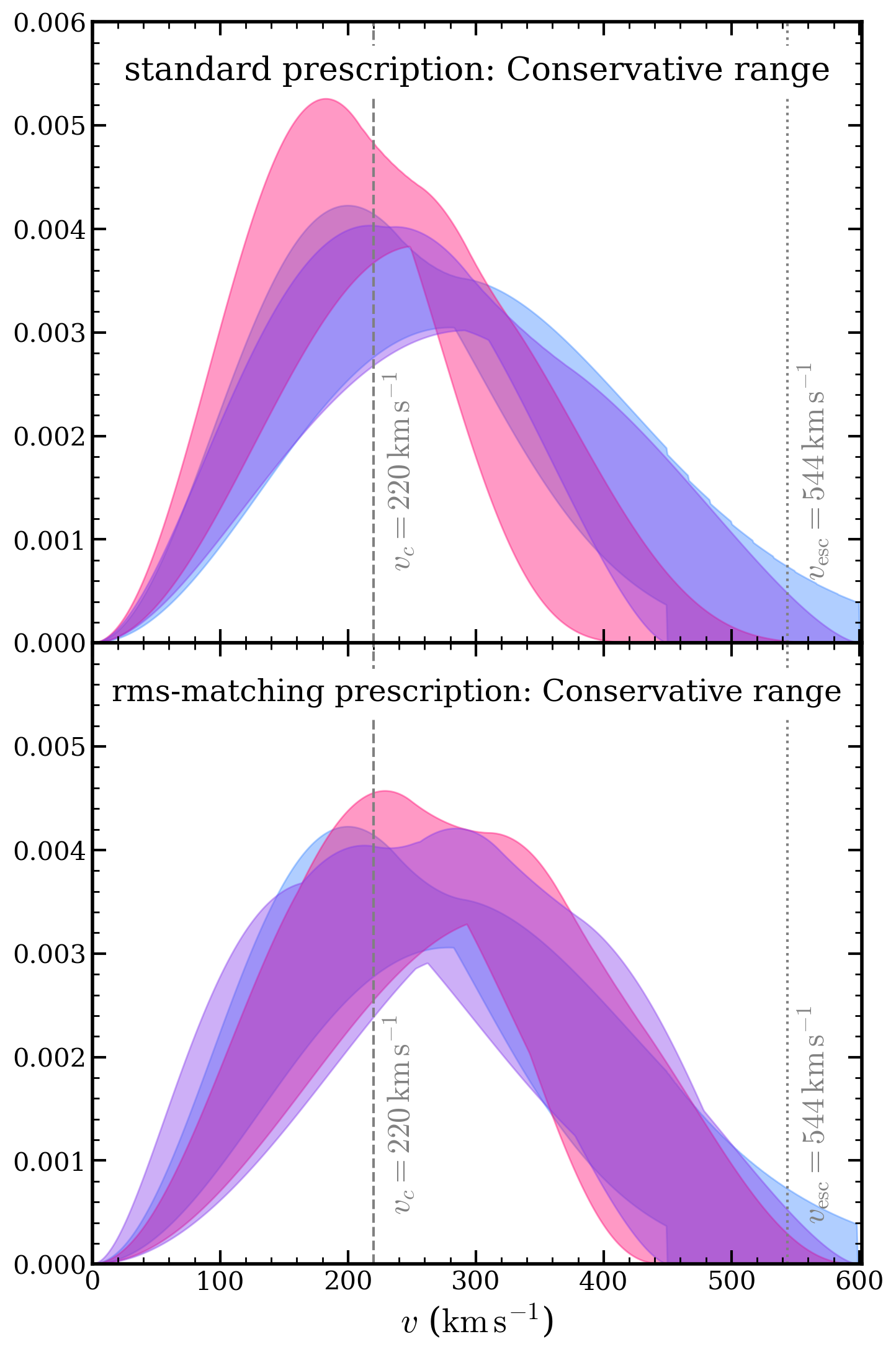}
    \label{fig:VDF_bands_conc}
}
\subfloat[]{%
    \includegraphics[width=0.26\textwidth]{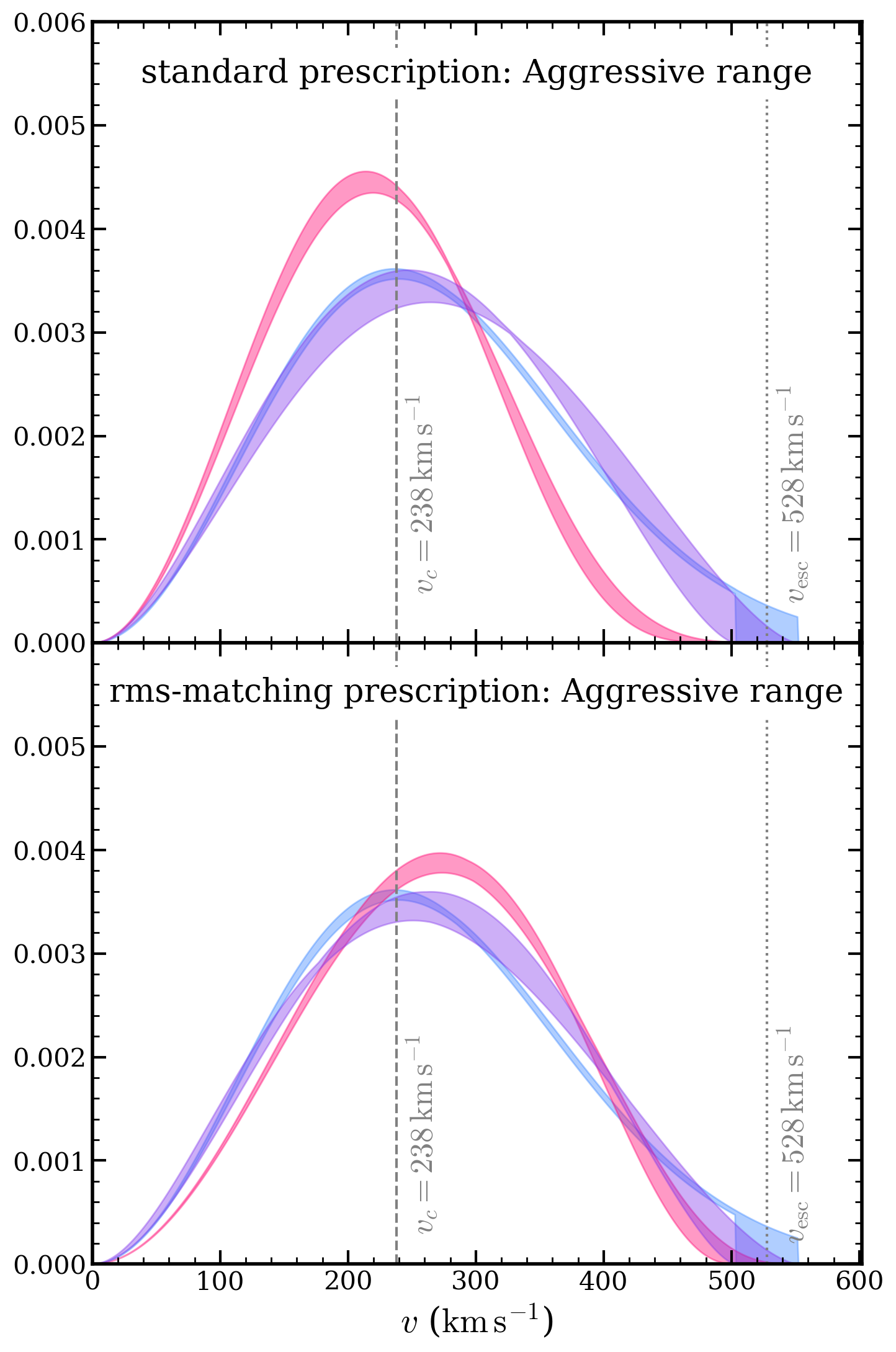}
    \label{fig:VDF_bands_aggr}
}
\caption{Speed distribution functions for different halo models and $v_0$ prescriptions, when the velocity parameters are (a) set to their central values in the conservative case, and (b,\,c) varied independently within their allowed ranges for conservative and aggressive parameter choices given in \cref{tab:Halo_Parameters}. In panel~(a), the solid (dashed) curves correspond to the rms-matching (standard) prescription.
\label{fig:VDF_conservative}}
\end{figure}

In \cref{fig:VDF_conservative} we compare the speed distribution $4\pi v^2 f_\mathrm{gal}(v)$ across the three halo models, for different parameter choices and prescriptions. Panel~(a) compares the three models at the central values of the velocity parameters $(v_\mathrm{c}, v_\mathrm{e}, v_\mathrm{esc}) = (220, 232, 544)~\mathrm{km\,s^{-1}}$ under the rms-matching (solid) and standard (dashed) prescriptions. Panels~(b) and~(c) show the spread when the velocity parameters are varied within their allowed ranges, for the conservative and aggressive parameter sets, respectively. Restricting the parameter ranges from the conservative to the aggressive set narrows the uncertainty bands, as expected. More notably, the rms-matching prescription substantially reduces the model-to-model spread, with the bands from different models largely overlapping, indicating that differences arising purely from the functional form are significantly mitigated once the overall energy scale is equalized.

For the results presented in~\cref{sec:randd}, we adopt the rms-matching prescription together with the conservative parameter ranges. This choice ensures a consistent comparison across halo models while retaining a conservative estimate of astrophysical uncertainties. For completeness, we present results obtained using the standard prescription and aggressive parameter choices in~\cref{App_B,App_C}.

\section{Results and discussion}
\label{sec:randd}

In this section, we present results on the projected reach and daily modulation signals for dark matter--single phonon scattering in four benchmark target materials---$\mathrm{GaAs}$, $\mathrm{Al}_2\mathrm{O}_3$, $\mathrm{SiO}_2$, and $\mathrm{CaWO}_4$---and discuss the impact of halo model assumptions and astrophysical parameter uncertainties. Our calculations are performed using the \texttt{PhonoDark} code~\cite{Trickle:2020oki}, which implements the formalism reviewed in \cref{sec:formalism}. We have adjusted the code to incorporate the different VDF models discussed in \cref{sec:VDF_astro}.

\subsection{Projected reach}

\begin{figure}[t]
\centering

\subfloat[$\mathrm{CaWO}_4$]{%
    \includegraphics[width=0.49\textwidth]{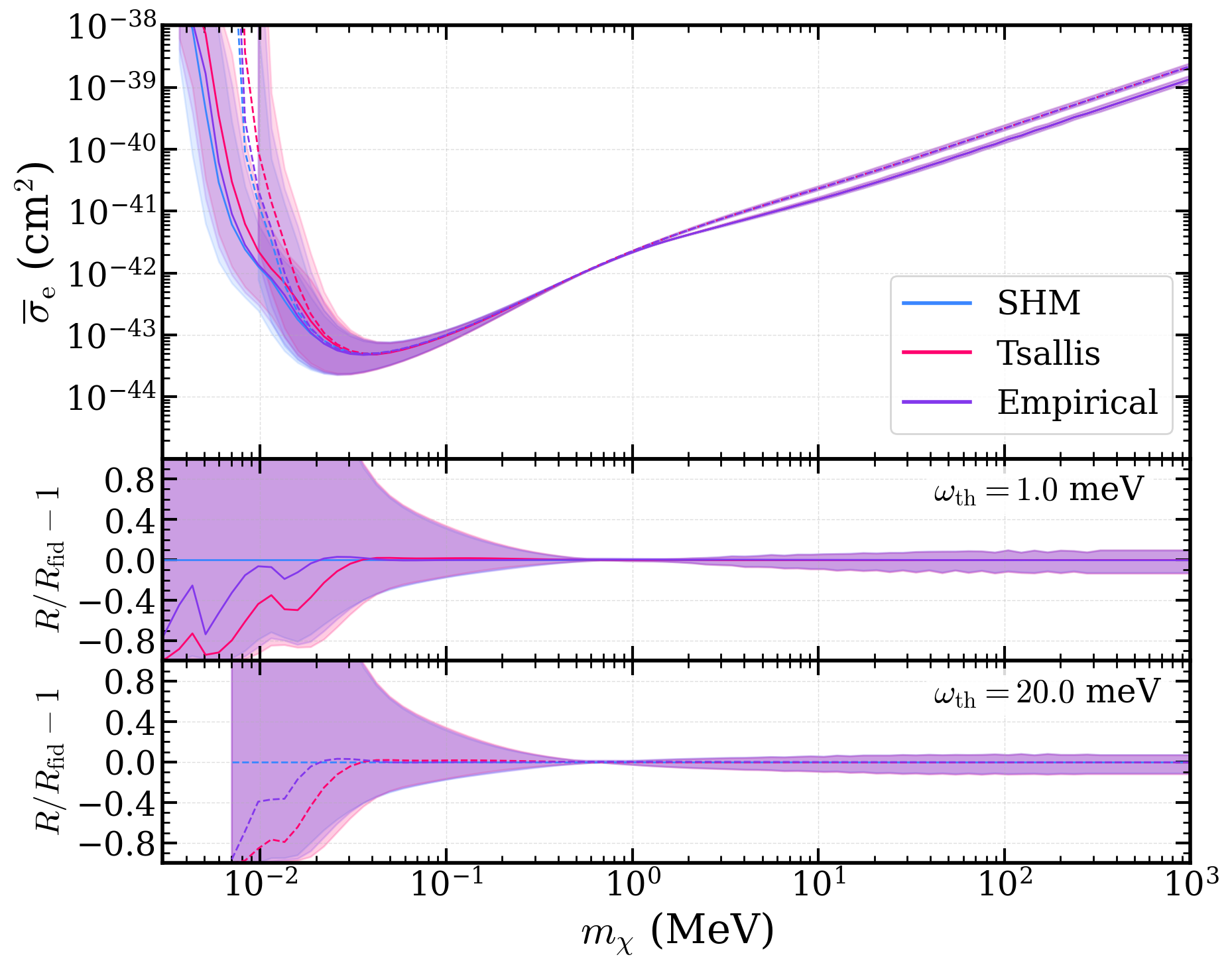}
}
\hfill
\subfloat[$\mathrm{SiO}_2$]{%
    \includegraphics[width=0.49\textwidth]{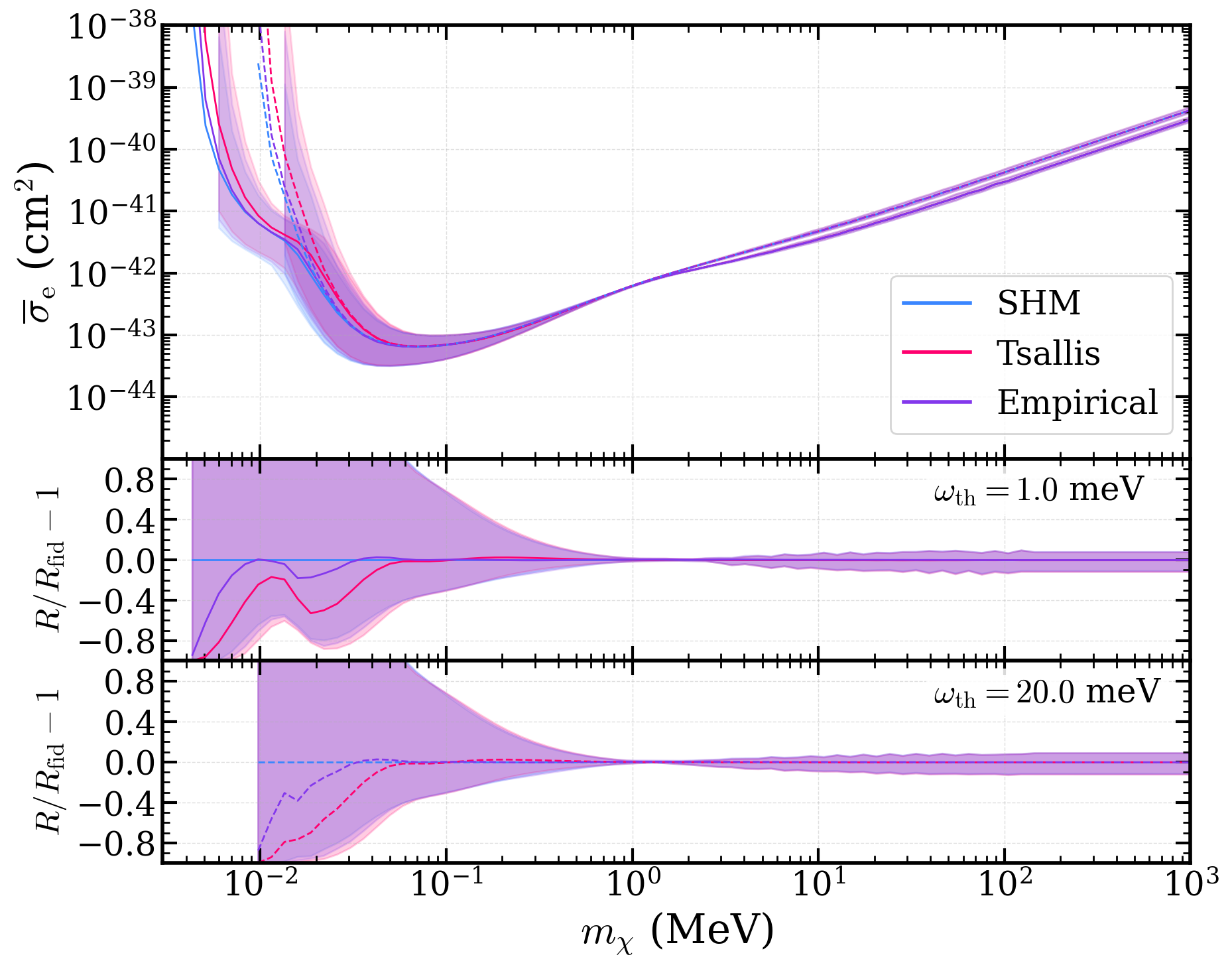}
}

\subfloat[$\mathrm{Al}_2\mathrm{O}_3$]{%
    \includegraphics[width=0.49\textwidth]{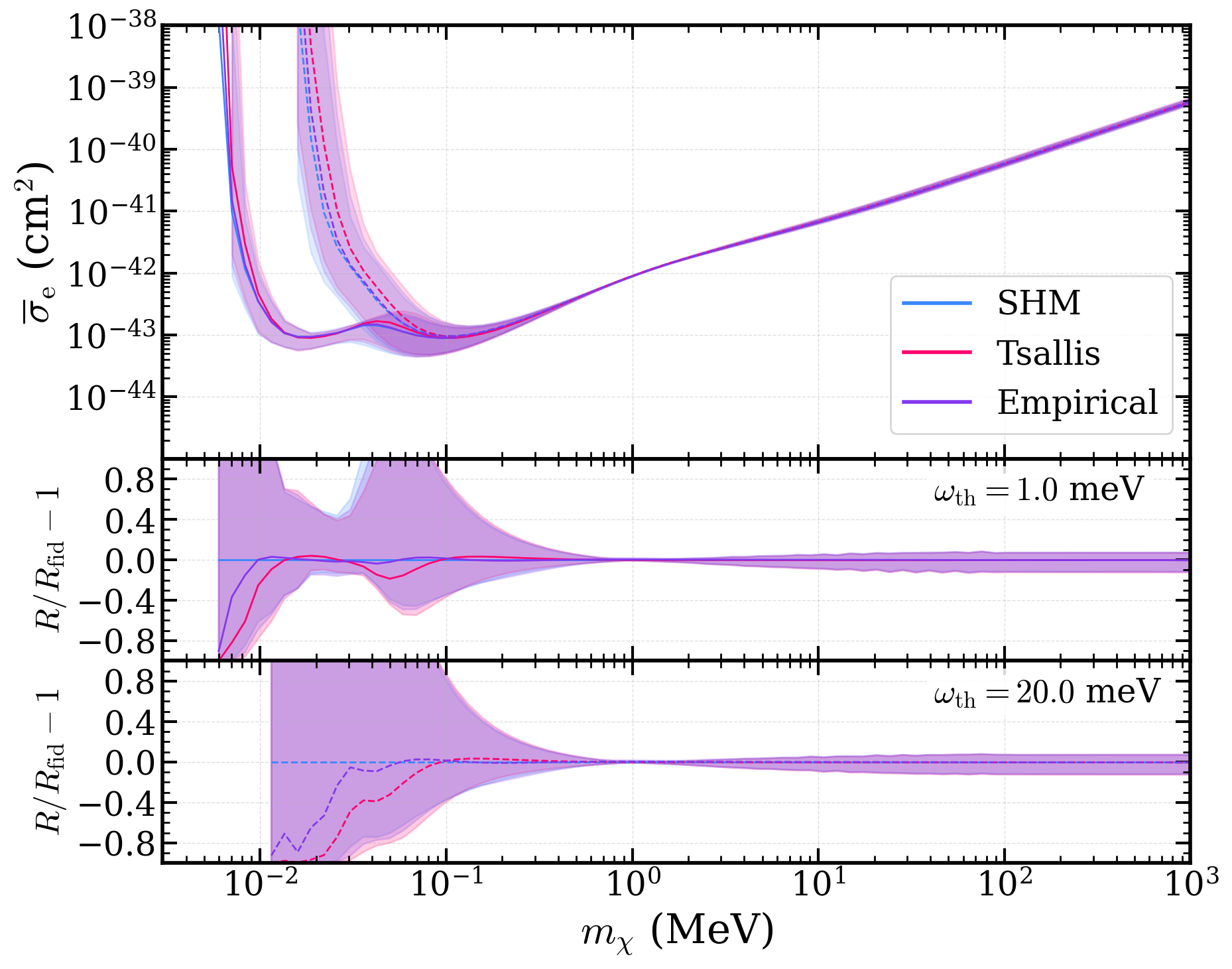}
}
\hfill
\subfloat[$\mathrm{GaAs}$]{%
    \includegraphics[width=0.49\textwidth]{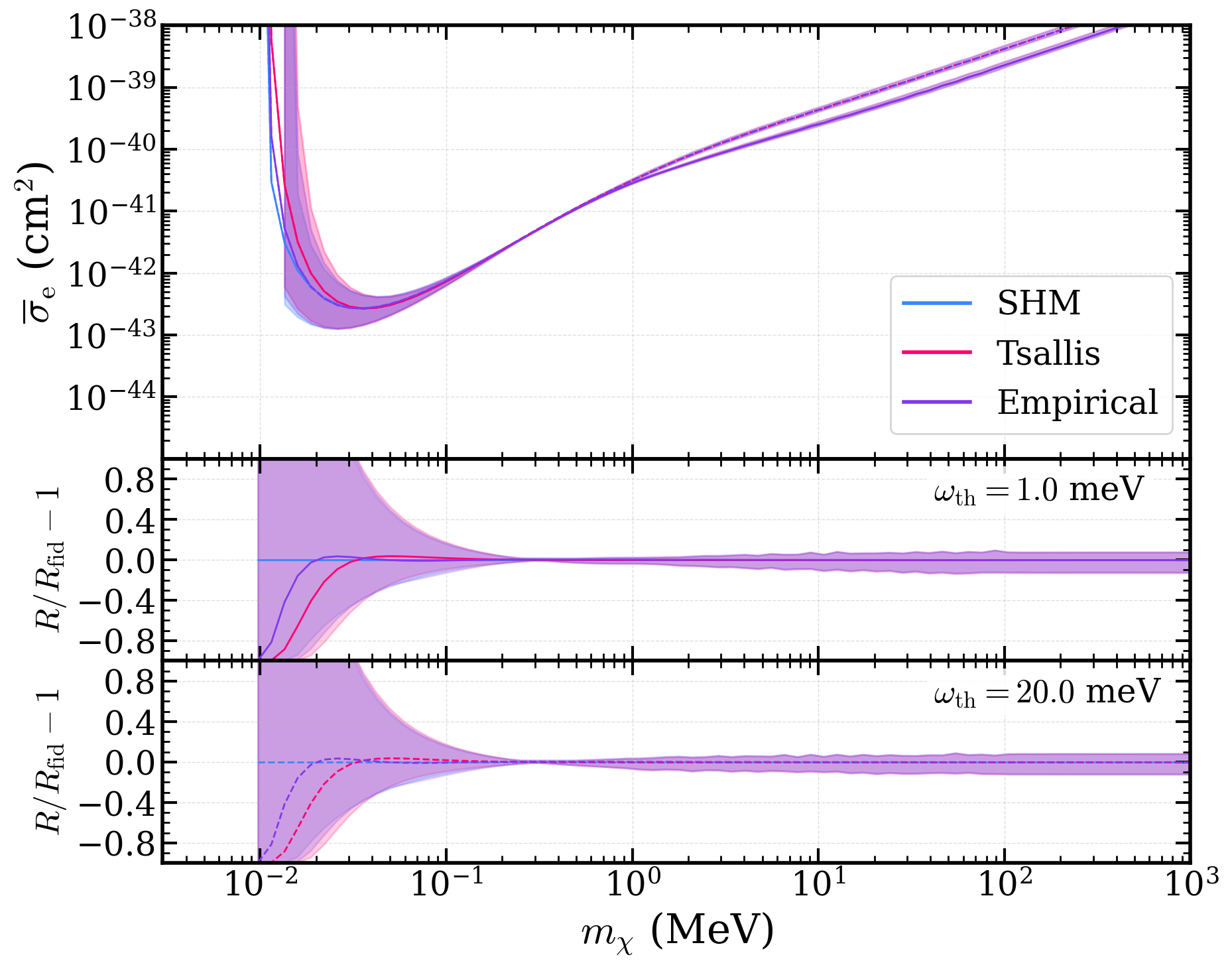}
}

\caption{Projected reach with uncertainty bands under different halo model assumptions (SHM, Tsallis, empirical) for light dark photon mediated scattering. Solid (dashed) curves correspond to central values of the velocity parameters and $\omega_\mathrm{min}=1~\mathrm{meV}$ ($20~\mathrm{meV}$). Shaded bands represent the uncertainties from varying the velocity parameters within the conservative ranges listed in~\cref{tab:Halo_Parameters}; bands for all three halo models are shown but largely overlap. Lower panels show the fractional deviation of the rate with respect to the SHM prediction at the central values of the velocity parameters.
\label{fig:LDP_reach_RMS_QQQ}}
\end{figure}

\begin{figure}[t]
\centering

\subfloat[$\mathrm{CaWO}_4$]{%
    \includegraphics[width=0.49\textwidth]{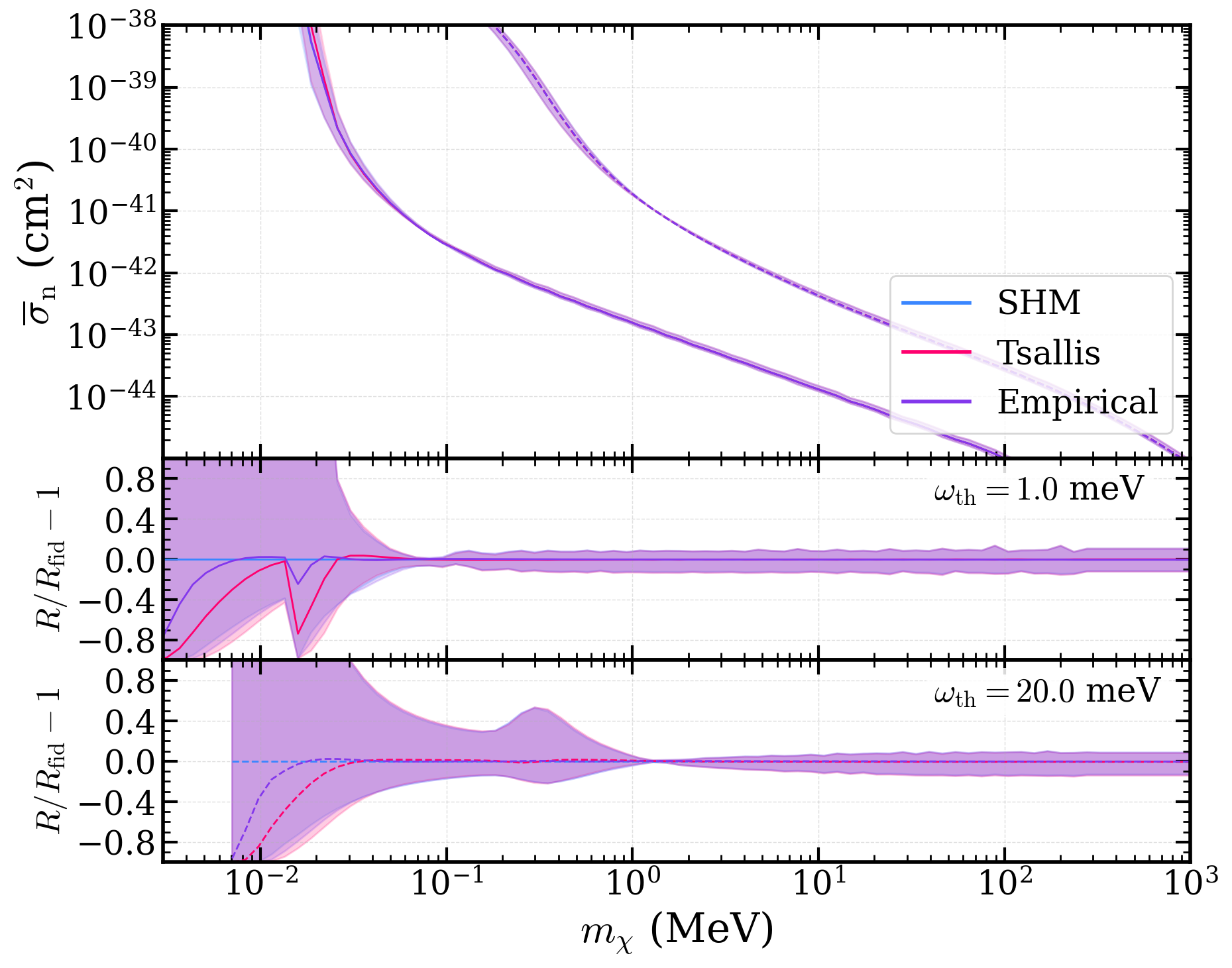}
}
\hfill
\subfloat[$\mathrm{SiO}_2$]{%
    \includegraphics[width=0.49\textwidth]{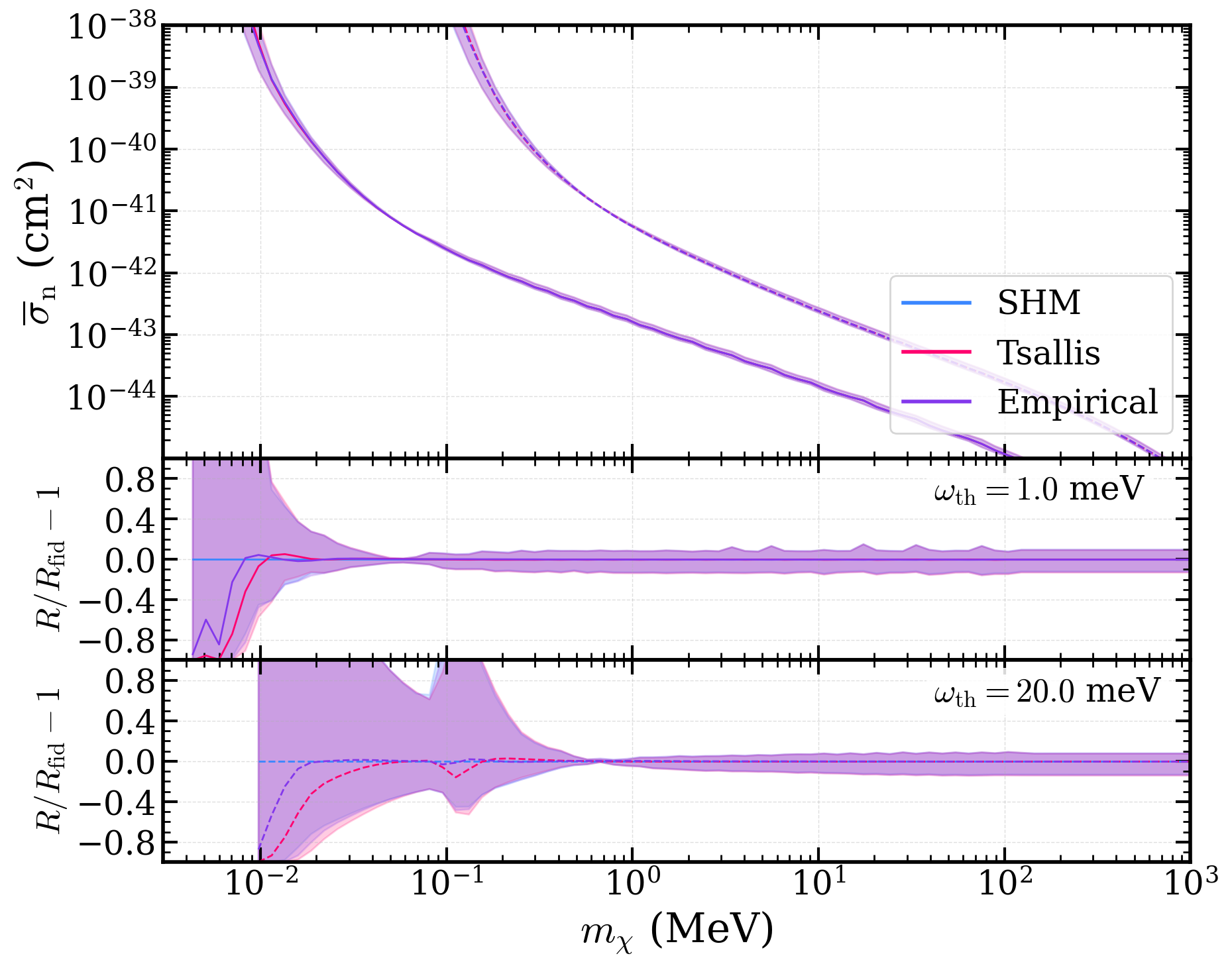}
}

\subfloat[$\mathrm{Al}_2\mathrm{O}_3$]{%
    \includegraphics[width=0.49\textwidth]{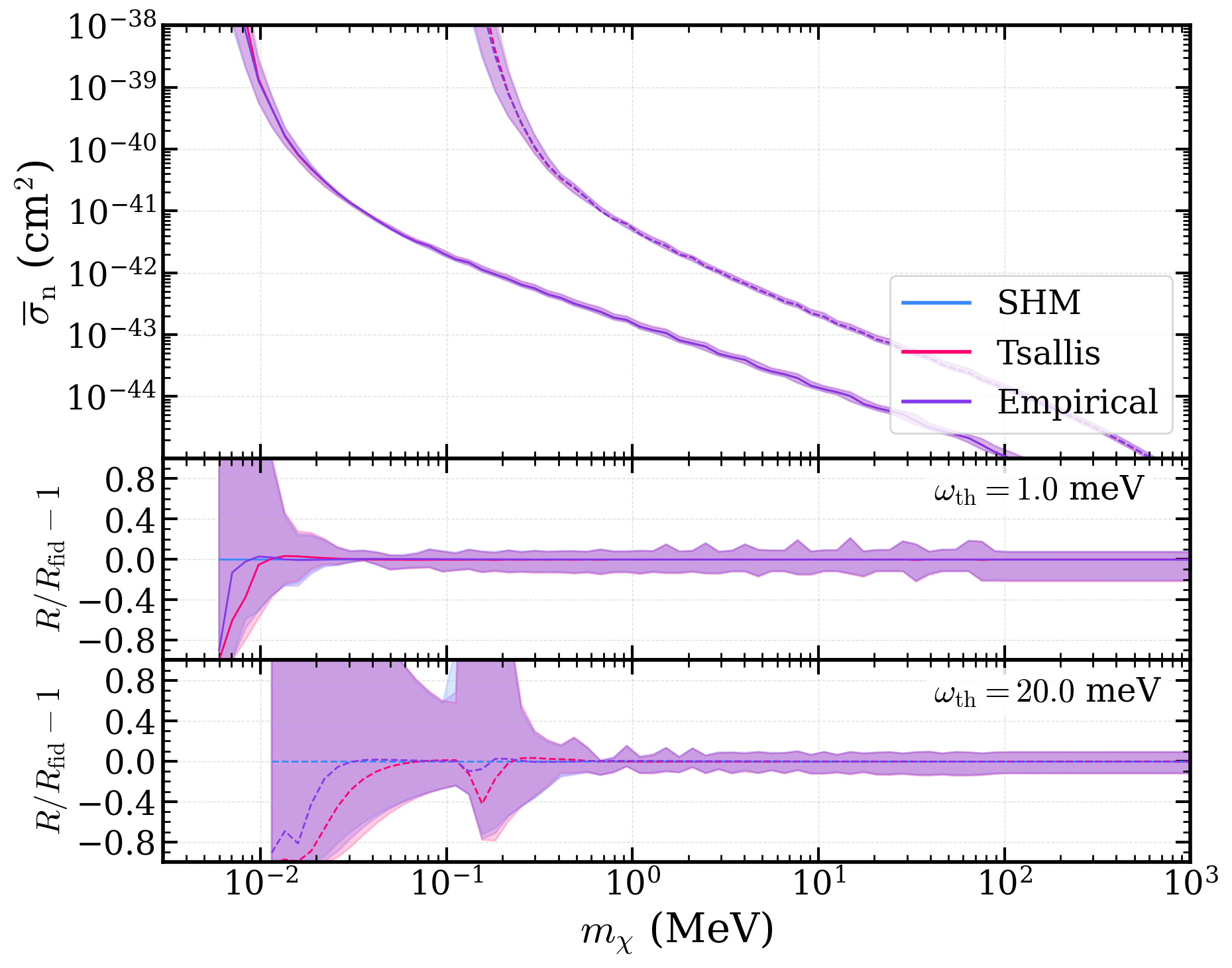}
}
\hfill
\subfloat[$\mathrm{GaAs}$]{%
    \includegraphics[width=0.49\textwidth]{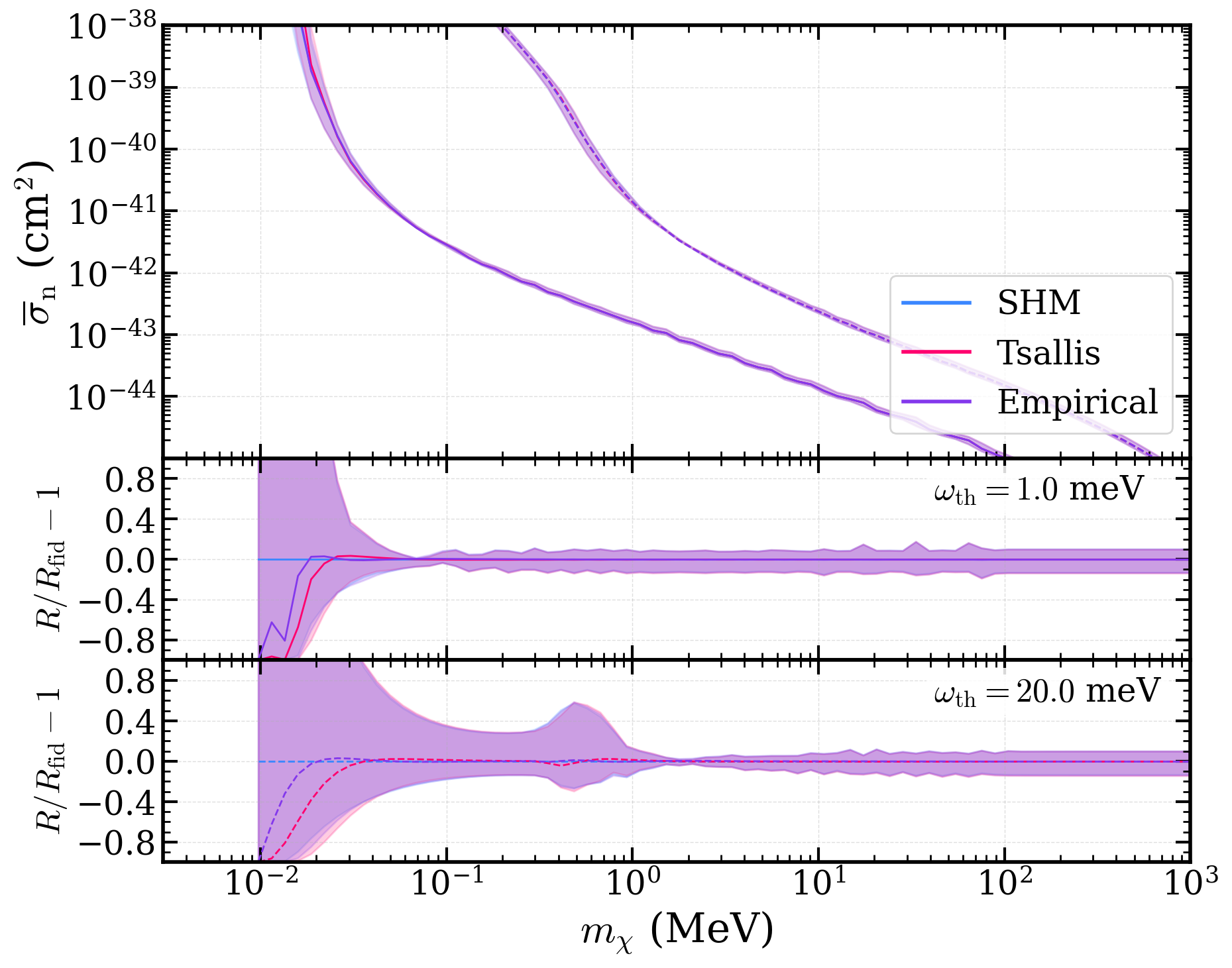}
}

\caption{Same as~\cref{fig:LDP_reach_RMS_QQQ}, for light hadrophilic scalar mediated scattering.
\label{fig:LM_reach_RMS_QQQ}}
\end{figure}

\begin{figure}[t]
\centering

\subfloat[$\mathrm{CaWO}_4$]{%
    \includegraphics[width=0.49\textwidth]{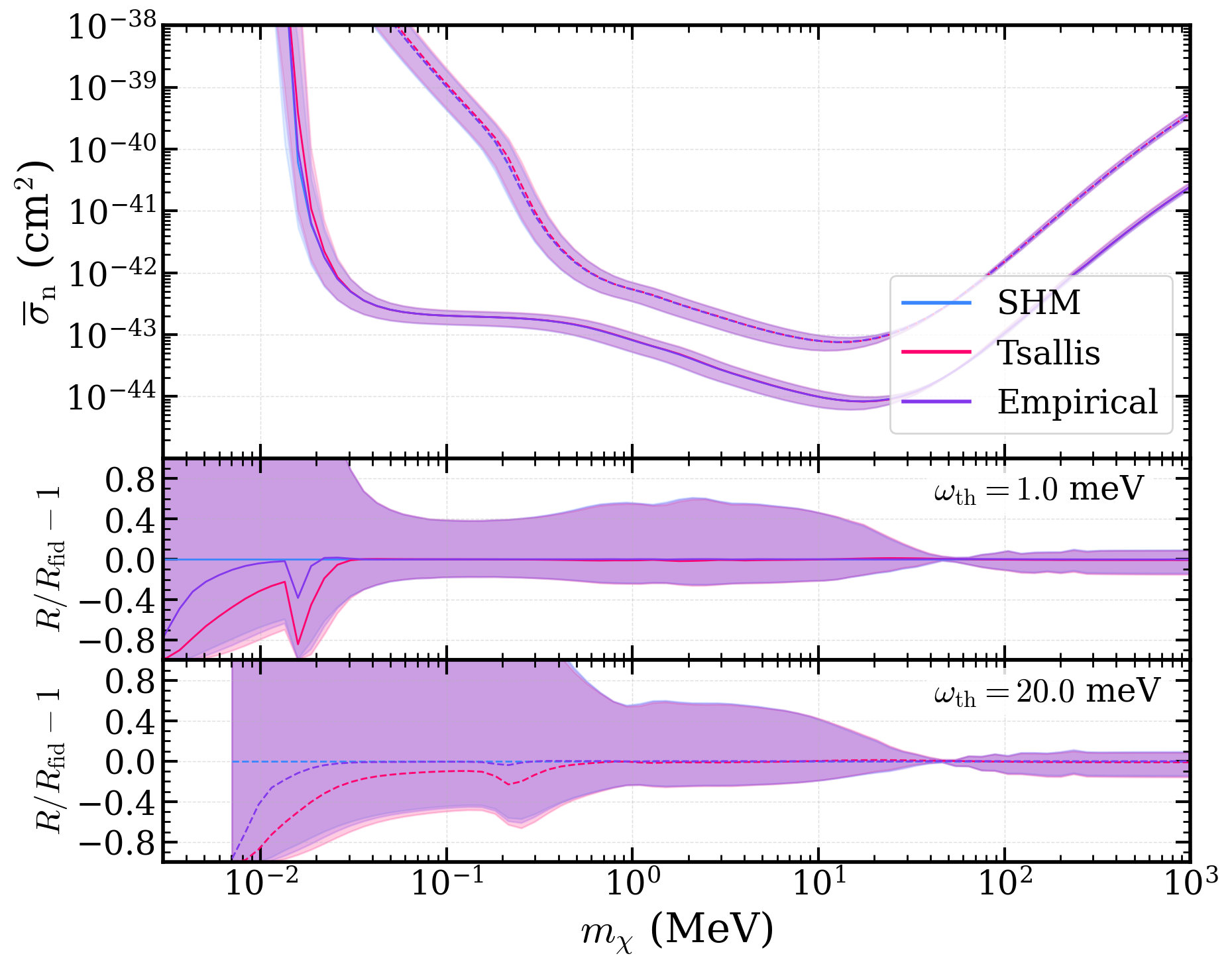}
}
\hfill
\subfloat[$\mathrm{SiO}_2$]{%
    \includegraphics[width=0.49\textwidth]{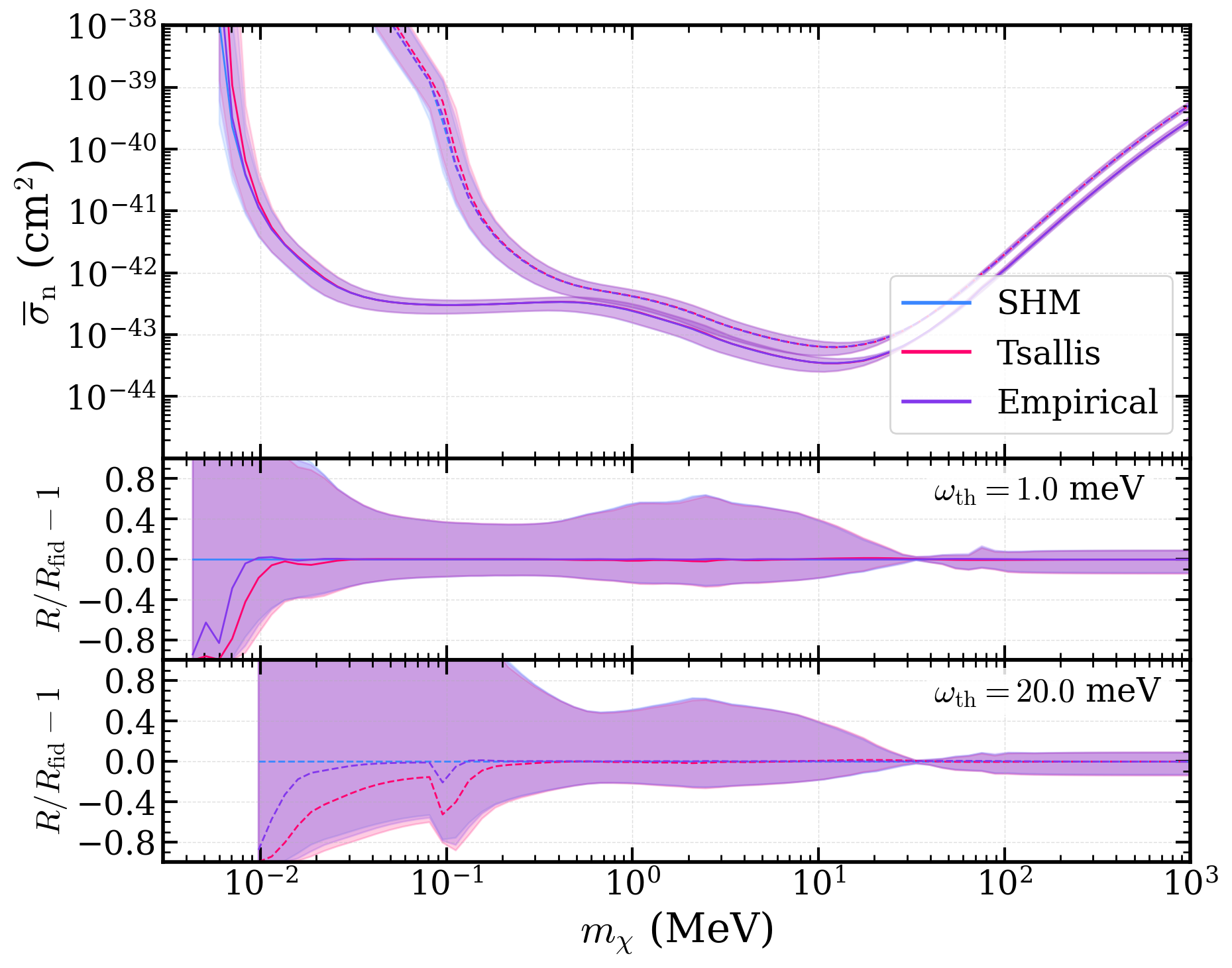}
}

\subfloat[$\mathrm{Al}_2\mathrm{O}_3$]{%
    \includegraphics[width=0.49\textwidth]{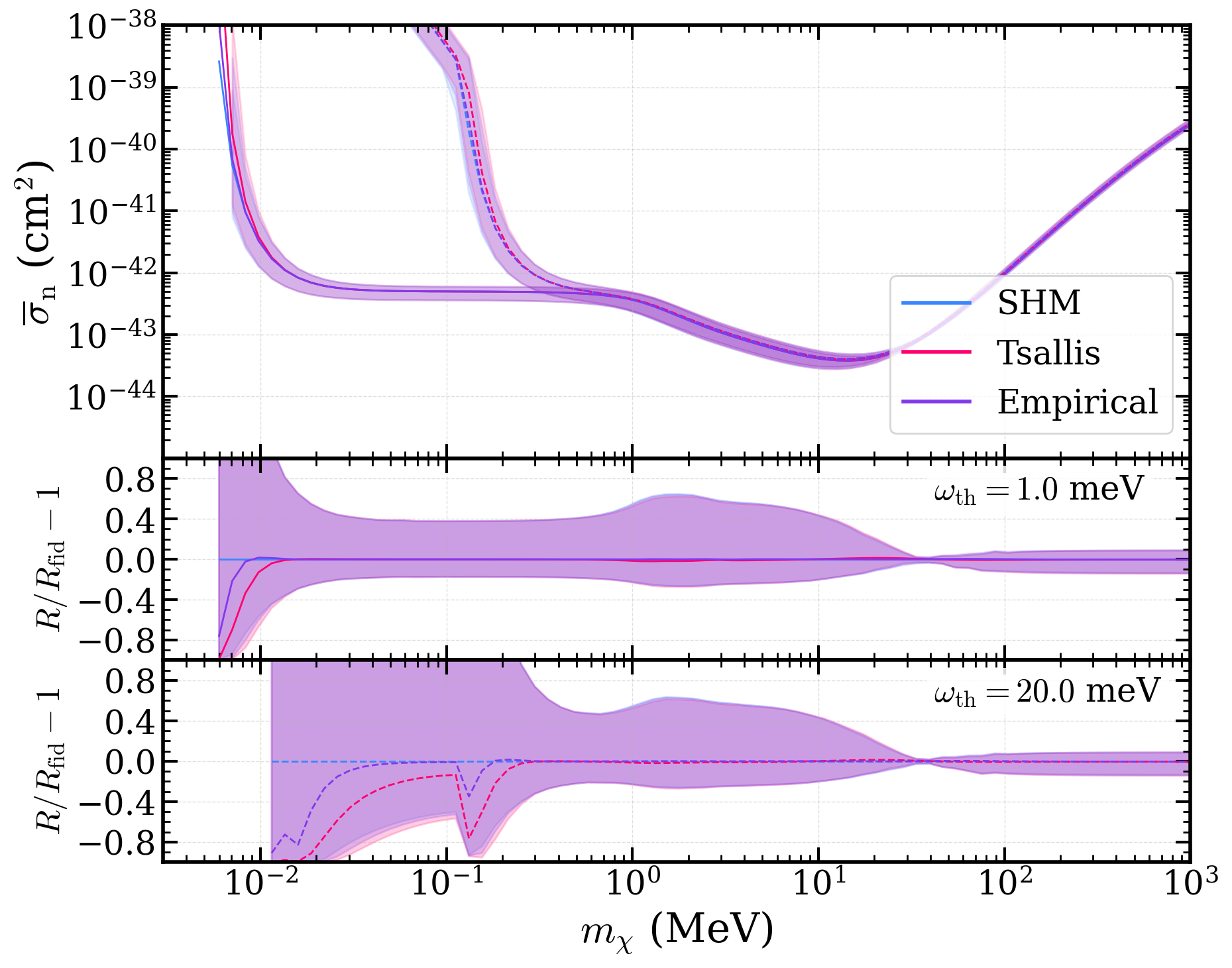}
}
\hfill
\subfloat[$\mathrm{GaAs}$]{%
    \includegraphics[width=0.49\textwidth]{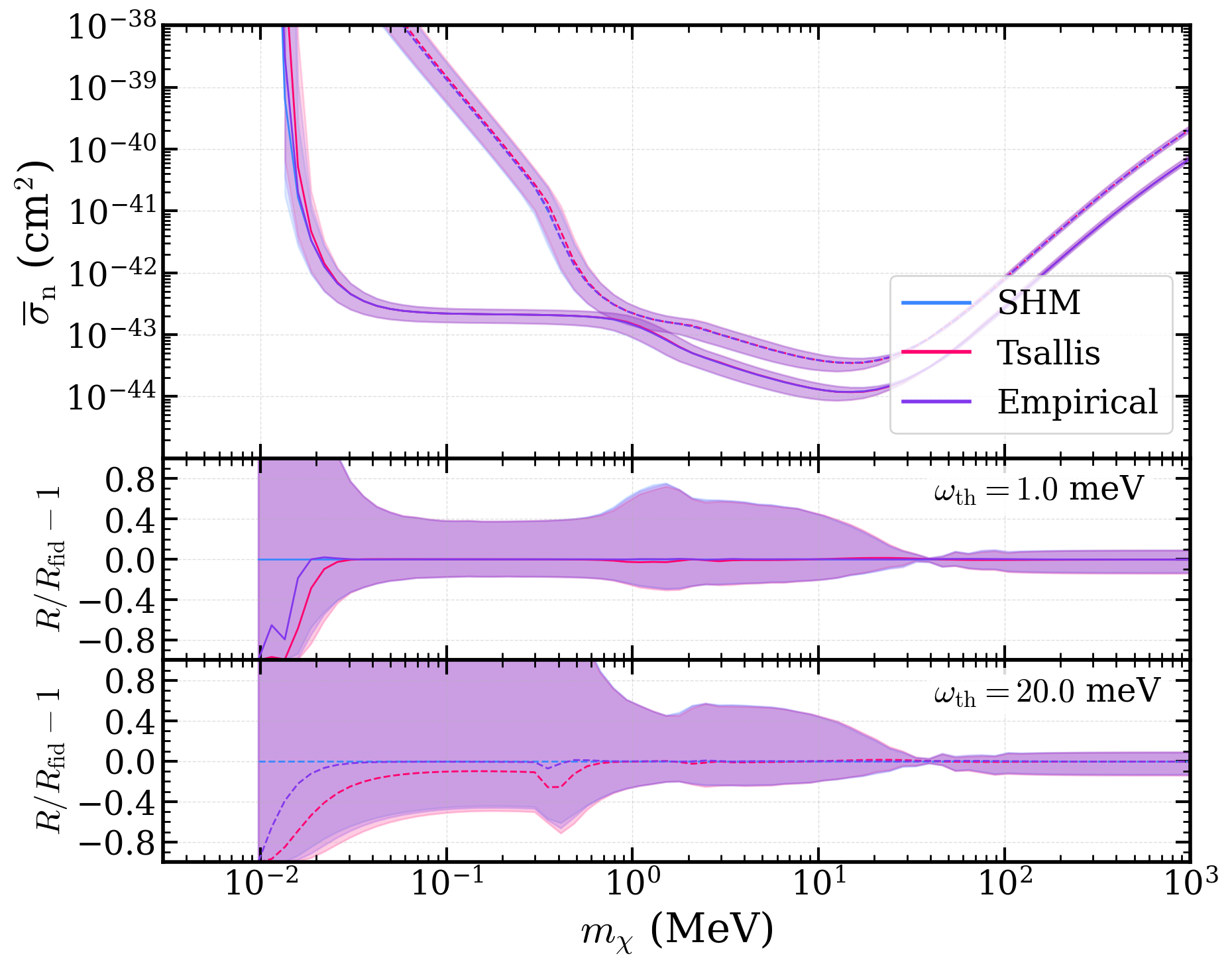}
}

\caption{Same as~\cref{fig:LDP_reach_RMS_QQQ}, for heavy hadrophilic scalar mediated scattering.
\label{fig:HM_reach_RMS_QQQ}}
\end{figure}

We present the projected reach for dark matter scattering under three benchmark mediator scenarios---light dark photon, light hadrophilic scalar, and heavy hadrophilic scalar---in \cref{fig:LDP_reach_RMS_QQQ,fig:LM_reach_RMS_QQQ,fig:HM_reach_RMS_QQQ}, respectively. In each plot, the upper panel shows the cross section reach ($\bar{\sigma}$) for a 1 kg detector assuming 3 events per year, with solid (dashed) curves corresponding to detector energy threshold $\omega_{\min}=1~\mathrm{meV}$ ($20~\mathrm{meV}$), evaluated at the central values of the velocity parameters $(v_\mathrm{c}, v_{\mathrm{e}}, v_{\mathrm{esc}}) = (220, 232, 544)~\mathrm{km}\,\mathrm{s}^{-1}$. For the Tsallis and empirical distributions, these correspond to $v_{0,\mathrm{Tsa}} = 280~\mathrm{km}\,\mathrm{s}^{-1}$ and $v_{0,\mathrm{Emp}} = 154~\mathrm{km}\,\mathrm{s}^{-1}$, respectively, following the rms-matching prescription discussed in~\cref{sec:prescriptions}. Shaded bands around the reach curves represent the uncertainties obtained by varying all three velocity parameters independently within their conservative ranges in \cref{tab:Halo_Parameters} and taking the envelope of the resulting rates at each dark matter mass (the bands for all three halo models are shown but, as discussed below, lie almost entirely on top of one another). The lower panels of the plots show the fractional deviation of the rate $R$ with respect to the fiducial rate $R_\mathrm{fid}$, defined as the rate computed using the central values of the velocity parameters in the SHM, for each threshold choice.

From the figures we see that the impact of variations in the velocity parameters is largest at the lowest dark matter masses accessible for a given phonon energy threshold, leading to $\mathcal{O}(1)$ fractional deviations in the rate. In this regime, the scattering kinematics impose a stringent minimum velocity~\cref{eq:v_minus}, which is larger for higher thresholds, such that only particles from the high-velocity tail of the distribution can produce above-threshold excitations. As the dark matter mass increases, the minimum velocity required for scattering decreases and a broader portion of the velocity distribution contributes to the rate, resulting in reduced sensitivity to variations in the velocity parameters and hence narrower uncertainty bands. The uncertainties remain tens of percent up to $m_\chi \sim 10~\mathrm{MeV}$ in the heavy hadrophilic scalar mediator scenario, while for the light mediator cases the uncertainty drops more rapidly as $m_\chi$ increases. This difference can be understood from the momentum dependence of the scattering process. A heavy mediator enhances contributions from large momentum transfers, thereby preferentially weighting scattering events that require higher incoming velocities. As a result, the heavy mediator case continues to emphasize the high-velocity tail of the distribution over a broad range of dark matter masses, maintaining sensitivity to variations in the halo parameters. In contrast, a light mediator shifts the weight toward smaller $q$, favoring processes that are kinematically accessible at lower velocities and reducing the relative importance of the high-velocity tail.

Another key observation is that, under the rms-matching prescription, the uncertainty bands in the projected reach for different halo models are nearly identical across the full dark matter mass range. This is consistent with the strong overlap of the corresponding speed distributions in \cref{fig:VDF_bands_conc}. The reach curves compare the three halo models at fixed rms velocity (i.e., fixed average dark matter kinetic energy), thereby isolating the effect of the VDF shape. Only small differences appear near the kinematic threshold, reflecting residual variations in the high-velocity tail. These differences are subdominant compared to the uncertainty bands obtained by varying the velocity parameters within each model. This demonstrates that, when departures from the SHM are compared at fixed average kinetic energy via the rms-matching prescription, the dominant source of uncertainty arises from variations in the velocity parameters rather than the functional form of the velocity distribution. In this sense, rms-matching absorbs differences between the halo models into a rescaling of the characteristic velocity scale, rendering the projected reach largely insensitive to the choice of halo model.

Focusing on the SHM, we show in \cref{tab:QQQ_SHM} the relative differences in the scattering rate when each velocity parameter is varied independently within its conservative range, for $\omega_\mathrm{min}=1~\mathrm{meV}$. As in Ref.~\cite{Radick:2020qip}, the relative difference is defined as:
\begin{equation}
     \text{rel.\ diff.\ (halo parameter)}  = \frac{|R_{\rm max}-R_{\rm min}|}{R_{\rm fid}} \,,
     \label{eq:relative_difference}
\end{equation}
where $R_{\rm max}$ ($R_{\rm min}$) is the maximum (minimum) scattering rate when the halo parameter is allowed to vary within the conservative range given in~\cref{tab:Halo_Parameters}. We see that the rate is most sensitive to variations in the circular speed $v_\mathrm{c}$, with subleading dependence on $v_{\mathrm{e}}$ and $v_{\mathrm{esc}}$. This can be attributed to the fact that, away from kinematic thresholds, the scattering rate is primarily governed by the bulk of the velocity distribution. We include additional tables in \cref{App_A} that show that $v_\mathrm{c}$ remains the leading source of uncertainty at the higher threshold $\omega_{\min}=20~\mathrm{meV}$ (\cref{tab:QQQ_SHM_20meV}) and for the Tsallis and empirical models under the rms-matching prescription (\cref{tab:QQQ_RMS_TSA,tab:QQQ_RMS_EMP}).

\begin{table}[t]
\resizebox{\textwidth}{!}{%
\begin{tabular}{|c|c|c|c|c|c|c|c|c|c|c|c|c|c|}
\rowcolor[HTML]{FFFFFF} 
\multicolumn{14}{c}{Standard Halo Model, $\omega_\text{min}$ = 1 meV} \\ \hline
\multicolumn{1}{|c|}{}                           & Target                     & \multicolumn{3}{c|}{$\text{GaAs}$}       & \multicolumn{3}{c|}{$\text{Al}_2\text{O}_3$}  & \multicolumn{3}{c|}{$\text{SiO}_2$}       & \multicolumn{3}{c|}{$\text{CaWO}_4$}      \\ \hline
\multicolumn{1}{|c|}{$m_\chi$}                   & Mediator                   & \multicolumn{1}{c|}{LDP} & \multicolumn{1}{c|}{LHS} & \multicolumn{1}{c|}{HHS} & \multicolumn{1}{c|}{LDP} & \multicolumn{1}{c|}{LHS} & \multicolumn{1}{c|}{HHS} & \multicolumn{1}{c|}{LDP} & \multicolumn{1}{c|}{LHS} & \multicolumn{1}{c|}{HHS} & \multicolumn{1}{c|}{LDP} & \multicolumn{1}{c|}{LHS} & \multicolumn{1}{c|}{HHS} \\ \hline
\multicolumn{1}{|c|}{}                           & rel.diff ($v_\mathrm{c}$)  & \multicolumn{1}{c|}{}    & \multicolumn{1}{c|}{}    & \multicolumn{1}{c|}{}    & \multicolumn{1}{c|}{\cellcolor[HTML]{F1A983}1.3192} & \multicolumn{1}{c|}{\cellcolor[HTML]{F7C7AC}0.8714} & \multicolumn{1}{c|}{\cellcolor[HTML]{F1A983}1.0199} & \multicolumn{1}{c|}{\cellcolor[HTML]{E97132}1.5491} & \multicolumn{1}{c|}{\cellcolor[HTML]{F7C7AC}0.7858} & \multicolumn{1}{c|}{\cellcolor[HTML]{F1A983}1.1674} & \multicolumn{1}{c|}{\cellcolor[HTML]{BE5014}2.1114} & \multicolumn{1}{c|}{\cellcolor[HTML]{F1A983}1.143}  & \multicolumn{1}{c|}{\cellcolor[HTML]{E97132}1.7987} \\
\multicolumn{1}{|c|}{}                           & rel.diff ($v_\mathrm{e}$)  & \multicolumn{1}{c|}{}    & \multicolumn{1}{c|}{}    & \multicolumn{1}{c|}{}    & \multicolumn{1}{c|}{\cellcolor[HTML]{FBE2D5}0.2791} & \multicolumn{1}{c|}{\cellcolor[HTML]{FBE2D5}0.1912} & \multicolumn{1}{c|}{\cellcolor[HTML]{FBE2D5}0.2665} & \multicolumn{1}{c|}{\cellcolor[HTML]{FBE2D5}0.2881} & \multicolumn{1}{c|}{\cellcolor[HTML]{FBE2D5}0.2796} & \multicolumn{1}{c|}{\cellcolor[HTML]{FBE2D5}0.3181} & \multicolumn{1}{c|}{\cellcolor[HTML]{FBE2D5}0.493}  & \multicolumn{1}{c|}{\cellcolor[HTML]{FBE2D5}0.185}  & \multicolumn{1}{c|}{\cellcolor[HTML]{FBE2D5}0.2977} \\
\multicolumn{1}{|c|}{\multirow{-3}{*}{0.01 MeV}} & rel.diff ($v_\mathrm{esc}$) & \multicolumn{1}{c|}{}   & \multicolumn{1}{c|}{}    & \multicolumn{1}{c|}{}    & \multicolumn{1}{c|}{\cellcolor[HTML]{FBE2D5}0.2373} & \multicolumn{1}{c|}{\cellcolor[HTML]{FBE2D5}0.2777} & \multicolumn{1}{c|}{\cellcolor[HTML]{FBE2D5}0.1817} & \multicolumn{1}{c|}{\cellcolor[HTML]{FBE2D5}0.3972} & \multicolumn{1}{c|}{\cellcolor[HTML]{FFFFFF}0.0823} & \multicolumn{1}{c|}{\cellcolor[HTML]{FBE2D5}0.2122} & \multicolumn{1}{c|}{\cellcolor[HTML]{F7C7AC}0.8841} & \multicolumn{1}{c|}{\cellcolor[HTML]{FBE2D5}0.201}  & \multicolumn{1}{c|}{\cellcolor[HTML]{FBE2D5}0.4915} \\ \hline
\multicolumn{1}{|c|}{}                           & rel.diff ($v_\mathrm{c}$)  & \multicolumn{1}{c|}{\cellcolor[HTML]{FBE2D5}0.186}  & \multicolumn{1}{c|}{\cellcolor[HTML]{FFFFFF}0.0505} & \multicolumn{1}{c|}{\cellcolor[HTML]{FBE2D5}0.3625} & \multicolumn{1}{c|}{\cellcolor[HTML]{F7C7AC}0.7956} & \multicolumn{1}{c|}{\cellcolor[HTML]{FBE2D5}0.1035} & \multicolumn{1}{c|}{\cellcolor[HTML]{FBE2D5}0.3562} & \multicolumn{1}{c|}{\cellcolor[HTML]{F7C7AC}0.6595} & \multicolumn{1}{c|}{\cellcolor[HTML]{FFFFFF}0.0733} & \multicolumn{1}{c|}{\cellcolor[HTML]{FBE2D5}0.3512} & \multicolumn{1}{c|}{\cellcolor[HTML]{FBE2D5}0.3378} & \multicolumn{1}{c|}{\cellcolor[HTML]{FFFFFF}0.051}  & \multicolumn{1}{c|}{\cellcolor[HTML]{FBE2D5}0.3655} \\
\multicolumn{1}{|c|}{}                           & rel.diff ($v_\mathrm{e}$)  & \multicolumn{1}{c|}{\cellcolor[HTML]{FFFFFF}0.0893} & \multicolumn{1}{c|}{\cellcolor[HTML]{FFFFFF}0.0081} & \multicolumn{1}{c|}{\cellcolor[HTML]{FBE2D5}0.1212} & \multicolumn{1}{c|}{\cellcolor[HTML]{FBE2D5}0.2313} & \multicolumn{1}{c|}{\cellcolor[HTML]{FFFFFF}0.054}  & \multicolumn{1}{c|}{\cellcolor[HTML]{FBE2D5}0.1171} & \multicolumn{1}{c|}{\cellcolor[HTML]{FBE2D5}0.1983} & \multicolumn{1}{c|}{\cellcolor[HTML]{FFFFFF}0.034}  & \multicolumn{1}{c|}{\cellcolor[HTML]{FBE2D5}0.1162} & \multicolumn{1}{c|}{\cellcolor[HTML]{FBE2D5}0.1269} & \multicolumn{1}{c|}{\cellcolor[HTML]{FFFFFF}0.003}  & \multicolumn{1}{c|}{\cellcolor[HTML]{FBE2D5}0.1211} \\
\multicolumn{1}{|c|}{\multirow{-3}{*}{0.1 MeV}}  & rel.diff ($v_\mathrm{esc}$) & \multicolumn{1}{c|}{\cellcolor[HTML]{FFFFFF}0.0204} & \multicolumn{1}{c|}{\cellcolor[HTML]{FFFFFF}0.0506} & \multicolumn{1}{c|}{\cellcolor[HTML]{FFFFFF}0.0568} & \multicolumn{1}{c|}{\cellcolor[HTML]{FBE2D5}0.1141} & \multicolumn{1}{c|}{\cellcolor[HTML]{FFFFFF}0.0491} & \multicolumn{1}{c|}{\cellcolor[HTML]{FFFFFF}0.0553} & \multicolumn{1}{c|}{\cellcolor[HTML]{FBE2D5}0.107}  & \multicolumn{1}{c|}{\cellcolor[HTML]{FFFFFF}0.0403} & \multicolumn{1}{c|}{\cellcolor[HTML]{FFFFFF}0.0552} & \multicolumn{1}{c|}{\cellcolor[HTML]{FFFFFF}0.0442} & \multicolumn{1}{c|}{\cellcolor[HTML]{FFFFFF}0.0459} & \multicolumn{1}{c|}{\cellcolor[HTML]{FFFFFF}0.0566} \\ \hline
\multicolumn{1}{|c|}{}                           & rel.diff ($v_\mathrm{c}$)  & \multicolumn{1}{c|}{\cellcolor[HTML]{FFFFFF}0.0396} & \multicolumn{1}{c|}{\cellcolor[HTML]{FBE2D5}0.1303} & \multicolumn{1}{c|}{\cellcolor[HTML]{F7C7AC}0.5256} & \multicolumn{1}{c|}{\cellcolor[HTML]{FFFFFF}0.0034} & \multicolumn{1}{c|}{\cellcolor[HTML]{FBE2D5}0.1313} & \multicolumn{1}{c|}{\cellcolor[HTML]{FBE2D5}0.4647} & \multicolumn{1}{c|}{\cellcolor[HTML]{FFFFFF}0.0178} & \multicolumn{1}{c|}{\cellcolor[HTML]{FBE2D5}0.1302} & \multicolumn{1}{c|}{\cellcolor[HTML]{FBE2D5}0.4895} & \multicolumn{1}{c|}{\cellcolor[HTML]{FFFFFF}0.0143} & \multicolumn{1}{c|}{\cellcolor[HTML]{FBE2D5}0.1272} & \multicolumn{1}{c|}{\cellcolor[HTML]{F7C7AC}0.5064} \\
\multicolumn{1}{|c|}{}                           & rel.diff ($v_\mathrm{e}$)  & \multicolumn{1}{c|}{\cellcolor[HTML]{FFFFFF}0.0126} & \multicolumn{1}{c|}{\cellcolor[HTML]{FFFFFF}0.0675} & \multicolumn{1}{c|}{\cellcolor[HTML]{FBE2D5}0.1492} & \multicolumn{1}{c|}{\cellcolor[HTML]{FFFFFF}0.0115} & \multicolumn{1}{c|}{\cellcolor[HTML]{FFFFFF}0.0651} & \multicolumn{1}{c|}{\cellcolor[HTML]{FBE2D5}0.1433} & \multicolumn{1}{c|}{\cellcolor[HTML]{FFFFFF}0.0203} & \multicolumn{1}{c|}{\cellcolor[HTML]{FFFFFF}0.0688} & \multicolumn{1}{c|}{\cellcolor[HTML]{FBE2D5}0.1436} & \multicolumn{1}{c|}{\cellcolor[HTML]{FFFFFF}0.0026} & \multicolumn{1}{c|}{\cellcolor[HTML]{FFFFFF}0.0602} & \multicolumn{1}{c|}{\cellcolor[HTML]{FBE2D5}0.1626} \\
\multicolumn{1}{|c|}{\multirow{-3}{*}{1 MeV}}    & rel.diff ($v_\mathrm{esc}$) & \multicolumn{1}{c|}{\cellcolor[HTML]{FFFFFF}0.0053} & \multicolumn{1}{c|}{\cellcolor[HTML]{FFFFFF}0.0336} & \multicolumn{1}{c|}{\cellcolor[HTML]{FBE2D5}0.1023} & \multicolumn{1}{c|}{\cellcolor[HTML]{FFFFFF}0.0009} & \multicolumn{1}{c|}{\cellcolor[HTML]{FFFFFF}0.004}  & \multicolumn{1}{c|}{\cellcolor[HTML]{FFFFFF}0.0837} & \multicolumn{1}{c|}{\cellcolor[HTML]{FFFFFF}0.0017} & \multicolumn{1}{c|}{\cellcolor[HTML]{FFFFFF}0.0199} & \multicolumn{1}{c|}{\cellcolor[HTML]{FFFFFF}0.0886} & \multicolumn{1}{c|}{\cellcolor[HTML]{FFFFFF}0.0021} & \multicolumn{1}{c|}{\cellcolor[HTML]{FFFFFF}0.009}  & \multicolumn{1}{c|}{\cellcolor[HTML]{FFFFFF}0.0887} \\ \hline
\multicolumn{1}{|c|}{}                           & rel.diff ($v_\mathrm{c}$)  & \multicolumn{1}{c|}{\cellcolor[HTML]{FFFFFF}0.0999} & \multicolumn{1}{c|}{\cellcolor[HTML]{FBE2D5}0.1297} & \multicolumn{1}{c|}{\cellcolor[HTML]{FBE2D5}0.4112} & \multicolumn{1}{c|}{\cellcolor[HTML]{FFFFFF}0.0886} & \multicolumn{1}{c|}{\cellcolor[HTML]{FBE2D5}0.1329} & \multicolumn{1}{c|}{\cellcolor[HTML]{FBE2D5}0.3986} & \multicolumn{1}{c|}{\cellcolor[HTML]{FFFFFF}0.0913} & \multicolumn{1}{c|}{\cellcolor[HTML]{FBE2D5}0.1388} & \multicolumn{1}{c|}{\cellcolor[HTML]{FBE2D5}0.3575} & \multicolumn{1}{c|}{\cellcolor[HTML]{FFFFFF}0.0937} & \multicolumn{1}{c|}{\cellcolor[HTML]{FBE2D5}0.1304} & \multicolumn{1}{c|}{\cellcolor[HTML]{FBE2D5}0.4243} \\
\multicolumn{1}{|c|}{}                           & rel.diff ($v_\mathrm{e}$)  & \multicolumn{1}{c|}{\cellcolor[HTML]{FFFFFF}0.0402} & \multicolumn{1}{c|}{\cellcolor[HTML]{FFFFFF}0.0749} & \multicolumn{1}{c|}{\cellcolor[HTML]{FBE2D5}0.1332} & \multicolumn{1}{c|}{\cellcolor[HTML]{FFFFFF}0.0399} & \multicolumn{1}{c|}{\cellcolor[HTML]{FFFFFF}0.0599} & \multicolumn{1}{c|}{\cellcolor[HTML]{FBE2D5}0.1315} & \multicolumn{1}{c|}{\cellcolor[HTML]{FFFFFF}0.0386} & \multicolumn{1}{c|}{\cellcolor[HTML]{FFFFFF}0.0701} & \multicolumn{1}{c|}{\cellcolor[HTML]{FBE2D5}0.1204} & \multicolumn{1}{c|}{\cellcolor[HTML]{FFFFFF}0.0413} & \multicolumn{1}{c|}{\cellcolor[HTML]{FFFFFF}0.0663} & \multicolumn{1}{c|}{\cellcolor[HTML]{FBE2D5}0.1345} \\
\multicolumn{1}{|c|}{\multirow{-3}{*}{10 MeV}}   & rel.diff ($v_\mathrm{esc}$) & \multicolumn{1}{c|}{\cellcolor[HTML]{FFFFFF}0.01}   & \multicolumn{1}{c|}{\cellcolor[HTML]{FFFFFF}0.0559} & \multicolumn{1}{c|}{\cellcolor[HTML]{FFFFFF}0.0677} & \multicolumn{1}{c|}{\cellcolor[HTML]{FFFFFF}0.0161} & \multicolumn{1}{c|}{\cellcolor[HTML]{FFFFFF}0.0366} & \multicolumn{1}{c|}{\cellcolor[HTML]{FFFFFF}0.0621} & \multicolumn{1}{c|}{\cellcolor[HTML]{FFFFFF}0.0316} & \multicolumn{1}{c|}{\cellcolor[HTML]{FFFFFF}0.0313} & \multicolumn{1}{c|}{\cellcolor[HTML]{FFFFFF}0.0538} & \multicolumn{1}{c|}{\cellcolor[HTML]{FFFFFF}0.0102} & \multicolumn{1}{c|}{\cellcolor[HTML]{FFFFFF}0.013}  & \multicolumn{1}{c|}{\cellcolor[HTML]{FFFFFF}0.0666} \\ \hline
\multicolumn{1}{|c|}{}                           & rel.diff ($v_\mathrm{c}$)  & \multicolumn{1}{c|}{\cellcolor[HTML]{FBE2D5}0.1251} & \multicolumn{1}{c|}{\cellcolor[HTML]{FBE2D5}0.1374} & \multicolumn{1}{c|}{\cellcolor[HTML]{FBE2D5}0.1248} & \multicolumn{1}{c|}{\cellcolor[HTML]{FBE2D5}0.1199} & \multicolumn{1}{c|}{\cellcolor[HTML]{FBE2D5}0.1355} & \multicolumn{1}{c|}{\cellcolor[HTML]{FBE2D5}0.1236} & \multicolumn{1}{c|}{\cellcolor[HTML]{FBE2D5}0.1225} & \multicolumn{1}{c|}{\cellcolor[HTML]{FBE2D5}0.1377} & \multicolumn{1}{c|}{\cellcolor[HTML]{FBE2D5}0.126}  & \multicolumn{1}{c|}{\cellcolor[HTML]{FBE2D5}0.1213} & \multicolumn{1}{c|}{\cellcolor[HTML]{FBE2D5}0.1293} & \multicolumn{1}{c|}{\cellcolor[HTML]{FBE2D5}0.1088} \\
\multicolumn{1}{|c|}{}                           & rel.diff ($v_\mathrm{e}$)  & \multicolumn{1}{c|}{\cellcolor[HTML]{FFFFFF}0.0604} & \multicolumn{1}{c|}{\cellcolor[HTML]{FFFFFF}0.0485} & \multicolumn{1}{c|}{\cellcolor[HTML]{FFFFFF}0.0537} & \multicolumn{1}{c|}{\cellcolor[HTML]{FFFFFF}0.0594} & \multicolumn{1}{c|}{\cellcolor[HTML]{FFFFFF}0.0709} & \multicolumn{1}{c|}{\cellcolor[HTML]{FFFFFF}0.0548} & \multicolumn{1}{c|}{\cellcolor[HTML]{FFFFFF}0.0609} & \multicolumn{1}{c|}{\cellcolor[HTML]{FFFFFF}0.0657} & \multicolumn{1}{c|}{\cellcolor[HTML]{FFFFFF}0.055}  & \multicolumn{1}{c|}{\cellcolor[HTML]{FFFFFF}0.0623} & \multicolumn{1}{c|}{\cellcolor[HTML]{FFFFFF}0.0723} & \multicolumn{1}{c|}{\cellcolor[HTML]{FFFFFF}0.0442} \\
\multicolumn{1}{|c|}{\multirow{-3}{*}{100 MeV}}  & rel.diff ($v_\mathrm{esc}$) & \multicolumn{1}{c|}{\cellcolor[HTML]{FFFFFF}0.0098} & \multicolumn{1}{c|}{\cellcolor[HTML]{FFFFFF}0.067}  & \multicolumn{1}{c|}{\cellcolor[HTML]{FFFFFF}0.0158} & \multicolumn{1}{c|}{\cellcolor[HTML]{FFFFFF}0.0075} & \multicolumn{1}{c|}{\cellcolor[HTML]{FFFFFF}0.0612} & \multicolumn{1}{c|}{\cellcolor[HTML]{FFFFFF}0.0158} & \multicolumn{1}{c|}{\cellcolor[HTML]{FFFFFF}0.0075} & \multicolumn{1}{c|}{\cellcolor[HTML]{FFFFFF}0.0048} & \multicolumn{1}{c|}{\cellcolor[HTML]{FFFFFF}0.0201} & \multicolumn{1}{c|}{\cellcolor[HTML]{FFFFFF}0.0262} & \multicolumn{1}{c|}{\cellcolor[HTML]{FFFFFF}0.0115} & \multicolumn{1}{c|}{\cellcolor[HTML]{FFFFFF}0.0132} \\ \hline
\end{tabular}%
}
\caption{Relative difference defined in \cref{eq:relative_difference} from independently varying $v_\mathrm{c}$, $v_{\mathrm{e}}$, and $v_{\mathrm{esc}}$ within their conservative ranges for representative dark matter masses, assuming $\omega_\text{min}=1~\mathrm{meV}$ and the SHM. Results are shown for different mediator scenarios; light dark photon (LDP), light hadrophilic scalar (LHS), and heavy hadrophilic scalar (HHS).
\label{tab:QQQ_SHM}}
\end{table}

\subsection{Daily modulation}

\begin{figure}[t]
\centering

\small
\begin{center}
\begin{tabular}{cccccc}
    \raisebox{0.5ex}{\tikz{\draw[black, line width=1.2pt] (0,0) -- (0.8,0);}} & SHM \quad &
    \raisebox{0.5ex}{\tikz{\draw[black, dashed, line width=1.2pt] (0,0) -- (0.8,0);}} & Tsallis \quad &
    \raisebox{0.5ex}{\tikz{\draw[black, dotted, line width=1.2pt] (0,0) -- (0.8,0);}} & Empirical
\end{tabular}
\end{center}

\subfloat[]{%
    \includegraphics[width=0.49\textwidth]{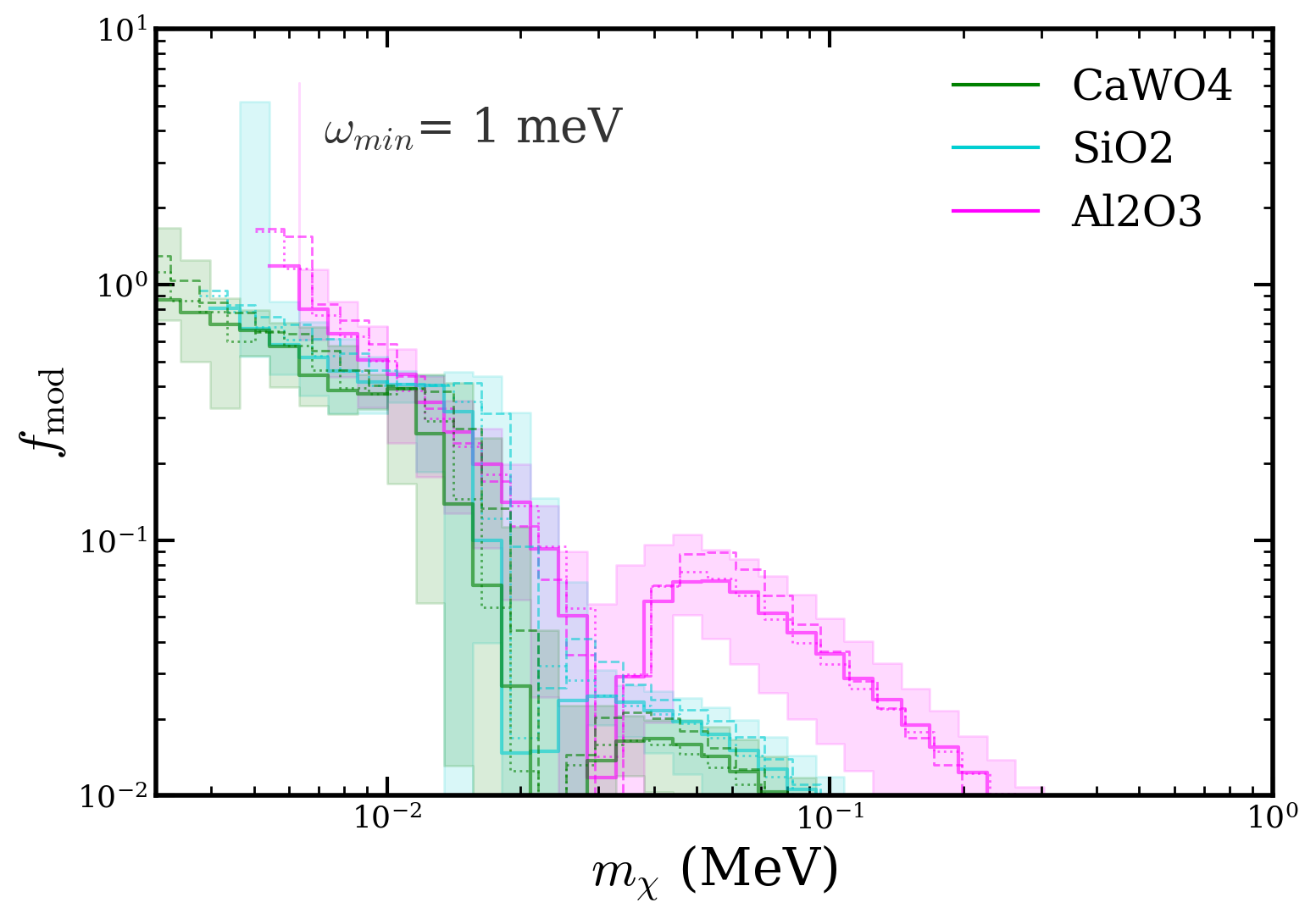}
}
\hfill
\subfloat[]{%
    \includegraphics[width=0.49\textwidth]{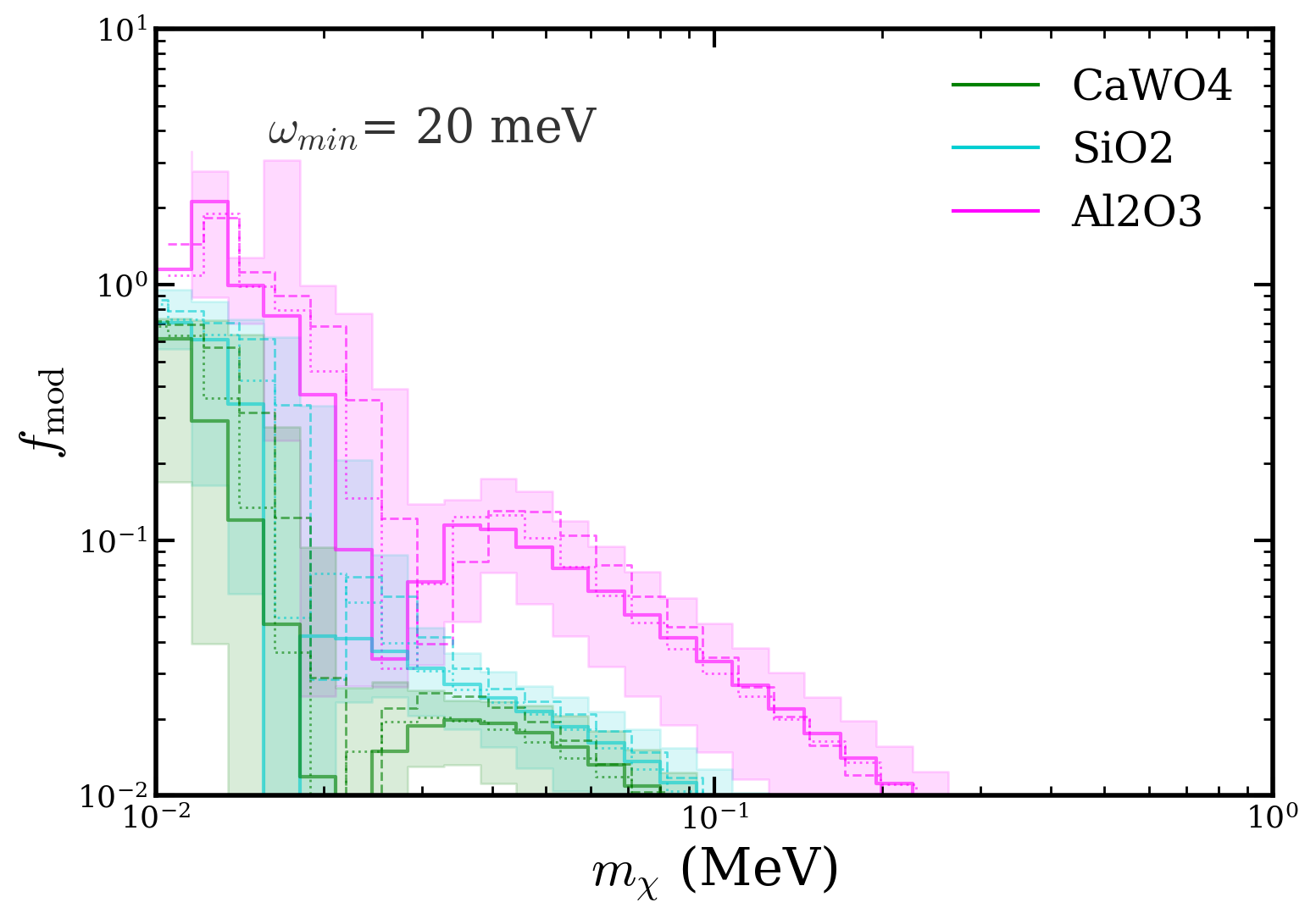}
}

\subfloat[]{%
    \includegraphics[width=0.49\textwidth]{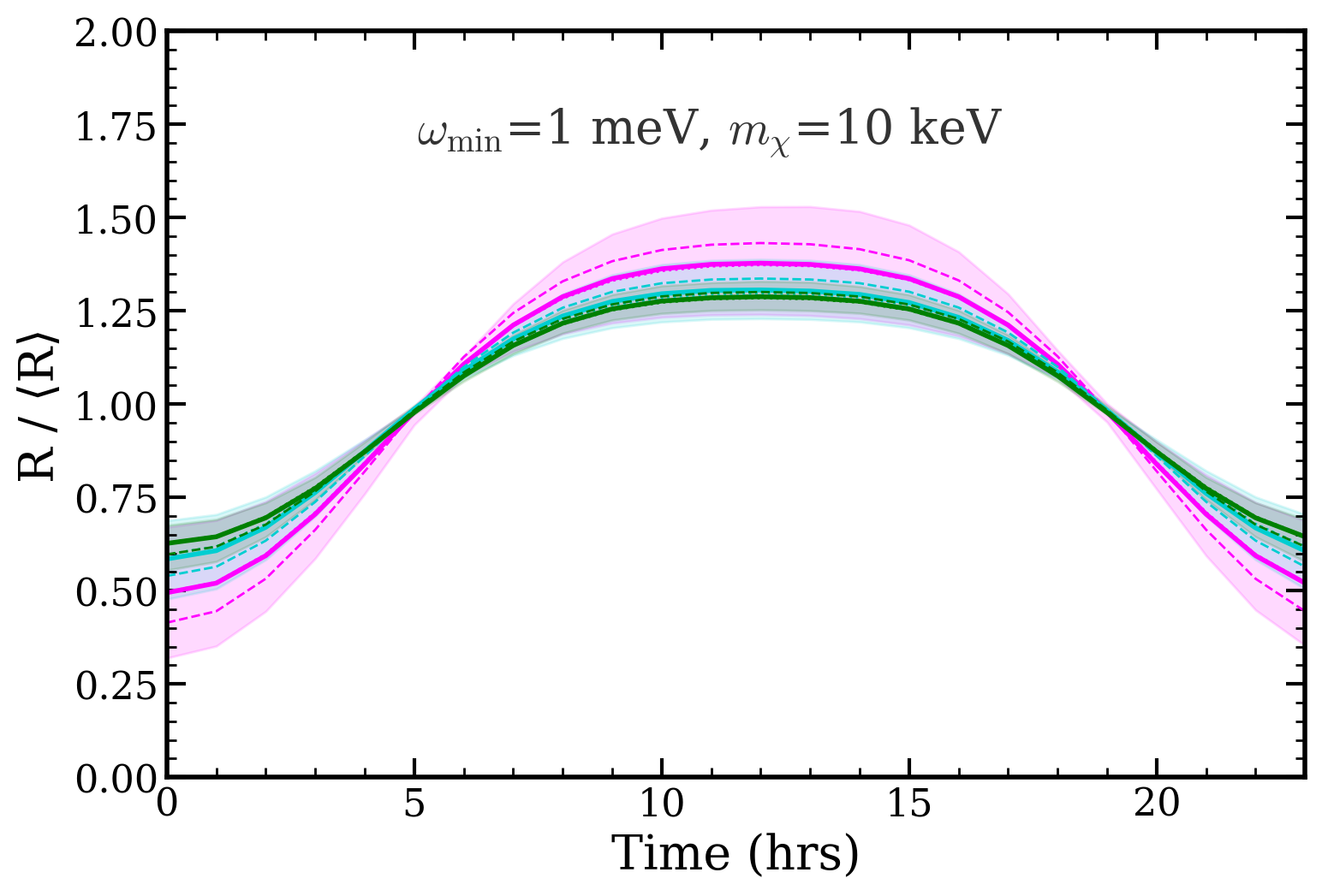}
}
\hfill
\subfloat[]{%
    \includegraphics[width=0.49\textwidth]{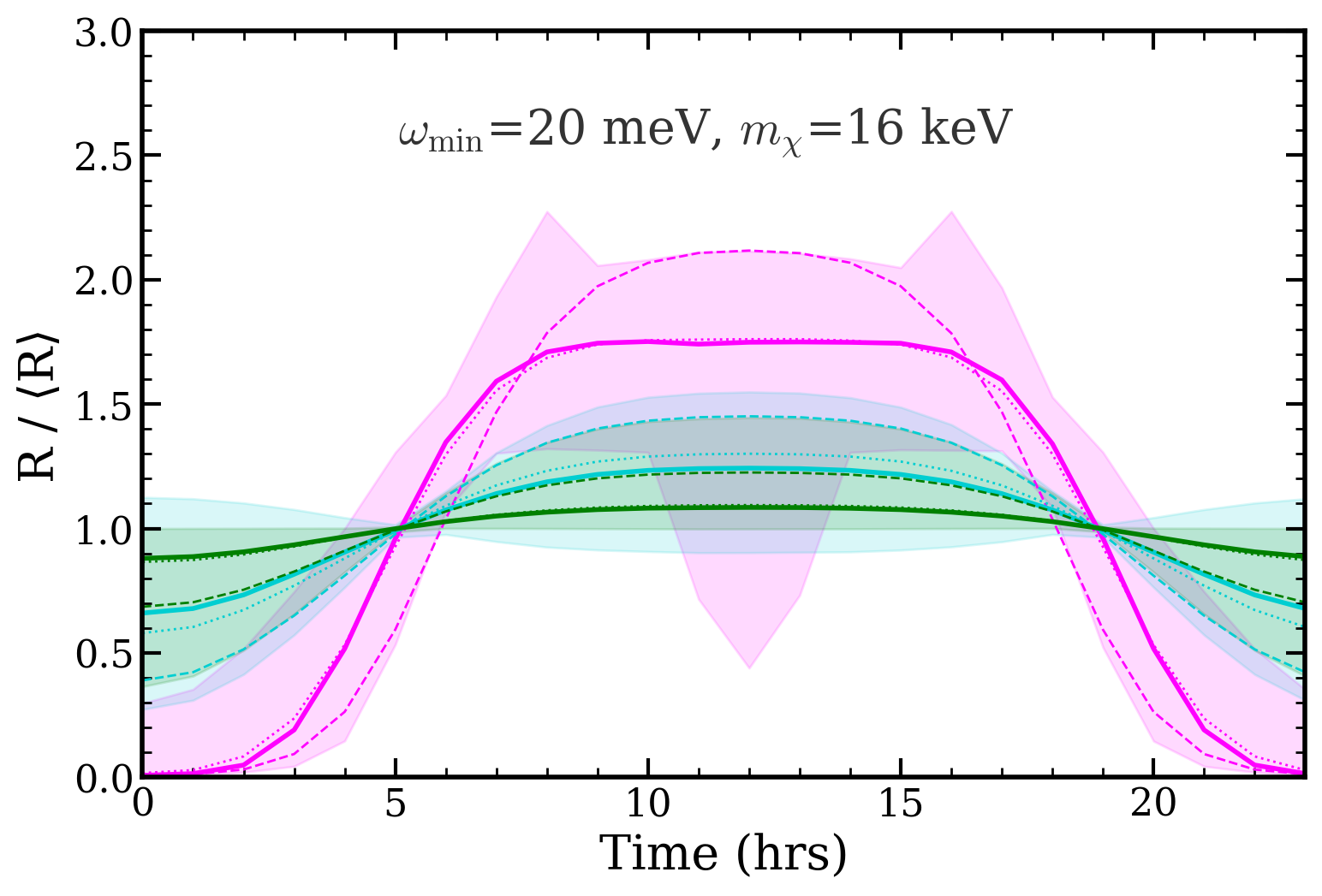}
}

\caption{Daily modulation amplitude $f_{\mathrm{mod}}$ defined in \cref{eq:f_mod} as a function of the dark matter mass for $\omega_\mathrm{min}=1~\mathrm{meV}$ and $20~\mathrm{meV}$ (upper panels), and time dependence of the rate normalized to its daily average at example dark matter masses (lower panels) for light dark photon mediated scattering. Solid, dashed, and dotted lines correspond to the SHM, Tsallis, and empirical velocity distributions, respectively, evaluated at the central values of the velocity parameters. The shaded bands show the effect of varying the halo velocity parameters within their allowed ranges under the SHM assumption.
\label{fig:LDP_DM_QQQ}}
\end{figure}

\begin{figure}[t]
\centering

\small
\begin{center}
\begin{tabular}{cccccc}
    \raisebox{0.5ex}{\tikz{\draw[black, line width=1.2pt] (0,0) -- (0.8,0);}} & SHM \quad &
    \raisebox{0.5ex}{\tikz{\draw[black, dashed, line width=1.2pt] (0,0) -- (0.8,0);}} & Tsallis \quad &
    \raisebox{0.5ex}{\tikz{\draw[black, dotted, line width=1.2pt] (0,0) -- (0.8,0);}} & Empirical
\end{tabular}
\end{center}

\subfloat[]{%
    \includegraphics[width=0.49\textwidth]{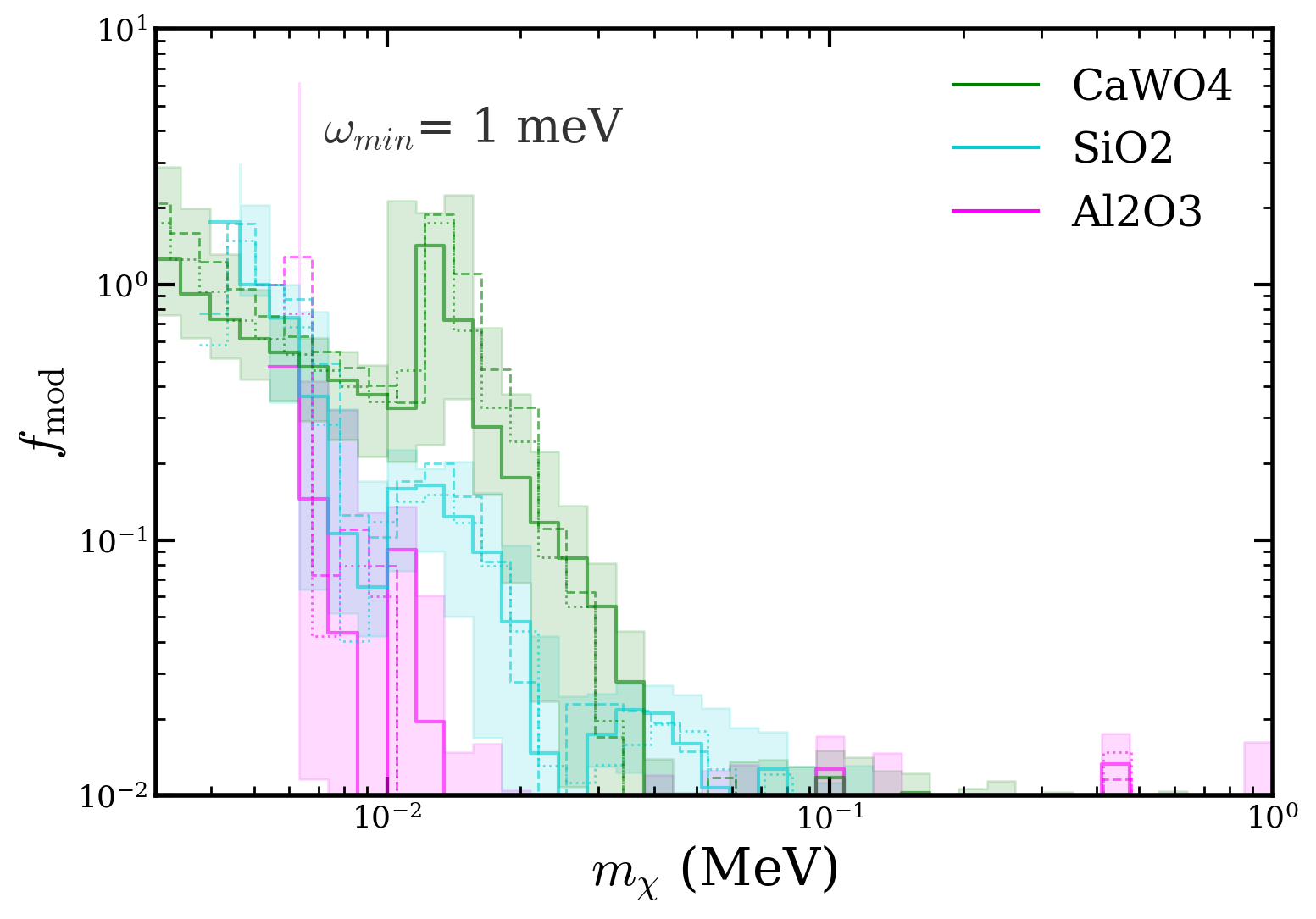}
}
\hfill
\subfloat[]{%
    \includegraphics[width=0.49\textwidth]{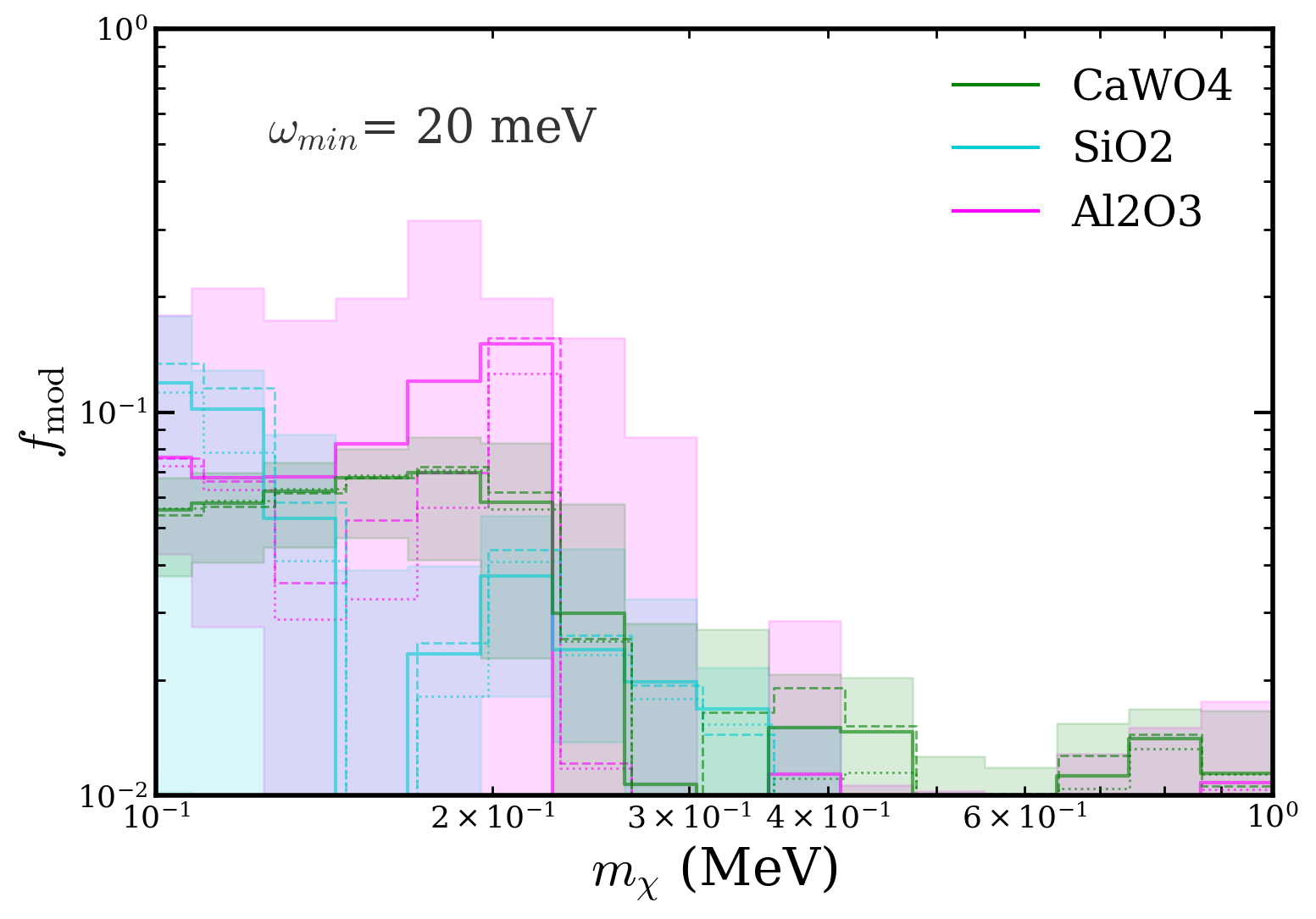}
}

\subfloat[]{%
    \includegraphics[width=0.49\textwidth]{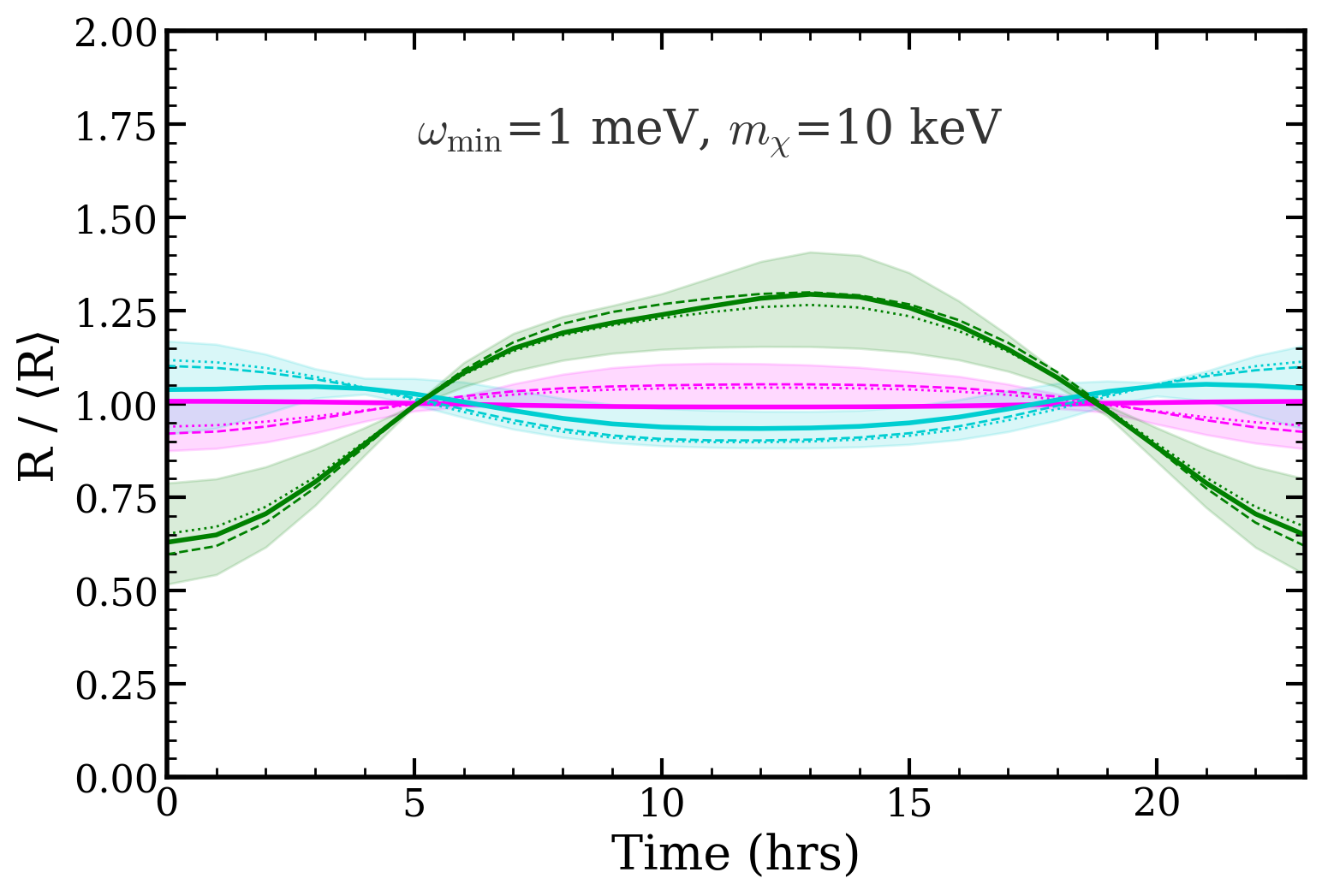}
}
\hfill
\subfloat[]{%
    \includegraphics[width=0.49\textwidth]{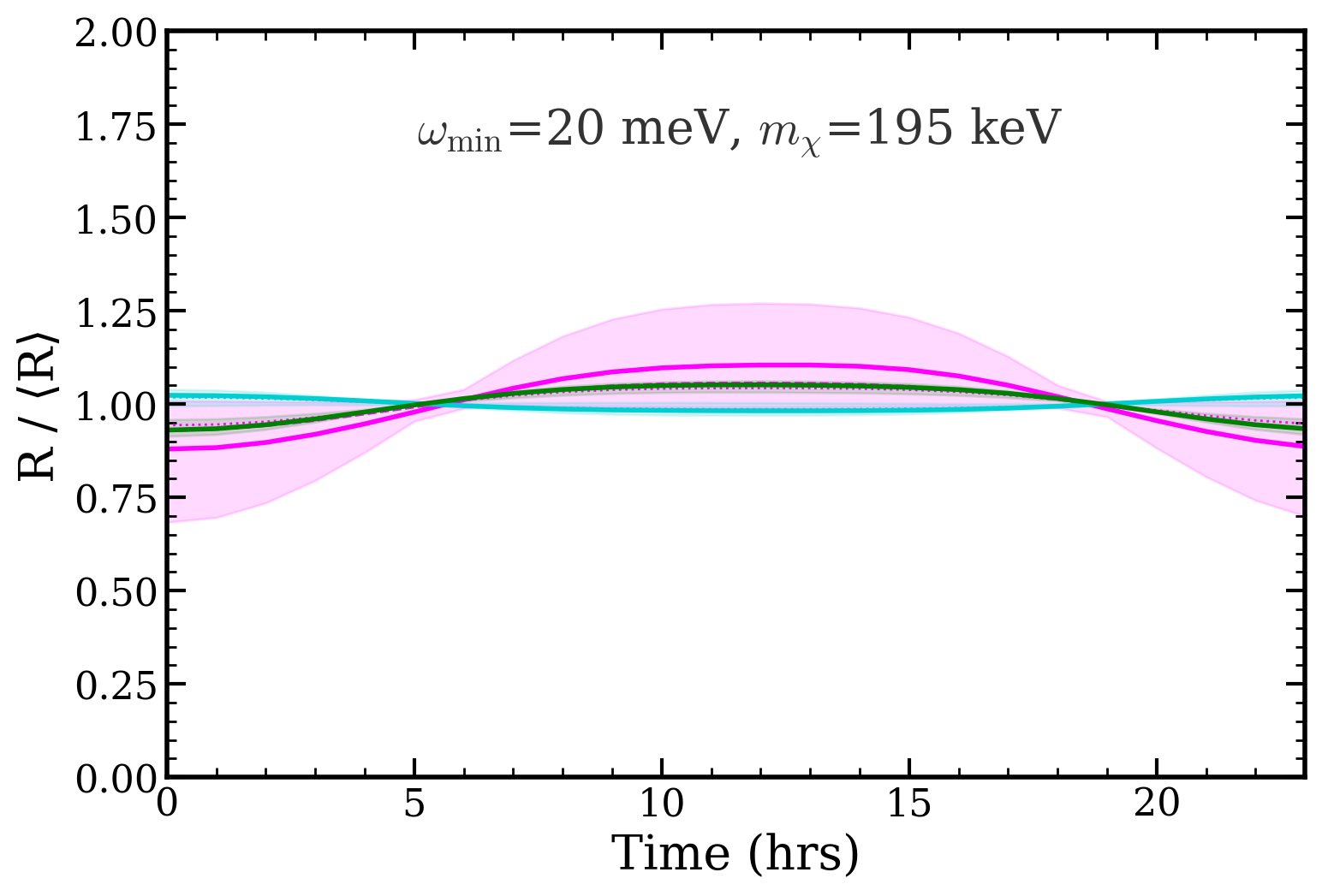}
}

\caption{Same as~\cref{fig:LDP_DM_QQQ}, for light hadrophilic scalar mediated scattering.
\label{fig:LM_DM_QQQ}}
\end{figure}

\begin{figure}[t]
\centering

\small
\begin{center}
\begin{tabular}{cccccc}
    \raisebox{0.5ex}{\tikz{\draw[black, line width=1.2pt] (0,0) -- (0.8,0);}} & SHM \quad &
    \raisebox{0.5ex}{\tikz{\draw[black, dashed, line width=1.2pt] (0,0) -- (0.8,0);}} & Tsallis \quad &
    \raisebox{0.5ex}{\tikz{\draw[black, dotted, line width=1.2pt] (0,0) -- (0.8,0);}} & Empirical
\end{tabular}
\end{center}

\subfloat[]{%
    \includegraphics[width=0.49\textwidth]{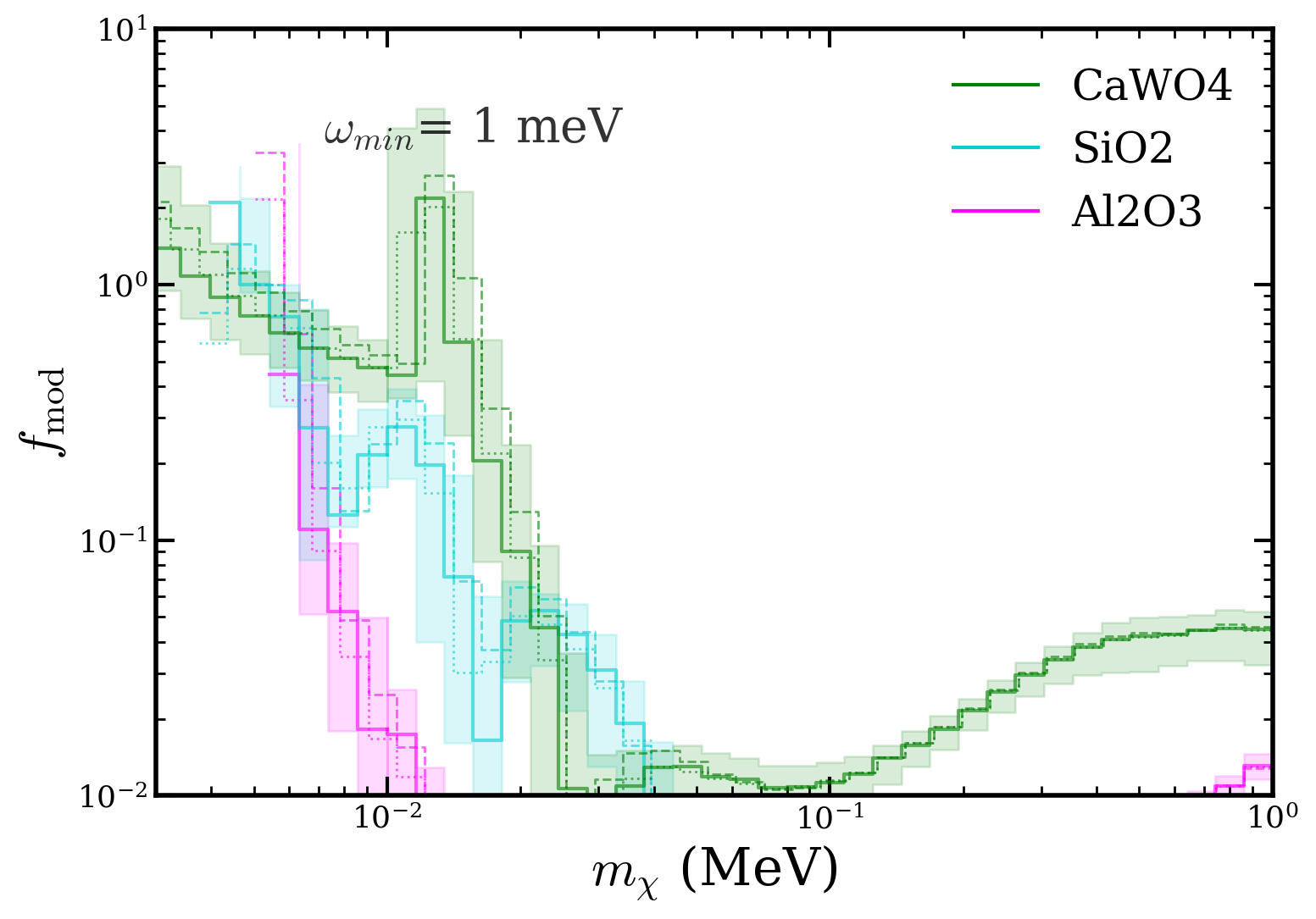}
}
\hfill
\subfloat[]{%
    \includegraphics[width=0.49\textwidth]{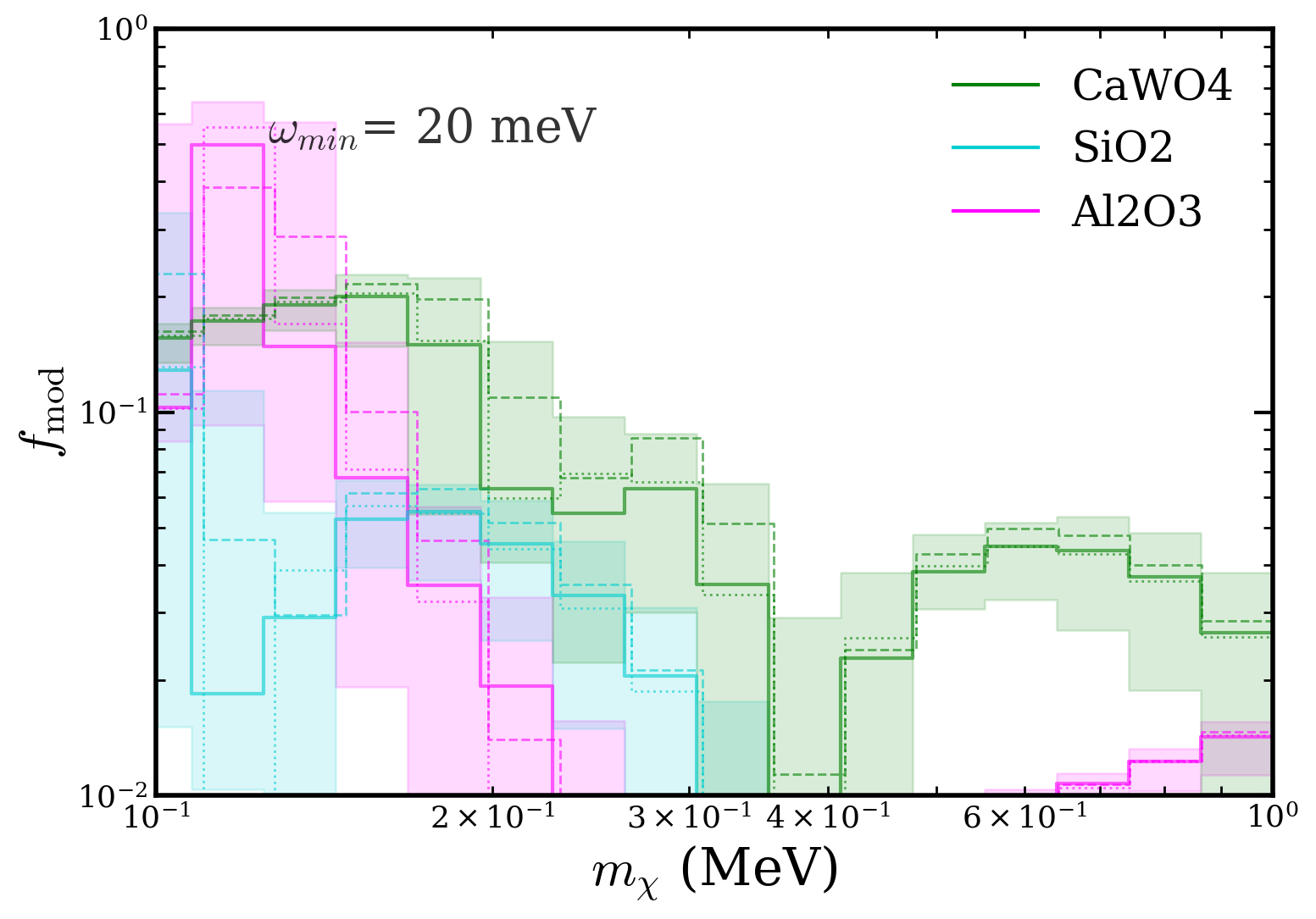}
}

\subfloat[]{%
    \includegraphics[width=0.49\textwidth]{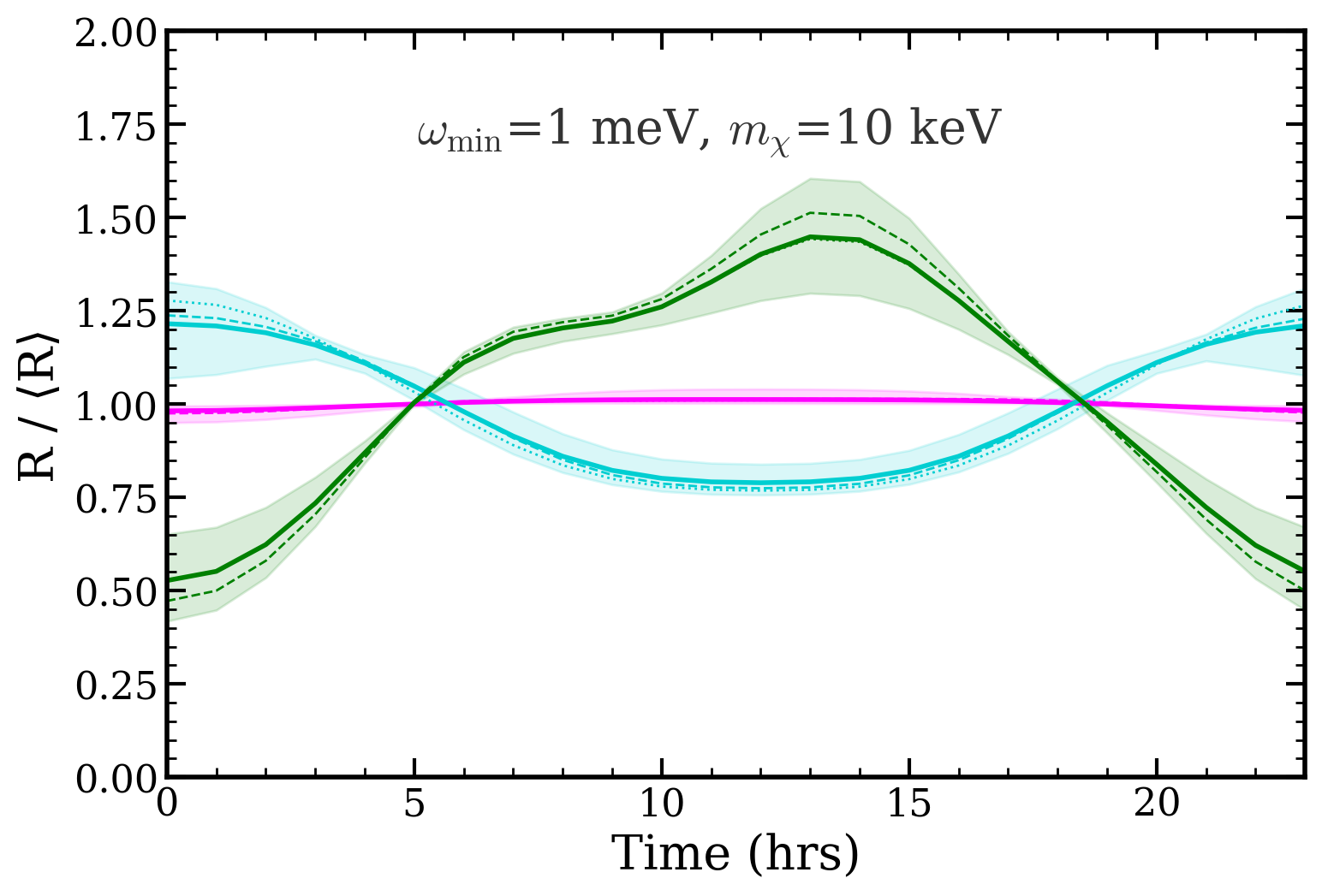}
}
\hfill
\subfloat[]{%
    \includegraphics[width=0.49\textwidth]{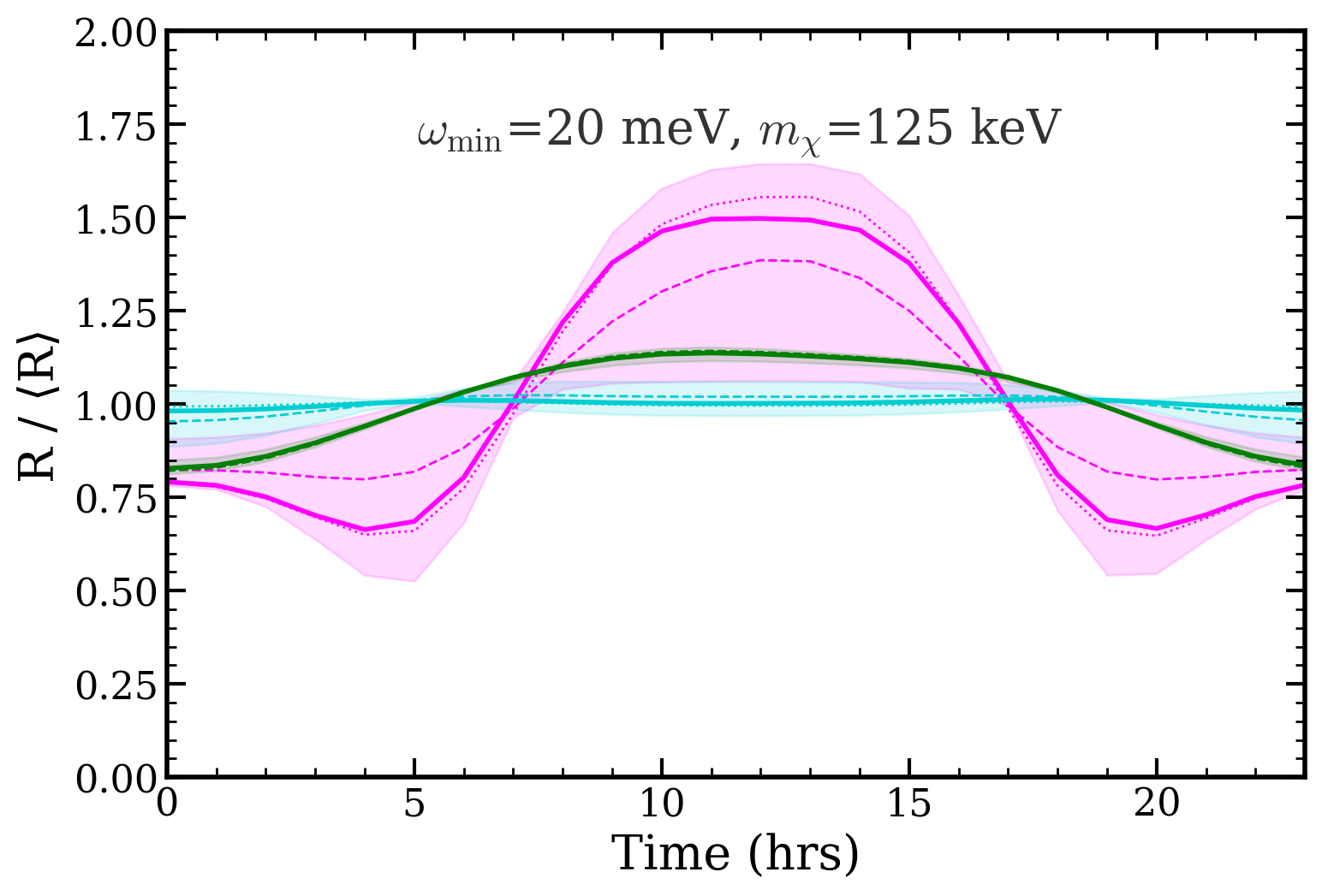}
}

\caption{Same as~\cref{fig:LDP_DM_QQQ}, for heavy hadrophilic scalar mediated scattering.
\label{fig:HM_DM_QQQ}}
\end{figure}

Moving on to daily modulation, we focus on the three anisotropic targets $\text{Al}_2\text{O}_3$, $\text{SiO}_2$, and $\text{CaWO}_4$; GaAs has a highly symmetric cubic zincblende structure and is effectively isotropic. In \cref{fig:LDP_DM_QQQ,fig:LM_DM_QQQ,fig:HM_DM_QQQ} we show, for the light dark photon, light hadrophilic scalar, and heavy hadrophilic scalar mediator scenarios respectively, the daily modulation amplitude (upper panels) defined as
\begin{equation}
     f_\mathrm{mod}  = \frac{\max\bigl|R(t)-\langle R\rangle\bigr|}{\langle R\rangle} \,,
     \label{eq:f_mod}
\end{equation}
and the time dependence of the rate normalized to its daily average $\langle R\rangle$ at representative dark matter masses (lower panels), for $\omega_{\min}=1~\mathrm{meV}$ and $20~\mathrm{meV}$.
The solid, dashed, and dotted curves correspond to the SHM, Tsallis, and empirical distributions, respectively, evaluated at the central values of the velocity parameters. The shaded bands represent uncertainties obtained by independently varying $(v_\mathrm{c}, v_\mathrm{e}, v_{\mathrm{esc}})$ within their conservative ranges and taking the envelope; for visual clarity we display only the SHM bands, having verified that the bands for the Tsallis and empirical models computed under the rms-matching prescription largely overlap with the SHM ones, as in the projected reach plots.

From the figures we see that the three halo models yield a similar qualitative $m_\chi$ dependence of $f_\mathrm{mod}$ and broadly preserve the relative ordering of target materials. From the $R(t)/\langle R\rangle$ plots, we see that, with the exception of the light dark photon mediator scenario near the lowest kinematically accessible mass ($m_\chi \simeq 16~\mathrm{keV}$ for $\omega_{\min} = 20~\mathrm{meV}$), the astrophysical uncertainties predominantly affect the amplitude of the daily modulation rather than its phase: in most cases, the locations of the maxima and minima remain nearly unchanged. This indicates that the phase of the modulation is primarily determined by the time-dependent direction of $\boldsymbol{v}_\mathrm{e}(t)$ relative to the crystal axes, while the velocity distribution mainly controls the relative weighting of contributing kinematic configurations, and hence the overall modulation amplitude.

Overall, these results show that daily modulation remains a robust and distinctive signature of anisotropic dark matter scattering across most of the parameter space, while also highlighting that its detailed interpretation requires consistently accounting for both halo model choices and velocity parameter uncertainties.

\section{Conclusions}
\label{sec:conclusion}

We have presented a systematic study of how uncertainties in the local dark matter velocity distribution propagate into direct detection observables utilizing single phonon excitations in crystal targets. Focusing on four representative targets ($\mathrm{GaAs}$, $\mathrm{Al}_2\mathrm{O}_3$, $\mathrm{SiO}_2$, and $\mathrm{CaWO}_4$) and three benchmark mediator scenarios (light dark photon, light hadrophilic scalar, and heavy hadrophilic scalar), we computed both the projected cross section reach and the daily modulation signal under three halo models (SHM, Tsallis, and empirical), while varying the astrophysical velocity parameters $(v_\mathrm{c}, v_\mathrm{e}, v_\mathrm{esc})$ within observationally motivated ranges. We found $\mathcal{O}(1\%)$ to $\mathcal{O}(100\%)$ fractional deviations in the predicted rates across the dark matter mass range of interest, with several robust trends emerging across all targets and mediator scenarios: (i)~the dominant source of uncertainty is the local circular velocity $v_\mathrm{c}$, which controls the bulk of the velocity distribution, with $v_\mathrm{e}$ and $v_\mathrm{esc}$ playing subleading roles; (ii)~once the three halo models are placed on equal dynamical footing via the rms-matching prescription, differences in the functional form of the VDF have a subdominant effect compared to parameter variations within any single model; (iii)~astrophysical uncertainties are most pronounced at the lowest dark matter masses kinematically accessible at a given energy threshold, and become milder at higher masses; (iv)~for the daily modulation signal, parameter variations primarily rescale the modulation amplitude while leaving the phase essentially intact.

A key methodological contribution of this work is the rms-matching prescription introduced in~\cref{sec:prescriptions}. It normalizes the characteristic velocity scale $v_0$ of each halo model so that the rms velocity---equivalently, the mean dark matter kinetic energy at the solar position---is held common across models. Any residual differences in predicted rates can then be attributed to the shape of the velocity distribution rather than to a mismatch in its overall energy scale, in contrast to the conventional identification $v_0 \equiv v_\mathrm{c}$ used in earlier analyses, which conflates shape and energy scale effects. We expect this prescription to be useful beyond the single phonon channel: it can be applied straightforwardly to studies of nuclear recoil, electron recoil, and dark matter absorption searches whenever non-Maxwellian halo models are compared to the SHM. The fact that the uncertainty bands from the SHM, Tsallis, and empirical models largely overlap under the rms-matching prescription also suggests that, at least within the family of analytic benchmark distributions considered here, the functional form of the VDF is not a dominant systematic for direct detection experiments, and astrophysical uncertainty can be assessed by varying the velocity parameters within the SHM.

This work can be extended in several directions. First, our analysis treats $v_\mathrm{c}$, $v_\mathrm{e}$, and $v_\mathrm{esc}$ as independent parameters, which by construction gives a conservative estimate of the uncertainty. Recent cosmological simulation analyses have begun to characterize the correlations among the parameters directly~\cite{Folsom:2025lly}, and incorporating them into rate predictions would tighten the astrophysical uncertainty. Second, the daily modulation results presented here adopt the benchmark detector orientation of Refs.~\cite{Griffin:2018bjn,Coskuner:2021qxo} in order to isolate astrophysical effects and enable direct comparison with previous work. A more comprehensive optimization over crystal orientations and geographic locations could enhance the modulation signal and remains an important direction for future study. Finally, our analysis could be extended to include multi-phonon excitations, which become relevant at higher energy thresholds. Together, these directions point toward a more complete characterization of how astrophysical inputs shape the discovery potential of phonon-based dark matter searches in the coming generation of experiments.

\acknowledgments
We thank Anirudhan A. Madathil and Yao-Yuan Mao for helpful discussions. We acknowledge the use of Claude Opus 4.6/4.7/4.8, Sonnet 4.6, and GPT 5.4/5.5 for literature search and manuscript revision. This work was supported in part by the U.S. National Science Foundation under grant PHY-2412880. This work was performed in part at the Aspen Center for Physics, which is supported by a grant from the Simons Foundation (1161654, Troyer).

\appendix
\crefalias{section}{appendix}

\section{Supplementary tables}
\label{App_A}

In this appendix, we present the relative differences in the signal rate defined in~\cref{eq:relative_difference} for the Tsallis and empirical velocity distributions under the rms-matching prescription (\cref{tab:QQQ_RMS_TSA,tab:QQQ_RMS_EMP}), along with results at the higher phonon threshold $\omega_{\min}=20~\mathrm{meV}$ for the SHM (\cref{tab:QQQ_SHM_20meV}). As in \cref{tab:QQQ_SHM} in the main text, we vary each astrophysical parameter within its conservative range given in \cref{tab:Halo_Parameters} with remaining parameters fixed to their central values. We see that, while the magnitude of the uncertainties varies across halo models and thresholds, their overall trend and parameter dependence remain consistent with the behavior observed in the main text. Under the rms-matching prescription, all three halo models show their strongest sensitivity to variations in the circular speed $v_\mathrm{c}$, with comparatively weaker dependence on $v_\mathrm{e}$ and $v_\mathrm{esc}$ except near kinematic thresholds. Increasing the phonon threshold amplifies the relative differences especially at low dark matter masses, but the hierarchy of uncertainties among the velocity parameters remains unchanged.

\begin{table}[t]
\resizebox{\textwidth}{!}{%
\begin{tabular}{|c|c|c|c|c|c|c|c|c|c|c|c|c|c|}
\rowcolor[HTML]{FFFFFF} 
\multicolumn{14}{c}{Tsallis Model, $\omega_{min}$ = 1 meV} \\ \hline
\multicolumn{1}{|c|}{}                           & Target                     & \multicolumn{3}{c|}{$\text{GaAs}$}       & \multicolumn{3}{c|}{$\text{Al}_2\text{O}_3$}  & \multicolumn{3}{c|}{$\text{SiO}_2$}       & \multicolumn{3}{c|}{$\text{CaWO}_4$}      \\ \hline
\multicolumn{1}{|c|}{$m_\chi$}                   & Mediator                   & \multicolumn{1}{c|}{LDP} & \multicolumn{1}{c|}{LHS} & \multicolumn{1}{c|}{HHS} & \multicolumn{1}{c|}{LDP} & \multicolumn{1}{c|}{LHS} & \multicolumn{1}{c|}{HHS} & \multicolumn{1}{c|}{LDP} & \multicolumn{1}{c|}{LHS} & \multicolumn{1}{c|}{HHS} & \multicolumn{1}{c|}{LDP} & \multicolumn{1}{c|}{LHS} & \multicolumn{1}{c|}{HHS} \\ \hline
\multicolumn{1}{|c|}{}                           & rel.diff ($v_\mathrm{c}$)  & \multicolumn{1}{c|}{}    & \multicolumn{1}{c|}{}    & \multicolumn{1}{c|}{}    & \multicolumn{1}{c|}{\cellcolor[HTML]{F1A983}1.337}  & \multicolumn{1}{c|}{\cellcolor[HTML]{F7C7AC}0.9093} & \multicolumn{1}{c|}{\cellcolor[HTML]{F1A983}1.0121} & \multicolumn{1}{c|}{\cellcolor[HTML]{F1A983}1.3564} & \multicolumn{1}{c|}{\cellcolor[HTML]{F7C7AC}0.8354} & \multicolumn{1}{c|}{\cellcolor[HTML]{F1A983}1.1681} & \multicolumn{1}{c|}{\cellcolor[HTML]{E97132}1.5964} & \multicolumn{1}{c|}{\cellcolor[HTML]{F1A983}1.1551} & \multicolumn{1}{c|}{\cellcolor[HTML]{E97132}1.5209} \\ 
\multicolumn{1}{|c|}{}                           & rel.diff ($v_\mathrm{e}$)  & \multicolumn{1}{c|}{}    & \multicolumn{1}{c|}{}    & \multicolumn{1}{c|}{}    & \multicolumn{1}{c|}{\cellcolor[HTML]{FBE2D5}0.2867} & \multicolumn{1}{c|}{\cellcolor[HTML]{FBE2D5}0.2043} & \multicolumn{1}{c|}{\cellcolor[HTML]{FBE2D5}0.2819} & \multicolumn{1}{c|}{\cellcolor[HTML]{FBE2D5}0.2567} & \multicolumn{1}{c|}{\cellcolor[HTML]{FBE2D5}0.3117} & \multicolumn{1}{c|}{\cellcolor[HTML]{FBE2D5}0.3329} & \multicolumn{1}{c|}{\cellcolor[HTML]{FBE2D5}0.3264} & \multicolumn{1}{c|}{\cellcolor[HTML]{FBE2D5}0.2045} & \multicolumn{1}{c|}{\cellcolor[HTML]{FBE2D5}0.2641} \\ 
\multicolumn{1}{|c|}{\multirow{-3}{*}{0.01 MeV}} & rel.diff ($v_\mathrm{esc}$) & \multicolumn{1}{c|}{}   & \multicolumn{1}{c|}{}    & \multicolumn{1}{c|}{}    & \multicolumn{1}{c|}{\cellcolor[HTML]{FBE2D5}0.2784} & \multicolumn{1}{c|}{\cellcolor[HTML]{FBE2D5}0.3115} & \multicolumn{1}{c|}{\cellcolor[HTML]{FBE2D5}0.1845} & \multicolumn{1}{c|}{\cellcolor[HTML]{FBE2D5}0.28}   & \multicolumn{1}{c|}{\cellcolor[HTML]{FBE2D5}0.1041} & \multicolumn{1}{c|}{\cellcolor[HTML]{FBE2D5}0.2224} & \multicolumn{1}{c|}{\cellcolor[HTML]{FBE2D5}0.3883} & \multicolumn{1}{c|}{\cellcolor[HTML]{FBE2D5}0.2219} & \multicolumn{1}{c|}{\cellcolor[HTML]{FBE2D5}0.3505} \\ \hline
\multicolumn{1}{|c|}{}                           & rel.diff ($v_\mathrm{c}$)  & \multicolumn{1}{c|}{\cellcolor[HTML]{FBE2D5}0.1882} & \multicolumn{1}{c|}{\cellcolor[HTML]{FFFFFF}0.0471} & \multicolumn{1}{c|}{\cellcolor[HTML]{FBE2D5}0.3609} & \multicolumn{1}{c|}{\cellcolor[HTML]{F7C7AC}0.825}  & \multicolumn{1}{c|}{\cellcolor[HTML]{FBE2D5}0.1053} & \multicolumn{1}{c|}{\cellcolor[HTML]{FBE2D5}0.3545} & \multicolumn{1}{c|}{\cellcolor[HTML]{F7C7AC}0.6588} & \multicolumn{1}{c|}{\cellcolor[HTML]{FFFFFF}0.0734} & \multicolumn{1}{c|}{\cellcolor[HTML]{FBE2D5}0.3501} & \multicolumn{1}{c|}{\cellcolor[HTML]{FBE2D5}0.3453} & \multicolumn{1}{c|}{\cellcolor[HTML]{FFFFFF}0.048}  & \multicolumn{1}{c|}{\cellcolor[HTML]{FBE2D5}0.3637} \\ 
\multicolumn{1}{|c|}{}                           & rel.diff ($v_\mathrm{e}$)  & \multicolumn{1}{c|}{\cellcolor[HTML]{FFFFFF}0.0827} & \multicolumn{1}{c|}{\cellcolor[HTML]{FFFFFF}0.0051} & \multicolumn{1}{c|}{\cellcolor[HTML]{FBE2D5}0.121}  & \multicolumn{1}{c|}{\cellcolor[HTML]{FBE2D5}0.2563} & \multicolumn{1}{c|}{\cellcolor[HTML]{FFFFFF}0.052}  & \multicolumn{1}{c|}{\cellcolor[HTML]{FBE2D5}0.117}  & \multicolumn{1}{c|}{\cellcolor[HTML]{FBE2D5}0.2057} & \multicolumn{1}{c|}{\cellcolor[HTML]{FFFFFF}0.034}  & \multicolumn{1}{c|}{\cellcolor[HTML]{FBE2D5}0.1161} & \multicolumn{1}{c|}{\cellcolor[HTML]{FBE2D5}0.1274} & \multicolumn{1}{c|}{\cellcolor[HTML]{FFFFFF}0.0019} & \multicolumn{1}{c|}{\cellcolor[HTML]{FBE2D5}0.121}  \\ 
\multicolumn{1}{|c|}{\multirow{-3}{*}{0.1 MeV}}  & rel.diff ($v_\mathrm{esc}$) & \multicolumn{1}{c|}{\cellcolor[HTML]{FFFFFF}0.02}   & \multicolumn{1}{c|}{\cellcolor[HTML]{FFFFFF}0.0507} & \multicolumn{1}{c|}{\cellcolor[HTML]{FFFFFF}0.0566} & \multicolumn{1}{c|}{\cellcolor[HTML]{FBE2D5}0.1317} & \multicolumn{1}{c|}{\cellcolor[HTML]{FFFFFF}0.0494} & \multicolumn{1}{c|}{\cellcolor[HTML]{FFFFFF}0.055}  & \multicolumn{1}{c|}{\cellcolor[HTML]{FBE2D5}0.1123} & \multicolumn{1}{c|}{\cellcolor[HTML]{FFFFFF}0.0395} & \multicolumn{1}{c|}{\cellcolor[HTML]{FFFFFF}0.0551} & \multicolumn{1}{c|}{\cellcolor[HTML]{FFFFFF}0.0477} & \multicolumn{1}{c|}{\cellcolor[HTML]{FFFFFF}0.0468} & \multicolumn{1}{c|}{\cellcolor[HTML]{FFFFFF}0.0564} \\ \hline
\multicolumn{1}{|c|}{}                           & rel.diff ($v_\mathrm{c}$)  & \multicolumn{1}{c|}{\cellcolor[HTML]{FFFFFF}0.0382} & \multicolumn{1}{c|}{\cellcolor[HTML]{FBE2D5}0.1377} & \multicolumn{1}{c|}{\cellcolor[HTML]{FBE2D5}0.4996} & \multicolumn{1}{c|}{\cellcolor[HTML]{FFFFFF}0.0013} & \multicolumn{1}{c|}{\cellcolor[HTML]{FBE2D5}0.1398} & \multicolumn{1}{c|}{\cellcolor[HTML]{FBE2D5}0.4508} & \multicolumn{1}{c|}{\cellcolor[HTML]{FFFFFF}0.0216} & \multicolumn{1}{c|}{\cellcolor[HTML]{FBE2D5}0.1385} & \multicolumn{1}{c|}{\cellcolor[HTML]{FBE2D5}0.4743} & \multicolumn{1}{c|}{\cellcolor[HTML]{FFFFFF}0.0115} & \multicolumn{1}{c|}{\cellcolor[HTML]{FBE2D5}0.1342} & \multicolumn{1}{c|}{\cellcolor[HTML]{FBE2D5}0.4938} \\ 
\multicolumn{1}{|c|}{}                           & rel.diff ($v_\mathrm{e}$)  & \multicolumn{1}{c|}{\cellcolor[HTML]{FFFFFF}0.013}  & \multicolumn{1}{c|}{\cellcolor[HTML]{FFFFFF}0.0618} & \multicolumn{1}{c|}{\cellcolor[HTML]{FBE2D5}0.1461} & \multicolumn{1}{c|}{\cellcolor[HTML]{FFFFFF}0.0075} & \multicolumn{1}{c|}{\cellcolor[HTML]{FFFFFF}0.0588} & \multicolumn{1}{c|}{\cellcolor[HTML]{FBE2D5}0.142}  & \multicolumn{1}{c|}{\cellcolor[HTML]{FFFFFF}0.0159} & \multicolumn{1}{c|}{\cellcolor[HTML]{FFFFFF}0.0626} & \multicolumn{1}{c|}{\cellcolor[HTML]{FBE2D5}0.1428} & \multicolumn{1}{c|}{\cellcolor[HTML]{FFFFFF}0.0002} & \multicolumn{1}{c|}{\cellcolor[HTML]{FFFFFF}0.0547} & \multicolumn{1}{c|}{\cellcolor[HTML]{FBE2D5}0.1633} \\ 
\multicolumn{1}{|c|}{\multirow{-3}{*}{1 MeV}}    & rel.diff ($v_\mathrm{esc}$) & \multicolumn{1}{c|}{\cellcolor[HTML]{FFFFFF}0.0044} & \multicolumn{1}{c|}{\cellcolor[HTML]{FFFFFF}0.0373} & \multicolumn{1}{c|}{\cellcolor[HTML]{FFFFFF}0.0906} & \multicolumn{1}{c|}{\cellcolor[HTML]{FFFFFF}0.0018} & \multicolumn{1}{c|}{\cellcolor[HTML]{FFFFFF}0.0079} & \multicolumn{1}{c|}{\cellcolor[HTML]{FFFFFF}0.0775} & \multicolumn{1}{c|}{\cellcolor[HTML]{FFFFFF}0.0039} & \multicolumn{1}{c|}{\cellcolor[HTML]{FFFFFF}0.0239} & \multicolumn{1}{c|}{\cellcolor[HTML]{FFFFFF}0.0825} & \multicolumn{1}{c|}{\cellcolor[HTML]{FFFFFF}0.0005} & \multicolumn{1}{c|}{\cellcolor[HTML]{FFFFFF}0.0117} & \multicolumn{1}{c|}{\cellcolor[HTML]{FFFFFF}0.0835} \\ \hline
\multicolumn{1}{|c|}{}                           & rel.diff ($v_\mathrm{c}$)  & \multicolumn{1}{c|}{\cellcolor[HTML]{FBE2D5}0.1039} & \multicolumn{1}{c|}{\cellcolor[HTML]{FBE2D5}0.1389} & \multicolumn{1}{c|}{\cellcolor[HTML]{FBE2D5}0.4121} & \multicolumn{1}{c|}{\cellcolor[HTML]{FFFFFF}0.0909} & \multicolumn{1}{c|}{\cellcolor[HTML]{FBE2D5}0.1422} & \multicolumn{1}{c|}{\cellcolor[HTML]{FBE2D5}0.3995} & \multicolumn{1}{c|}{\cellcolor[HTML]{FFFFFF}0.0936} & \multicolumn{1}{c|}{\cellcolor[HTML]{FBE2D5}0.1475} & \multicolumn{1}{c|}{\cellcolor[HTML]{FBE2D5}0.3605} & \multicolumn{1}{c|}{\cellcolor[HTML]{FFFFFF}0.0967} & \multicolumn{1}{c|}{\cellcolor[HTML]{FBE2D5}0.1387} & \multicolumn{1}{c|}{\cellcolor[HTML]{FBE2D5}0.4224} \\ 
\multicolumn{1}{|c|}{}                           & rel.diff ($v_\mathrm{e}$)  & \multicolumn{1}{c|}{\cellcolor[HTML]{FFFFFF}0.0371} & \multicolumn{1}{c|}{\cellcolor[HTML]{FFFFFF}0.0682} & \multicolumn{1}{c|}{\cellcolor[HTML]{FBE2D5}0.1349} & \multicolumn{1}{c|}{\cellcolor[HTML]{FFFFFF}0.0381} & \multicolumn{1}{c|}{\cellcolor[HTML]{FFFFFF}0.0532} & \multicolumn{1}{c|}{\cellcolor[HTML]{FBE2D5}0.1332} & \multicolumn{1}{c|}{\cellcolor[HTML]{FFFFFF}0.0369} & \multicolumn{1}{c|}{\cellcolor[HTML]{FFFFFF}0.0633} & \multicolumn{1}{c|}{\cellcolor[HTML]{FBE2D5}0.1212} & \multicolumn{1}{c|}{\cellcolor[HTML]{FFFFFF}0.0388} & \multicolumn{1}{c|}{\cellcolor[HTML]{FFFFFF}0.06}   & \multicolumn{1}{c|}{\cellcolor[HTML]{FBE2D5}0.1362} \\ 
\multicolumn{1}{|c|}{\multirow{-3}{*}{10 MeV}}   & rel.diff ($v_\mathrm{esc}$) & \multicolumn{1}{c|}{\cellcolor[HTML]{FFFFFF}0.0101} & \multicolumn{1}{c|}{\cellcolor[HTML]{FFFFFF}0.0603} & \multicolumn{1}{c|}{\cellcolor[HTML]{FFFFFF}0.0692} & \multicolumn{1}{c|}{\cellcolor[HTML]{FFFFFF}0.0174} & \multicolumn{1}{c|}{\cellcolor[HTML]{FFFFFF}0.0318} & \multicolumn{1}{c|}{\cellcolor[HTML]{FFFFFF}0.0633} & \multicolumn{1}{c|}{\cellcolor[HTML]{FFFFFF}0.0324} & \multicolumn{1}{c|}{\cellcolor[HTML]{FFFFFF}0.0354} & \multicolumn{1}{c|}{\cellcolor[HTML]{FFFFFF}0.0556} & \multicolumn{1}{c|}{\cellcolor[HTML]{FFFFFF}0.0115} & \multicolumn{1}{c|}{\cellcolor[HTML]{FFFFFF}0.0166} & \multicolumn{1}{c|}{\cellcolor[HTML]{FFFFFF}0.0662} \\ \hline
\multicolumn{1}{|c|}{}                           & rel.diff ($v_\mathrm{c}$)  & \multicolumn{1}{c|}{\cellcolor[HTML]{FBE2D5}0.1326} & \multicolumn{1}{c|}{\cellcolor[HTML]{FBE2D5}0.1466} & \multicolumn{1}{c|}{\cellcolor[HTML]{FBE2D5}0.1273} & \multicolumn{1}{c|}{\cellcolor[HTML]{FBE2D5}0.127}  & \multicolumn{1}{c|}{\cellcolor[HTML]{FBE2D5}0.1449} & \multicolumn{1}{c|}{\cellcolor[HTML]{FBE2D5}0.1266} & \multicolumn{1}{c|}{\cellcolor[HTML]{FBE2D5}0.1301} & \multicolumn{1}{c|}{\cellcolor[HTML]{FBE2D5}0.1469} & \multicolumn{1}{c|}{\cellcolor[HTML]{FBE2D5}0.1292} & \multicolumn{1}{c|}{\cellcolor[HTML]{FBE2D5}0.1288} & \multicolumn{1}{c|}{\cellcolor[HTML]{FBE2D5}0.1392} & \multicolumn{1}{c|}{\cellcolor[HTML]{FBE2D5}0.1106} \\ 
\multicolumn{1}{|c|}{}                           & rel.diff ($v_\mathrm{e}$)  & \multicolumn{1}{c|}{\cellcolor[HTML]{FFFFFF}0.0547} & \multicolumn{1}{c|}{\cellcolor[HTML]{FFFFFF}0.0418} & \multicolumn{1}{c|}{\cellcolor[HTML]{FFFFFF}0.0506} & \multicolumn{1}{c|}{\cellcolor[HTML]{FFFFFF}0.0541} & \multicolumn{1}{c|}{\cellcolor[HTML]{FFFFFF}0.065}  & \multicolumn{1}{c|}{\cellcolor[HTML]{FFFFFF}0.0516} & \multicolumn{1}{c|}{\cellcolor[HTML]{FFFFFF}0.0553} & \multicolumn{1}{c|}{\cellcolor[HTML]{FFFFFF}0.0591} & \multicolumn{1}{c|}{\cellcolor[HTML]{FFFFFF}0.0515} & \multicolumn{1}{c|}{\cellcolor[HTML]{FFFFFF}0.0567} & \multicolumn{1}{c|}{\cellcolor[HTML]{FFFFFF}0.0656} & \multicolumn{1}{c|}{\cellcolor[HTML]{FFFFFF}0.0429} \\ 
\multicolumn{1}{|c|}{\multirow{-3}{*}{100 MeV}}  & rel.diff ($v_\mathrm{esc}$) & \multicolumn{1}{c|}{\cellcolor[HTML]{FFFFFF}0.0135} & \multicolumn{1}{c|}{\cellcolor[HTML]{FFFFFF}0.0633} & \multicolumn{1}{c|}{\cellcolor[HTML]{FFFFFF}0.0169} & \multicolumn{1}{c|}{\cellcolor[HTML]{FFFFFF}0.0108} & \multicolumn{1}{c|}{\cellcolor[HTML]{FFFFFF}0.0647} & \multicolumn{1}{c|}{\cellcolor[HTML]{FFFFFF}0.0173} & \multicolumn{1}{c|}{\cellcolor[HTML]{FFFFFF}0.008}  & \multicolumn{1}{c|}{\cellcolor[HTML]{FFFFFF}0.0052} & \multicolumn{1}{c|}{\cellcolor[HTML]{FFFFFF}0.0217} & \multicolumn{1}{c|}{\cellcolor[HTML]{FFFFFF}0.0299} & \multicolumn{1}{c|}{\cellcolor[HTML]{FFFFFF}0.0093} & \multicolumn{1}{c|}{\cellcolor[HTML]{FFFFFF}0.014}  \\ \hline
\end{tabular}%
}
\caption{Same as~\cref{tab:QQQ_SHM} for the Tsallis model.
\label{tab:QQQ_RMS_TSA}}
\end{table}
\begin{table}[t]
\resizebox{\textwidth}{!}{%
\begin{tabular}{|c|c|c|c|c|c|c|c|c|c|c|c|c|c|}
\rowcolor[HTML]{FFFFFF} 
\multicolumn{14}{c}{Empirical Model, $\omega_{min}$ = 1 meV} \\ \hline
\multicolumn{1}{|c|}{}                           & Target                     & \multicolumn{3}{c|}{$\text{GaAs}$}       & \multicolumn{3}{c|}{$\text{Al}_2\text{O}_3$}  & \multicolumn{3}{c|}{$\text{SiO}_2$}       & \multicolumn{3}{c|}{$\text{CaWO}_4$}      \\ \hline
\multicolumn{1}{|c|}{$m_\chi$}                   & Mediator                   & \multicolumn{1}{c|}{LDP} & \multicolumn{1}{c|}{LHS} & \multicolumn{1}{c|}{HHS} & \multicolumn{1}{c|}{LDP} & \multicolumn{1}{c|}{LHS} & \multicolumn{1}{c|}{HHS} & \multicolumn{1}{c|}{LDP} & \multicolumn{1}{c|}{LHS} & \multicolumn{1}{c|}{HHS} & \multicolumn{1}{c|}{LDP} & \multicolumn{1}{c|}{LHS} & \multicolumn{1}{c|}{HHS} \\ \hline
\multicolumn{1}{|c|}{}                           & rel.diff ($v_\mathrm{c}$)  & \multicolumn{1}{c|}{}    & \multicolumn{1}{c|}{}    & \multicolumn{1}{c|}{}    & \multicolumn{1}{c|}{\cellcolor[HTML]{F1A983}1.2247} & \multicolumn{1}{c|}{\cellcolor[HTML]{F7C7AC}0.8821} & \multicolumn{1}{c|}{\cellcolor[HTML]{F7C7AC}0.9673} & \multicolumn{1}{c|}{\cellcolor[HTML]{F1A983}1.2562} & \multicolumn{1}{c|}{\cellcolor[HTML]{F7C7AC}0.826}  & \multicolumn{1}{c|}{\cellcolor[HTML]{F1A983}1.0935} & \multicolumn{1}{c|}{\cellcolor[HTML]{F1A983}1.465}  & \multicolumn{1}{c|}{\cellcolor[HTML]{F1A983}1.0803} & \multicolumn{1}{c|}{\cellcolor[HTML]{F1A983}1.4002} \\ 
\multicolumn{1}{|c|}{}                           & rel.diff ($v_\mathrm{e}$)  & \multicolumn{1}{c|}{}    & \multicolumn{1}{c|}{}    & \multicolumn{1}{c|}{}    & \multicolumn{1}{c|}{\cellcolor[HTML]{FBE2D5}0.2881} & \multicolumn{1}{c|}{\cellcolor[HTML]{FBE2D5}0.1905} & \multicolumn{1}{c|}{\cellcolor[HTML]{FBE2D5}0.2648} & \multicolumn{1}{c|}{\cellcolor[HTML]{FBE2D5}0.269}  & \multicolumn{1}{c|}{\cellcolor[HTML]{FBE2D5}0.2741} & \multicolumn{1}{c|}{\cellcolor[HTML]{FBE2D5}0.3219} & \multicolumn{1}{c|}{\cellcolor[HTML]{FBE2D5}0.4052} & \multicolumn{1}{c|}{\cellcolor[HTML]{FBE2D5}0.1847} & \multicolumn{1}{c|}{\cellcolor[HTML]{FBE2D5}0.2996} \\ 
\multicolumn{1}{|c|}{\multirow{-3}{*}{0.01 MeV}} & rel.diff ($v_\mathrm{esc}$) & \multicolumn{1}{c|}{}   & \multicolumn{1}{c|}{}    & \multicolumn{1}{c|}{}    & \multicolumn{1}{c|}{\cellcolor[HTML]{FBE2D5}0.3616} & \multicolumn{1}{c|}{\cellcolor[HTML]{FBE2D5}0.2816} & \multicolumn{1}{c|}{\cellcolor[HTML]{FBE2D5}0.2301} & \multicolumn{1}{c|}{\cellcolor[HTML]{F7C7AC}0.5118} & \multicolumn{1}{c|}{\cellcolor[HTML]{FFFFFF}0.0962} & \multicolumn{1}{c|}{\cellcolor[HTML]{FBE2D5}0.2916} & \multicolumn{1}{c|}{\cellcolor[HTML]{F7C7AC}0.9642} & \multicolumn{1}{c|}{\cellcolor[HTML]{FBE2D5}0.2845} & \multicolumn{1}{c|}{\cellcolor[HTML]{F7C7AC}0.6784} \\ \hline
\multicolumn{1}{|c|}{}                           & rel.diff ($v_\mathrm{c}$)  & \multicolumn{1}{c|}{\cellcolor[HTML]{FBE2D5}0.2181} & \multicolumn{1}{c|}{\cellcolor[HTML]{FFFFFF}0.0584} & \multicolumn{1}{c|}{\cellcolor[HTML]{FBE2D5}0.3639} & \multicolumn{1}{c|}{\cellcolor[HTML]{F7C7AC}0.7995} & \multicolumn{1}{c|}{\cellcolor[HTML]{FBE2D5}0.1116} & \multicolumn{1}{c|}{\cellcolor[HTML]{FBE2D5}0.3573} & \multicolumn{1}{c|}{\cellcolor[HTML]{F7C7AC}0.6583} & \multicolumn{1}{c|}{\cellcolor[HTML]{FFFFFF}0.0773} & \multicolumn{1}{c|}{\cellcolor[HTML]{FBE2D5}0.3545} & \multicolumn{1}{c|}{\cellcolor[HTML]{FBE2D5}0.3646} & \multicolumn{1}{c|}{\cellcolor[HTML]{FFFFFF}0.0581} & \multicolumn{1}{c|}{\cellcolor[HTML]{FBE2D5}0.3667} \\ 
\multicolumn{1}{|c|}{}                           & rel.diff ($v_\mathrm{e}$)  & \multicolumn{1}{c|}{\cellcolor[HTML]{FFFFFF}0.0912} & \multicolumn{1}{c|}{\cellcolor[HTML]{FFFFFF}0.0084} & \multicolumn{1}{c|}{\cellcolor[HTML]{FBE2D5}0.1211} & \multicolumn{1}{c|}{\cellcolor[HTML]{FBE2D5}0.2271} & \multicolumn{1}{c|}{\cellcolor[HTML]{FFFFFF}0.0548} & \multicolumn{1}{c|}{\cellcolor[HTML]{FBE2D5}0.1169} & \multicolumn{1}{c|}{\cellcolor[HTML]{FBE2D5}0.1983} & \multicolumn{1}{c|}{\cellcolor[HTML]{FFFFFF}0.0342} & \multicolumn{1}{c|}{\cellcolor[HTML]{FBE2D5}0.116}  & \multicolumn{1}{c|}{\cellcolor[HTML]{FBE2D5}0.1273} & \multicolumn{1}{c|}{\cellcolor[HTML]{FFFFFF}0.0033} & \multicolumn{1}{c|}{\cellcolor[HTML]{FBE2D5}0.121}  \\ 
\multicolumn{1}{|c|}{\multirow{-3}{*}{0.1 MeV}}  & rel.diff ($v_\mathrm{esc}$) & \multicolumn{1}{c|}{\cellcolor[HTML]{FFFFFF}0.0088} & \multicolumn{1}{c|}{\cellcolor[HTML]{FFFFFF}0.054}  & \multicolumn{1}{c|}{\cellcolor[HTML]{FFFFFF}0.0552} & \multicolumn{1}{c|}{\cellcolor[HTML]{FBE2D5}0.1291} & \multicolumn{1}{c|}{\cellcolor[HTML]{FFFFFF}0.0426} & \multicolumn{1}{c|}{\cellcolor[HTML]{FFFFFF}0.0541} & \multicolumn{1}{c|}{\cellcolor[HTML]{FBE2D5}0.1167} & \multicolumn{1}{c|}{\cellcolor[HTML]{FFFFFF}0.0371} & \multicolumn{1}{c|}{\cellcolor[HTML]{FFFFFF}0.0522} & \multicolumn{1}{c|}{\cellcolor[HTML]{FFFFFF}0.0284} & \multicolumn{1}{c|}{\cellcolor[HTML]{FFFFFF}0.0511} & \multicolumn{1}{c|}{\cellcolor[HTML]{FFFFFF}0.0552} \\ \hline
\multicolumn{1}{|c|}{}                           & rel.diff ($v_\mathrm{c}$)  & \multicolumn{1}{c|}{\cellcolor[HTML]{FFFFFF}0.0447} & \multicolumn{1}{c|}{\cellcolor[HTML]{FBE2D5}0.1363} & \multicolumn{1}{c|}{\cellcolor[HTML]{FBE2D5}0.4898} & \multicolumn{1}{c|}{\cellcolor[HTML]{FFFFFF}0.0075} & \multicolumn{1}{c|}{\cellcolor[HTML]{FBE2D5}0.1352} & \multicolumn{1}{c|}{\cellcolor[HTML]{FBE2D5}0.4469} & \multicolumn{1}{c|}{\cellcolor[HTML]{FFFFFF}0.0165} & \multicolumn{1}{c|}{\cellcolor[HTML]{FBE2D5}0.134}  & \multicolumn{1}{c|}{\cellcolor[HTML]{FBE2D5}0.4706} & \multicolumn{1}{c|}{\cellcolor[HTML]{FFFFFF}0.0183} & \multicolumn{1}{c|}{\cellcolor[HTML]{FBE2D5}0.1331} & \multicolumn{1}{c|}{\cellcolor[HTML]{FBE2D5}0.4908} \\ 
\multicolumn{1}{|c|}{}                           & rel.diff ($v_\mathrm{e}$)  & \multicolumn{1}{c|}{\cellcolor[HTML]{FFFFFF}0.013}  & \multicolumn{1}{c|}{\cellcolor[HTML]{FFFFFF}0.0677} & \multicolumn{1}{c|}{\cellcolor[HTML]{FBE2D5}0.1495} & \multicolumn{1}{c|}{\cellcolor[HTML]{FFFFFF}0.0118} & \multicolumn{1}{c|}{\cellcolor[HTML]{FFFFFF}0.0653} & \multicolumn{1}{c|}{\cellcolor[HTML]{FBE2D5}0.143}  & \multicolumn{1}{c|}{\cellcolor[HTML]{FFFFFF}0.0213} & \multicolumn{1}{c|}{\cellcolor[HTML]{FFFFFF}0.069}  & \multicolumn{1}{c|}{\cellcolor[HTML]{FBE2D5}0.1435} & \multicolumn{1}{c|}{\cellcolor[HTML]{FFFFFF}0.0027} & \multicolumn{1}{c|}{\cellcolor[HTML]{FFFFFF}0.0606} & \multicolumn{1}{c|}{\cellcolor[HTML]{FBE2D5}0.162}  \\ 
\multicolumn{1}{|c|}{\multirow{-3}{*}{1 MeV}}    & rel.diff ($v_\mathrm{esc}$) & \multicolumn{1}{c|}{\cellcolor[HTML]{FFFFFF}0.0011} & \multicolumn{1}{c|}{\cellcolor[HTML]{FFFFFF}0.0321} & \multicolumn{1}{c|}{\cellcolor[HTML]{FBE2D5}0.1207} & \multicolumn{1}{c|}{\cellcolor[HTML]{FFFFFF}0.0052} & \multicolumn{1}{c|}{\cellcolor[HTML]{FFFFFF}0.0065} & \multicolumn{1}{c|}{\cellcolor[HTML]{FFFFFF}0.0923} & \multicolumn{1}{c|}{\cellcolor[HTML]{FFFFFF}0.0049} & \multicolumn{1}{c|}{\cellcolor[HTML]{FFFFFF}0.0207} & \multicolumn{1}{c|}{\cellcolor[HTML]{FFFFFF}0.099}  & \multicolumn{1}{c|}{\cellcolor[HTML]{FFFFFF}0.0028} & \multicolumn{1}{c|}{\cellcolor[HTML]{FFFFFF}0.0074} & \multicolumn{1}{c|}{\cellcolor[HTML]{FFFFFF}0.0959} \\ \hline
\multicolumn{1}{|c|}{}                           & rel.diff ($v_\mathrm{c}$)  & \multicolumn{1}{c|}{\cellcolor[HTML]{FBE2D5}0.1054} & \multicolumn{1}{c|}{\cellcolor[HTML]{FBE2D5}0.1321} & \multicolumn{1}{c|}{\cellcolor[HTML]{FBE2D5}0.4159} & \multicolumn{1}{c|}{\cellcolor[HTML]{FFFFFF}0.0944} & \multicolumn{1}{c|}{\cellcolor[HTML]{FBE2D5}0.136}  & \multicolumn{1}{c|}{\cellcolor[HTML]{FBE2D5}0.4047} & \multicolumn{1}{c|}{\cellcolor[HTML]{FFFFFF}0.0969} & \multicolumn{1}{c|}{\cellcolor[HTML]{FBE2D5}0.1455} & \multicolumn{1}{c|}{\cellcolor[HTML]{FBE2D5}0.3688} & \multicolumn{1}{c|}{\cellcolor[HTML]{FFFFFF}0.099}  & \multicolumn{1}{c|}{\cellcolor[HTML]{FBE2D5}0.1352} & \multicolumn{1}{c|}{\cellcolor[HTML]{FBE2D5}0.4247} \\ 
\multicolumn{1}{|c|}{}                           & rel.diff ($v_\mathrm{e}$)  & \multicolumn{1}{c|}{\cellcolor[HTML]{FFFFFF}0.0405} & \multicolumn{1}{c|}{\cellcolor[HTML]{FFFFFF}0.0749} & \multicolumn{1}{c|}{\cellcolor[HTML]{FBE2D5}0.1328} & \multicolumn{1}{c|}{\cellcolor[HTML]{FFFFFF}0.0403} & \multicolumn{1}{c|}{\cellcolor[HTML]{FFFFFF}0.06}   & \multicolumn{1}{c|}{\cellcolor[HTML]{FBE2D5}0.1309} & \multicolumn{1}{c|}{\cellcolor[HTML]{FFFFFF}0.0391} & \multicolumn{1}{c|}{\cellcolor[HTML]{FFFFFF}0.0703} & \multicolumn{1}{c|}{\cellcolor[HTML]{FBE2D5}0.1199} & \multicolumn{1}{c|}{\cellcolor[HTML]{FFFFFF}0.0416} & \multicolumn{1}{c|}{\cellcolor[HTML]{FFFFFF}0.0664} & \multicolumn{1}{c|}{\cellcolor[HTML]{FBE2D5}0.134}  \\ 
\multicolumn{1}{|c|}{\multirow{-3}{*}{10 MeV}}   & rel.diff ($v_\mathrm{esc}$) & \multicolumn{1}{c|}{\cellcolor[HTML]{FFFFFF}0.0078} & \multicolumn{1}{c|}{\cellcolor[HTML]{FFFFFF}0.0583} & \multicolumn{1}{c|}{\cellcolor[HTML]{FFFFFF}0.0647} & \multicolumn{1}{c|}{\cellcolor[HTML]{FFFFFF}0.0132} & \multicolumn{1}{c|}{\cellcolor[HTML]{FFFFFF}0.0345} & \multicolumn{1}{c|}{\cellcolor[HTML]{FFFFFF}0.0583} & \multicolumn{1}{c|}{\cellcolor[HTML]{FFFFFF}0.0282} & \multicolumn{1}{c|}{\cellcolor[HTML]{FFFFFF}0.03}   & \multicolumn{1}{c|}{\cellcolor[HTML]{FFFFFF}0.0463} & \multicolumn{1}{c|}{\cellcolor[HTML]{FFFFFF}0.009}  & \multicolumn{1}{c|}{\cellcolor[HTML]{FFFFFF}0.0148} & \multicolumn{1}{c|}{\cellcolor[HTML]{FFFFFF}0.066}  \\ \hline
\multicolumn{1}{|c|}{}                           & rel.diff ($v_\mathrm{c}$)  & \multicolumn{1}{c|}{\cellcolor[HTML]{FBE2D5}0.1299} & \multicolumn{1}{c|}{\cellcolor[HTML]{FBE2D5}0.1424} & \multicolumn{1}{c|}{\cellcolor[HTML]{FBE2D5}0.1356} & \multicolumn{1}{c|}{\cellcolor[HTML]{FBE2D5}0.124}  & \multicolumn{1}{c|}{\cellcolor[HTML]{FBE2D5}0.1392} & \multicolumn{1}{c|}{\cellcolor[HTML]{FBE2D5}0.1332} & \multicolumn{1}{c|}{\cellcolor[HTML]{FBE2D5}0.1261} & \multicolumn{1}{c|}{\cellcolor[HTML]{FBE2D5}0.1428} & \multicolumn{1}{c|}{\cellcolor[HTML]{FBE2D5}0.1363} & \multicolumn{1}{c|}{\cellcolor[HTML]{FBE2D5}0.125}  & \multicolumn{1}{c|}{\cellcolor[HTML]{FBE2D5}0.1302} & \multicolumn{1}{c|}{\cellcolor[HTML]{FBE2D5}0.1162} \\ 
\multicolumn{1}{|c|}{}                           & rel.diff ($v_\mathrm{e}$)  & \multicolumn{1}{c|}{\cellcolor[HTML]{FFFFFF}0.0606} & \multicolumn{1}{c|}{\cellcolor[HTML]{FFFFFF}0.0486} & \multicolumn{1}{c|}{\cellcolor[HTML]{FFFFFF}0.0549} & \multicolumn{1}{c|}{\cellcolor[HTML]{FFFFFF}0.0595} & \multicolumn{1}{c|}{\cellcolor[HTML]{FFFFFF}0.0702} & \multicolumn{1}{c|}{\cellcolor[HTML]{FFFFFF}0.0559} & \multicolumn{1}{c|}{\cellcolor[HTML]{FFFFFF}0.061}  & \multicolumn{1}{c|}{\cellcolor[HTML]{FFFFFF}0.0656} & \multicolumn{1}{c|}{\cellcolor[HTML]{FFFFFF}0.056}  & \multicolumn{1}{c|}{\cellcolor[HTML]{FFFFFF}0.0623} & \multicolumn{1}{c|}{\cellcolor[HTML]{FFFFFF}0.0718} & \multicolumn{1}{c|}{\cellcolor[HTML]{FFFFFF}0.0449} \\ 
\multicolumn{1}{|c|}{\multirow{-3}{*}{100 MeV}}  & rel.diff ($v_\mathrm{esc}$) & \multicolumn{1}{c|}{\cellcolor[HTML]{FFFFFF}0.0095} & \multicolumn{1}{c|}{\cellcolor[HTML]{FFFFFF}0.0648} & \multicolumn{1}{c|}{\cellcolor[HTML]{FFFFFF}0.0092} & \multicolumn{1}{c|}{\cellcolor[HTML]{FFFFFF}0.0092} & \multicolumn{1}{c|}{\cellcolor[HTML]{FFFFFF}0.0566} & \multicolumn{1}{c|}{\cellcolor[HTML]{FFFFFF}0.0103} & \multicolumn{1}{c|}{\cellcolor[HTML]{FFFFFF}0.0058} & \multicolumn{1}{c|}{\cellcolor[HTML]{FFFFFF}0.0025} & \multicolumn{1}{c|}{\cellcolor[HTML]{FFFFFF}0.013}  & \multicolumn{1}{c|}{\cellcolor[HTML]{FFFFFF}0.0285} & \multicolumn{1}{c|}{\cellcolor[HTML]{FFFFFF}0.0063} & \multicolumn{1}{c|}{\cellcolor[HTML]{FFFFFF}0.0104} \\ \hline
\end{tabular}%
}
\caption{Same as~\cref{tab:QQQ_SHM} for the empirical model.
\label{tab:QQQ_RMS_EMP}}
\end{table}
\begin{table}[t]
\resizebox{\textwidth}{!}{%
\begin{tabular}{|c|c|c|c|c|c|c|c|c|c|c|c|c|c|}
\rowcolor[HTML]{FFFFFF} 
\multicolumn{14}{c}{Standard Halo Model, $\omega_{min}$ = 20 meV} \\ \hline
\multicolumn{1}{|c|}{}                           & Target                     & \multicolumn{3}{c|}{$\text{GaAs}$}       & \multicolumn{3}{c|}{$\text{Al}_2\text{O}_3$}  & \multicolumn{3}{c|}{$\text{SiO}_2$}       & \multicolumn{3}{c|}{$\text{CaWO}_4$}      \\ \hline
\multicolumn{1}{|c|}{$m_\chi$}                   & Mediator                   & \multicolumn{1}{c|}{LDP} & \multicolumn{1}{c|}{LHS} & \multicolumn{1}{c|}{HHS} & \multicolumn{1}{c|}{LDP} & \multicolumn{1}{c|}{LHS} & \multicolumn{1}{c|}{HHS} & \multicolumn{1}{c|}{LDP} & \multicolumn{1}{c|}{LHS} & \multicolumn{1}{c|}{HHS} & \multicolumn{1}{c|}{LDP} & \multicolumn{1}{c|}{LHS} & \multicolumn{1}{c|}{HHS} \\ \hline
\multicolumn{1}{|c|}{}                           & rel.diff ($v_\mathrm{c}$)  & \multicolumn{1}{c|}{}    & \multicolumn{1}{c|}{}    & \multicolumn{1}{c|}{}    & \multicolumn{1}{c|}{}    & \multicolumn{1}{c|}{}    & \multicolumn{1}{c|}{}    & \multicolumn{1}{c|}{\cellcolor[HTML]{BE5014}3.899}  & \multicolumn{1}{c|}{\cellcolor[HTML]{BE5014}3.8351} & \multicolumn{1}{c|}{\cellcolor[HTML]{BE5014}3.9204} & \multicolumn{1}{c|}{\cellcolor[HTML]{BE5014}3.3432} & \multicolumn{1}{c|}{\cellcolor[HTML]{BE5014}2.856}  & \multicolumn{1}{c|}{\cellcolor[HTML]{BE5014}3.0987} \\ 
\multicolumn{1}{|c|}{}                           & rel.diff ($v_\mathrm{e}$)  & \multicolumn{1}{c|}{}    & \multicolumn{1}{c|}{}    & \multicolumn{1}{c|}{}    & \multicolumn{1}{c|}{}    & \multicolumn{1}{c|}{}    & \multicolumn{1}{c|}{}    & \multicolumn{1}{c|}{\cellcolor[HTML]{F1A983}1.0885} & \multicolumn{1}{c|}{\cellcolor[HTML]{F1A983}1.032}  & \multicolumn{1}{c|}{\cellcolor[HTML]{F1A983}1.1704} & \multicolumn{1}{c|}{\cellcolor[HTML]{F1A983}1.0172} & \multicolumn{1}{c|}{\cellcolor[HTML]{F7C7AC}0.5786} & \multicolumn{1}{c|}{\cellcolor[HTML]{F7C7AC}0.6963} \\ 
\multicolumn{1}{|c|}{\multirow{-3}{*}{0.01 MeV}} & rel.diff ($v_\mathrm{esc}$) & \multicolumn{1}{c|}{}   & \multicolumn{1}{c|}{}    & \multicolumn{1}{c|}{}    & \multicolumn{1}{c|}{}    & \multicolumn{1}{c|}{}    & \multicolumn{1}{c|}{}    & \multicolumn{1}{c|}{\cellcolor[HTML]{BE5014}2.4032} & \multicolumn{1}{c|}{\cellcolor[HTML]{BE5014}2.0544} & \multicolumn{1}{c|}{\cellcolor[HTML]{BE5014}2.5377} & \multicolumn{1}{c|}{\cellcolor[HTML]{BE5014}2.5201} & \multicolumn{1}{c|}{\cellcolor[HTML]{F1A983}1.1136} & \multicolumn{1}{c|}{\cellcolor[HTML]{F1A983}1.4276} \\ \hline
\multicolumn{1}{|c|}{}                           & rel.diff ($v_\mathrm{c}$)  & \multicolumn{1}{c|}{\cellcolor[HTML]{FBE2D5}0.1865} & \multicolumn{1}{c|}{\cellcolor[HTML]{FBE2D5}0.3511} & \multicolumn{1}{c|}{\cellcolor[HTML]{F1A983}1.1601} & \multicolumn{1}{c|}{\cellcolor[HTML]{F7C7AC}0.8639} & \multicolumn{1}{c|}{\cellcolor[HTML]{F7C7AC}0.5819} & \multicolumn{1}{c|}{\cellcolor[HTML]{F1A983}1.3296} & \multicolumn{1}{c|}{\cellcolor[HTML]{F7C7AC}0.6621} & \multicolumn{1}{c|}{\cellcolor[HTML]{F7C7AC}0.7505} & \multicolumn{1}{c|}{\cellcolor[HTML]{BE5014}2.4849} & \multicolumn{1}{c|}{\cellcolor[HTML]{FBE2D5}0.3416} & \multicolumn{1}{c|}{\cellcolor[HTML]{FBE2D5}0.3583} & \multicolumn{1}{c|}{\cellcolor[HTML]{F1A983}1.1317} \\ 
\multicolumn{1}{|c|}{}                           & rel.diff ($v_\mathrm{e}$)  & \multicolumn{1}{c|}{\cellcolor[HTML]{FFFFFF}0.0895} & \multicolumn{1}{c|}{\cellcolor[HTML]{FBE2D5}0.1292} & \multicolumn{1}{c|}{\cellcolor[HTML]{FBE2D5}0.2815} & \multicolumn{1}{c|}{\cellcolor[HTML]{FBE2D5}0.2508} & \multicolumn{1}{c|}{\cellcolor[HTML]{FBE2D5}0.1611} & \multicolumn{1}{c|}{\cellcolor[HTML]{FBE2D5}0.3124} & \multicolumn{1}{c|}{\cellcolor[HTML]{FBE2D5}0.1992} & \multicolumn{1}{c|}{\cellcolor[HTML]{FBE2D5}0.2245} & \multicolumn{1}{c|}{\cellcolor[HTML]{F7C7AC}0.5671} & \multicolumn{1}{c|}{\cellcolor[HTML]{FBE2D5}0.1282} & \multicolumn{1}{c|}{\cellcolor[HTML]{FBE2D5}0.1092} & \multicolumn{1}{c|}{\cellcolor[HTML]{FBE2D5}0.2622} \\ 
\multicolumn{1}{|c|}{\multirow{-3}{*}{0.1 MeV}}  & rel.diff ($v_\mathrm{esc}$) & \multicolumn{1}{c|}{\cellcolor[HTML]{FFFFFF}0.0202} & \multicolumn{1}{c|}{\cellcolor[HTML]{FFFFFF}0.0499} & \multicolumn{1}{c|}{\cellcolor[HTML]{FBE2D5}0.2616} & \multicolumn{1}{c|}{\cellcolor[HTML]{FBE2D5}0.1261} & \multicolumn{1}{c|}{\cellcolor[HTML]{FFFFFF}0.0864} & \multicolumn{1}{c|}{\cellcolor[HTML]{FBE2D5}0.3194} & \multicolumn{1}{c|}{\cellcolor[HTML]{FBE2D5}0.1075} & \multicolumn{1}{c|}{\cellcolor[HTML]{FBE2D5}0.2325} & \multicolumn{1}{c|}{\cellcolor[HTML]{F1A983}1.0045} & \multicolumn{1}{c|}{\cellcolor[HTML]{FFFFFF}0.0448} & \multicolumn{1}{c|}{\cellcolor[HTML]{FFFFFF}0.0506} & \multicolumn{1}{c|}{\cellcolor[HTML]{FBE2D5}0.2547} \\ \hline
\multicolumn{1}{|c|}{}                           & rel.diff ($v_\mathrm{c}$)  & \multicolumn{1}{c|}{\cellcolor[HTML]{FFFFFF}0.0655} & \multicolumn{1}{c|}{\cellcolor[HTML]{FBE2D5}0.1793} & \multicolumn{1}{c|}{\cellcolor[HTML]{F7C7AC}0.5737} & \multicolumn{1}{c|}{\cellcolor[HTML]{FFFFFF}0.0025} & \multicolumn{1}{c|}{\cellcolor[HTML]{FFFFFF}0.043}  & \multicolumn{1}{c|}{\cellcolor[HTML]{FBE2D5}0.4729} & \multicolumn{1}{c|}{\cellcolor[HTML]{FFFFFF}0.0132} & \multicolumn{1}{c|}{\cellcolor[HTML]{FFFFFF}0.0465} & \multicolumn{1}{c|}{\cellcolor[HTML]{FBE2D5}0.4607} & \multicolumn{1}{c|}{\cellcolor[HTML]{FFFFFF}0.0254} & \multicolumn{1}{c|}{\cellcolor[HTML]{FFFFFF}0.073}  & \multicolumn{1}{c|}{\cellcolor[HTML]{F7C7AC}0.5116} \\ 
\multicolumn{1}{|c|}{}                           & rel.diff ($v_\mathrm{e}$)  & \multicolumn{1}{c|}{\cellcolor[HTML]{FFFFFF}0.0209} & \multicolumn{1}{c|}{\cellcolor[HTML]{FFFFFF}0.0423} & \multicolumn{1}{c|}{\cellcolor[HTML]{FBE2D5}0.1744} & \multicolumn{1}{c|}{\cellcolor[HTML]{FFFFFF}0.0117} & \multicolumn{1}{c|}{\cellcolor[HTML]{FFFFFF}0.0407} & \multicolumn{1}{c|}{\cellcolor[HTML]{FBE2D5}0.1447} & \multicolumn{1}{c|}{\cellcolor[HTML]{FFFFFF}0.0191} & \multicolumn{1}{c|}{\cellcolor[HTML]{FFFFFF}0.011}  & \multicolumn{1}{c|}{\cellcolor[HTML]{FBE2D5}0.1403} & \multicolumn{1}{c|}{\cellcolor[HTML]{FFFFFF}0.0014} & \multicolumn{1}{c|}{\cellcolor[HTML]{FFFFFF}0.0365} & \multicolumn{1}{c|}{\cellcolor[HTML]{FBE2D5}0.1498} \\ 
\multicolumn{1}{|c|}{\multirow{-3}{*}{1 MeV}}    & rel.diff ($v_\mathrm{esc}$) & \multicolumn{1}{c|}{\cellcolor[HTML]{FFFFFF}0.0089} & \multicolumn{1}{c|}{\cellcolor[HTML]{FFFFFF}0.0376} & \multicolumn{1}{c|}{\cellcolor[HTML]{FFFFFF}0.0942} & \multicolumn{1}{c|}{\cellcolor[HTML]{FFFFFF}0.0009} & \multicolumn{1}{c|}{\cellcolor[HTML]{FBE2D5}0.1289} & \multicolumn{1}{c|}{\cellcolor[HTML]{FFFFFF}0.0845} & \multicolumn{1}{c|}{\cellcolor[HTML]{FFFFFF}0.001}  & \multicolumn{1}{c|}{\cellcolor[HTML]{FFFFFF}0.0093} & \multicolumn{1}{c|}{\cellcolor[HTML]{FFFFFF}0.0787} & \multicolumn{1}{c|}{\cellcolor[HTML]{FFFFFF}0.004}  & \multicolumn{1}{c|}{\cellcolor[HTML]{FFFFFF}0.0053} & \multicolumn{1}{c|}{\cellcolor[HTML]{FFFFFF}0.0858} \\ \hline
\multicolumn{1}{|c|}{}                           & rel.diff ($v_\mathrm{c}$)  & \multicolumn{1}{c|}{\cellcolor[HTML]{FBE2D5}0.1008} & \multicolumn{1}{c|}{\cellcolor[HTML]{FBE2D5}0.1041} & \multicolumn{1}{c|}{\cellcolor[HTML]{FBE2D5}0.4077} & \multicolumn{1}{c|}{\cellcolor[HTML]{FFFFFF}0.0889} & \multicolumn{1}{c|}{\cellcolor[HTML]{FBE2D5}0.1085} & \multicolumn{1}{c|}{\cellcolor[HTML]{FBE2D5}0.3978} & \multicolumn{1}{c|}{\cellcolor[HTML]{FFFFFF}0.0916} & \multicolumn{1}{c|}{\cellcolor[HTML]{FBE2D5}0.1131} & \multicolumn{1}{c|}{\cellcolor[HTML]{FBE2D5}0.3615} & \multicolumn{1}{c|}{\cellcolor[HTML]{FFFFFF}0.0962} & \multicolumn{1}{c|}{\cellcolor[HTML]{FBE2D5}0.1103} & \multicolumn{1}{c|}{\cellcolor[HTML]{FBE2D5}0.3861} \\ 
\multicolumn{1}{|c|}{}                           & rel.diff ($v_\mathrm{e}$)  & \multicolumn{1}{c|}{\cellcolor[HTML]{FFFFFF}0.0463} & \multicolumn{1}{c|}{\cellcolor[HTML]{FFFFFF}0.0628} & \multicolumn{1}{c|}{\cellcolor[HTML]{FBE2D5}0.135}  & \multicolumn{1}{c|}{\cellcolor[HTML]{FFFFFF}0.0406} & \multicolumn{1}{c|}{\cellcolor[HTML]{FFFFFF}0.0603} & \multicolumn{1}{c|}{\cellcolor[HTML]{FBE2D5}0.133}  & \multicolumn{1}{c|}{\cellcolor[HTML]{FFFFFF}0.0376} & \multicolumn{1}{c|}{\cellcolor[HTML]{FFFFFF}0.0578} & \multicolumn{1}{c|}{\cellcolor[HTML]{FBE2D5}0.1213} & \multicolumn{1}{c|}{\cellcolor[HTML]{FFFFFF}0.0425} & \multicolumn{1}{c|}{\cellcolor[HTML]{FFFFFF}0.044}  & \multicolumn{1}{c|}{\cellcolor[HTML]{FBE2D5}0.1228} \\ 
\multicolumn{1}{|c|}{\multirow{-3}{*}{10 MeV}}   & rel.diff ($v_\mathrm{esc}$) & \multicolumn{1}{c|}{\cellcolor[HTML]{FFFFFF}0.007}  & \multicolumn{1}{c|}{\cellcolor[HTML]{FFFFFF}0.0603} & \multicolumn{1}{c|}{\cellcolor[HTML]{FFFFFF}0.0658} & \multicolumn{1}{c|}{\cellcolor[HTML]{FFFFFF}0.0114} & \multicolumn{1}{c|}{\cellcolor[HTML]{FFFFFF}0.0454} & \multicolumn{1}{c|}{\cellcolor[HTML]{FFFFFF}0.0607} & \multicolumn{1}{c|}{\cellcolor[HTML]{FFFFFF}0.0022} & \multicolumn{1}{c|}{\cellcolor[HTML]{FFFFFF}0.0246} & \multicolumn{1}{c|}{\cellcolor[HTML]{FFFFFF}0.0531} & \multicolumn{1}{c|}{\cellcolor[HTML]{FFFFFF}0.0098} & \multicolumn{1}{c|}{\cellcolor[HTML]{FFFFFF}0.0134} & \multicolumn{1}{c|}{\cellcolor[HTML]{FFFFFF}0.0555} \\ \hline
\multicolumn{1}{|c|}{}                           & rel.diff ($v_\mathrm{c}$)  & \multicolumn{1}{c|}{\cellcolor[HTML]{FBE2D5}0.116}  & \multicolumn{1}{c|}{\cellcolor[HTML]{FBE2D5}0.137}  & \multicolumn{1}{c|}{\cellcolor[HTML]{FBE2D5}0.1237} & \multicolumn{1}{c|}{\cellcolor[HTML]{FBE2D5}0.1186} & \multicolumn{1}{c|}{\cellcolor[HTML]{FBE2D5}0.1331} & \multicolumn{1}{c|}{\cellcolor[HTML]{FBE2D5}0.1234} & \multicolumn{1}{c|}{\cellcolor[HTML]{FBE2D5}0.1191} & \multicolumn{1}{c|}{\cellcolor[HTML]{FBE2D5}0.1362} & \multicolumn{1}{c|}{\cellcolor[HTML]{FBE2D5}0.1246} & \multicolumn{1}{c|}{\cellcolor[HTML]{FBE2D5}0.1189} & \multicolumn{1}{c|}{\cellcolor[HTML]{FBE2D5}0.143}  & \multicolumn{1}{c|}{\cellcolor[HTML]{FBE2D5}0.1191} \\ 
\multicolumn{1}{|c|}{}                           & rel.diff ($v_\mathrm{e}$)  & \multicolumn{1}{c|}{\cellcolor[HTML]{FFFFFF}0.0573} & \multicolumn{1}{c|}{\cellcolor[HTML]{FFFFFF}0.0725} & \multicolumn{1}{c|}{\cellcolor[HTML]{FFFFFF}0.0547} & \multicolumn{1}{c|}{\cellcolor[HTML]{FFFFFF}0.061}  & \multicolumn{1}{c|}{\cellcolor[HTML]{FFFFFF}0.0612} & \multicolumn{1}{c|}{\cellcolor[HTML]{FFFFFF}0.055}  & \multicolumn{1}{c|}{\cellcolor[HTML]{FFFFFF}0.0614} & \multicolumn{1}{c|}{\cellcolor[HTML]{FFFFFF}0.07}   & \multicolumn{1}{c|}{\cellcolor[HTML]{FFFFFF}0.0601} & \multicolumn{1}{c|}{\cellcolor[HTML]{FFFFFF}0.0563} & \multicolumn{1}{c|}{\cellcolor[HTML]{FFFFFF}0.0662} & \multicolumn{1}{c|}{\cellcolor[HTML]{FFFFFF}0.0437} \\ 
\multicolumn{1}{|c|}{\multirow{-3}{*}{100 MeV}}  & rel.diff ($v_\mathrm{esc}$) & \multicolumn{1}{c|}{\cellcolor[HTML]{FFFFFF}0.0255} & \multicolumn{1}{c|}{\cellcolor[HTML]{FFFFFF}0.0345} & \multicolumn{1}{c|}{\cellcolor[HTML]{FFFFFF}0.0158} & \multicolumn{1}{c|}{\cellcolor[HTML]{FFFFFF}0.0121} & \multicolumn{1}{c|}{\cellcolor[HTML]{FFFFFF}0.0366} & \multicolumn{1}{c|}{\cellcolor[HTML]{FFFFFF}0.0158} & \multicolumn{1}{c|}{\cellcolor[HTML]{FFFFFF}0.0224} & \multicolumn{1}{c|}{\cellcolor[HTML]{FFFFFF}0.019}  & \multicolumn{1}{c|}{\cellcolor[HTML]{FFFFFF}0.022}  & \multicolumn{1}{c|}{\cellcolor[HTML]{FFFFFF}0.0139} & \multicolumn{1}{c|}{\cellcolor[HTML]{FFFFFF}0.0202} & \multicolumn{1}{c|}{\cellcolor[HTML]{FFFFFF}0.0199} \\ \hline
\end{tabular}%
}
\caption{Same as~\cref{tab:QQQ_SHM} for the SHM and $\omega_\text{min} = 20 \mathrm{meV}$.
\label{tab:QQQ_SHM_20meV}}
\end{table}

\section{Results using the standard prescription}
\label{App_B}

\begin{figure}[t]
\centering

\subfloat[$\mathrm{CaWO}_4$]{%
    \includegraphics[width=0.49\textwidth]{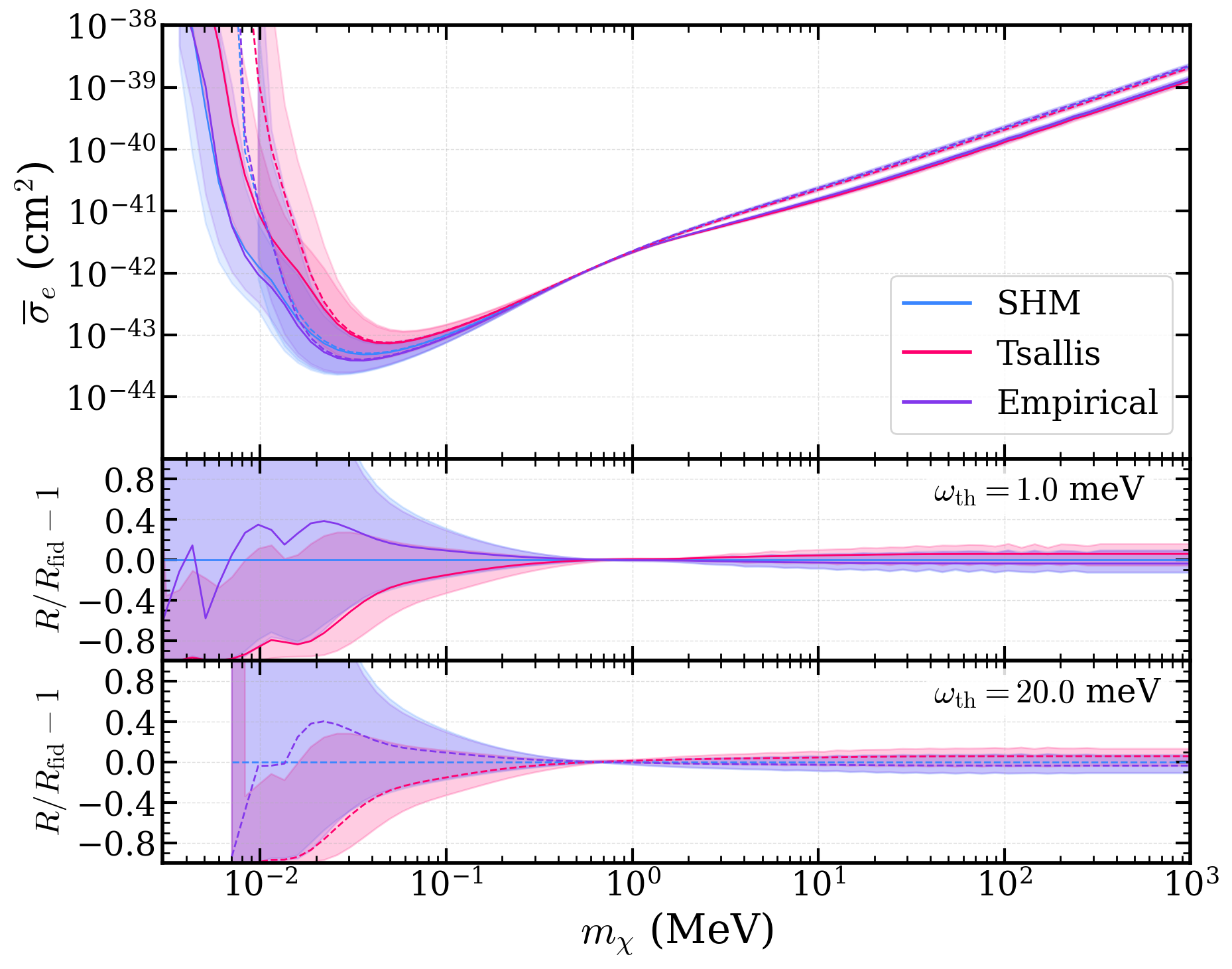}
}
\hfill
\subfloat[$\mathrm{SiO}_2$]{%
    \includegraphics[width=0.49\textwidth]{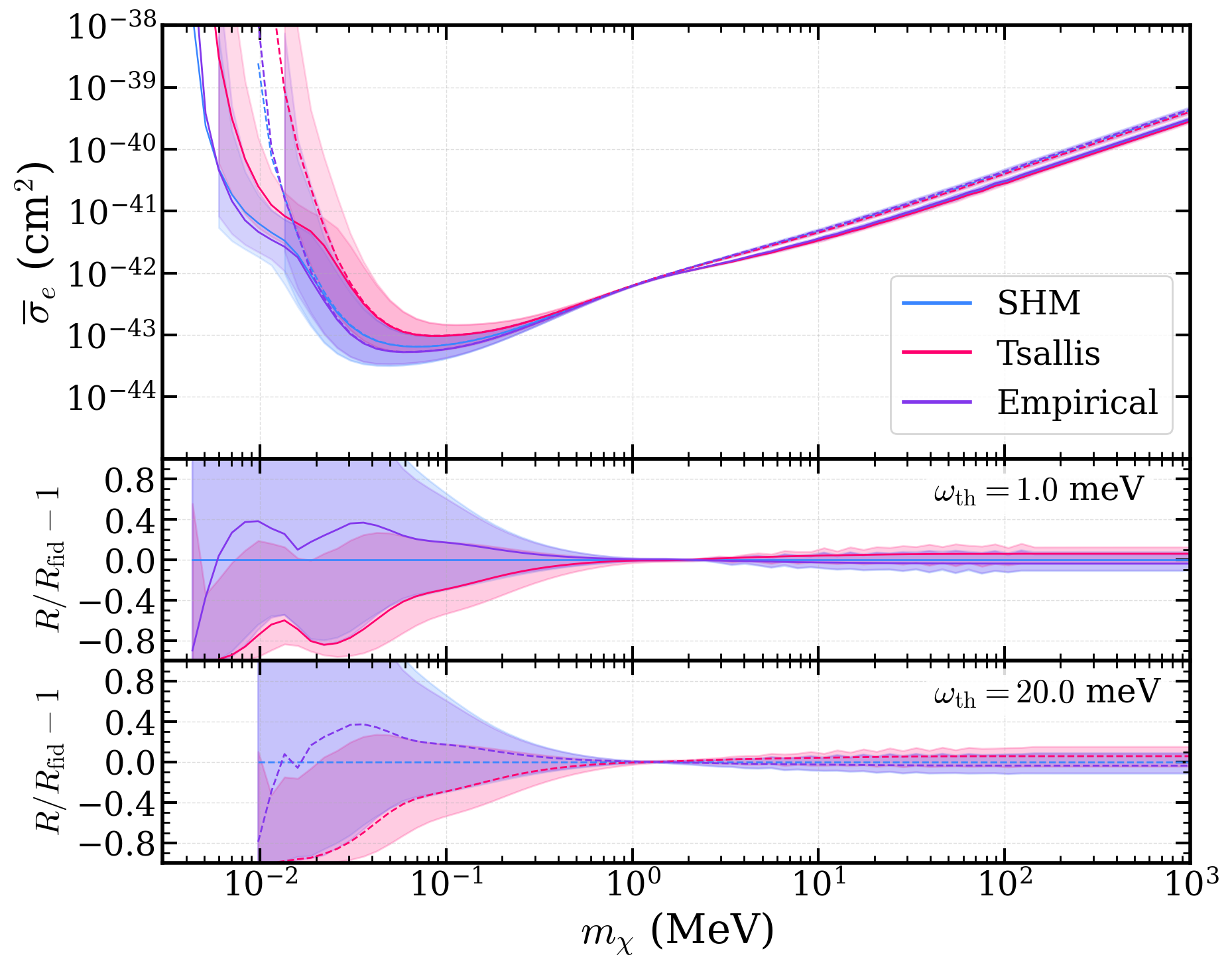}
}

\subfloat[$\mathrm{Al}_2\mathrm{O}_3$]{%
    \includegraphics[width=0.49\textwidth]{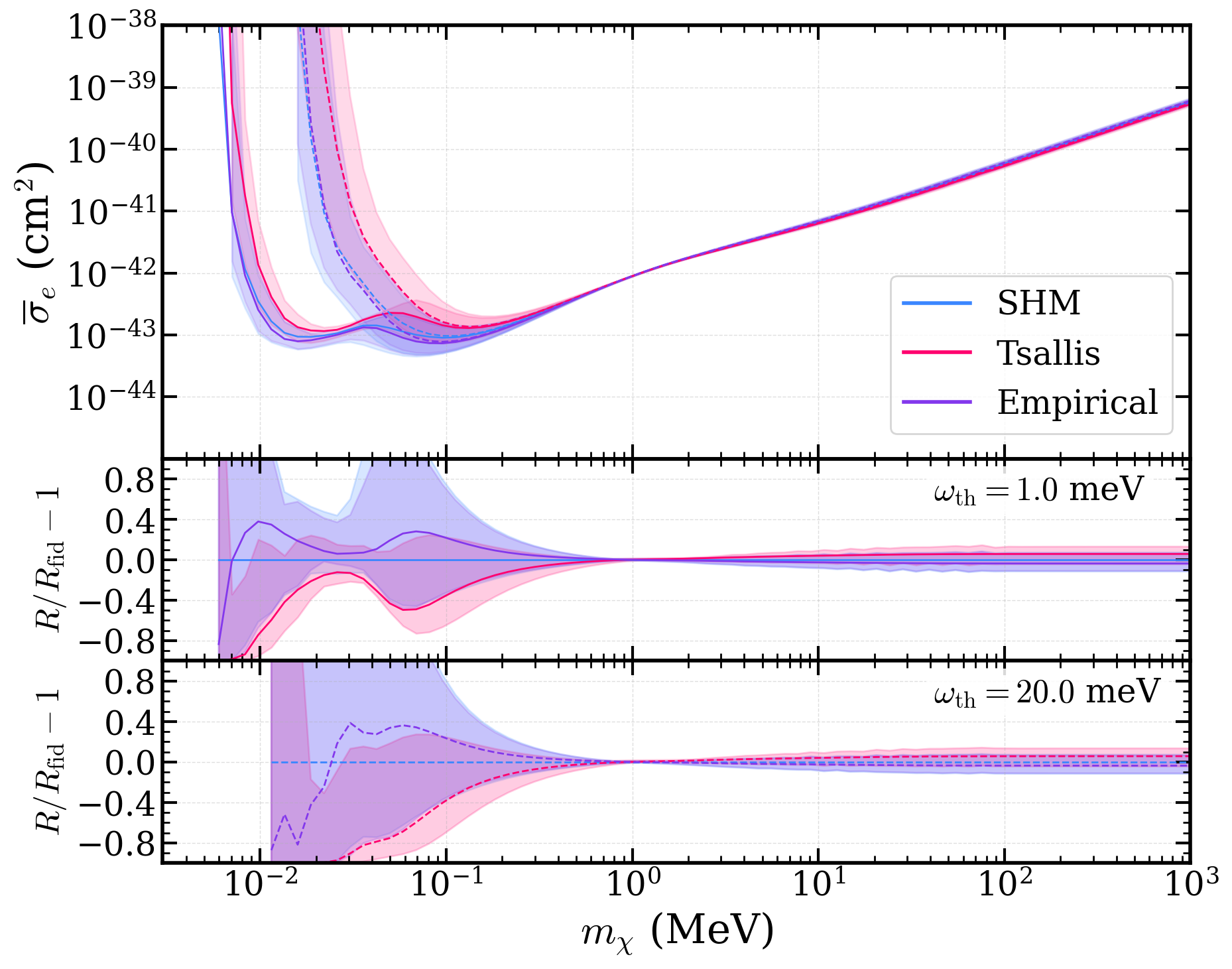}
}
\hfill
\subfloat[$\mathrm{GaAs}$]{%
    \includegraphics[width=0.49\textwidth]{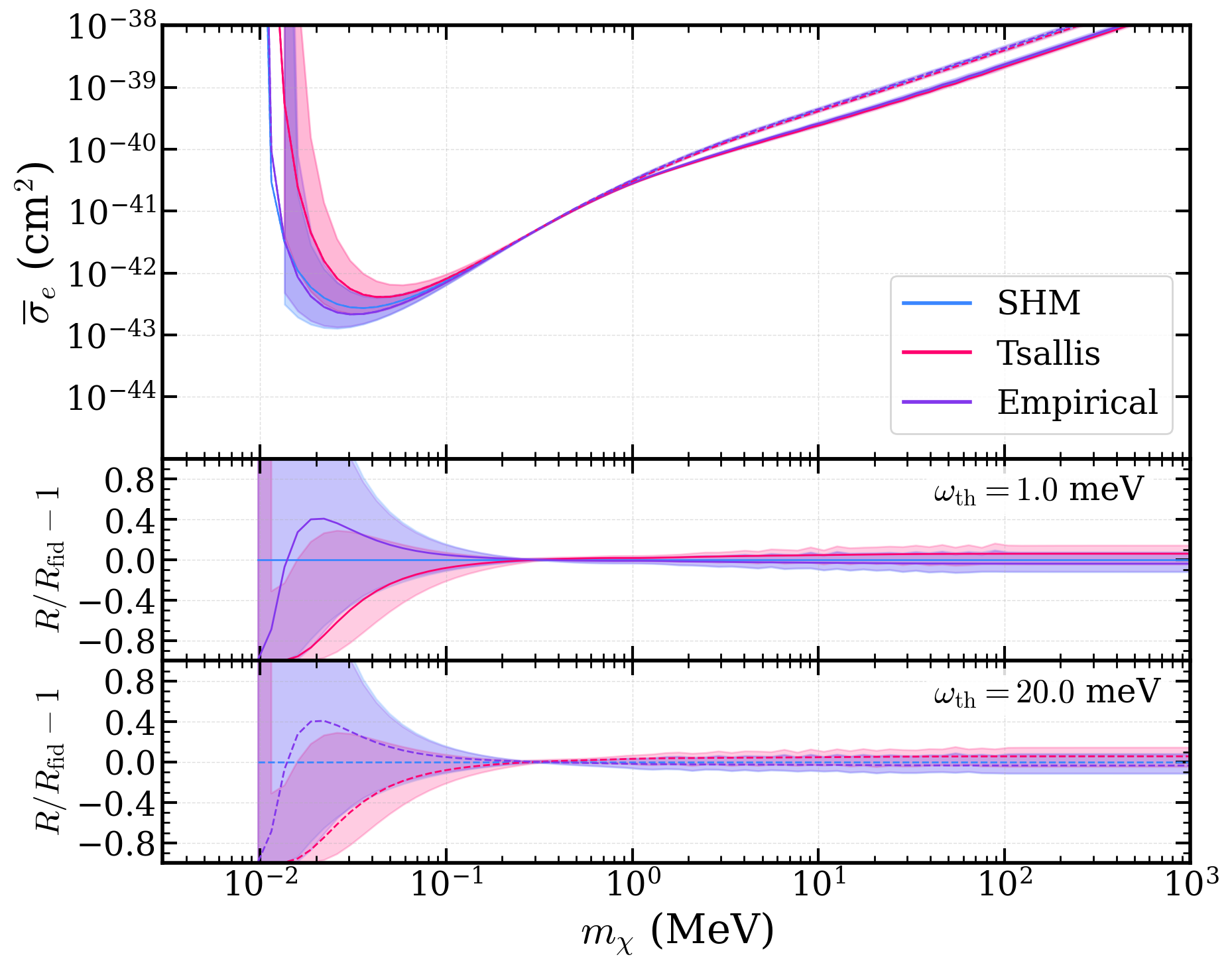}
}

\caption{Projected reach with uncertainty bands under different halo model assumptions (SHM, Tsallis, empirical) following the standard prescription (which sets $v_0=v_\mathrm{c}$ in all models) for light dark photon mediated scattering. Solid (dashed) curves correspond to central values of the velocity parameters and $\omega_\mathrm{min}=1~\mathrm{meV}$ ($20~\mathrm{meV}$). Shaded bands represent the uncertainties from varying the velocity parameters within the conservative ranges listed in~\cref{tab:Halo_Parameters}. Lower panels show the fractional deviation of the rate with respect to the SHM prediction at the central values of the velocity parameters.
\label{fig:LDP_reach_QQQ}}
\end{figure}

\begin{figure}[t]
\centering

\subfloat[$\mathrm{CaWO}_4$]{%
    \includegraphics[width=0.49\textwidth]{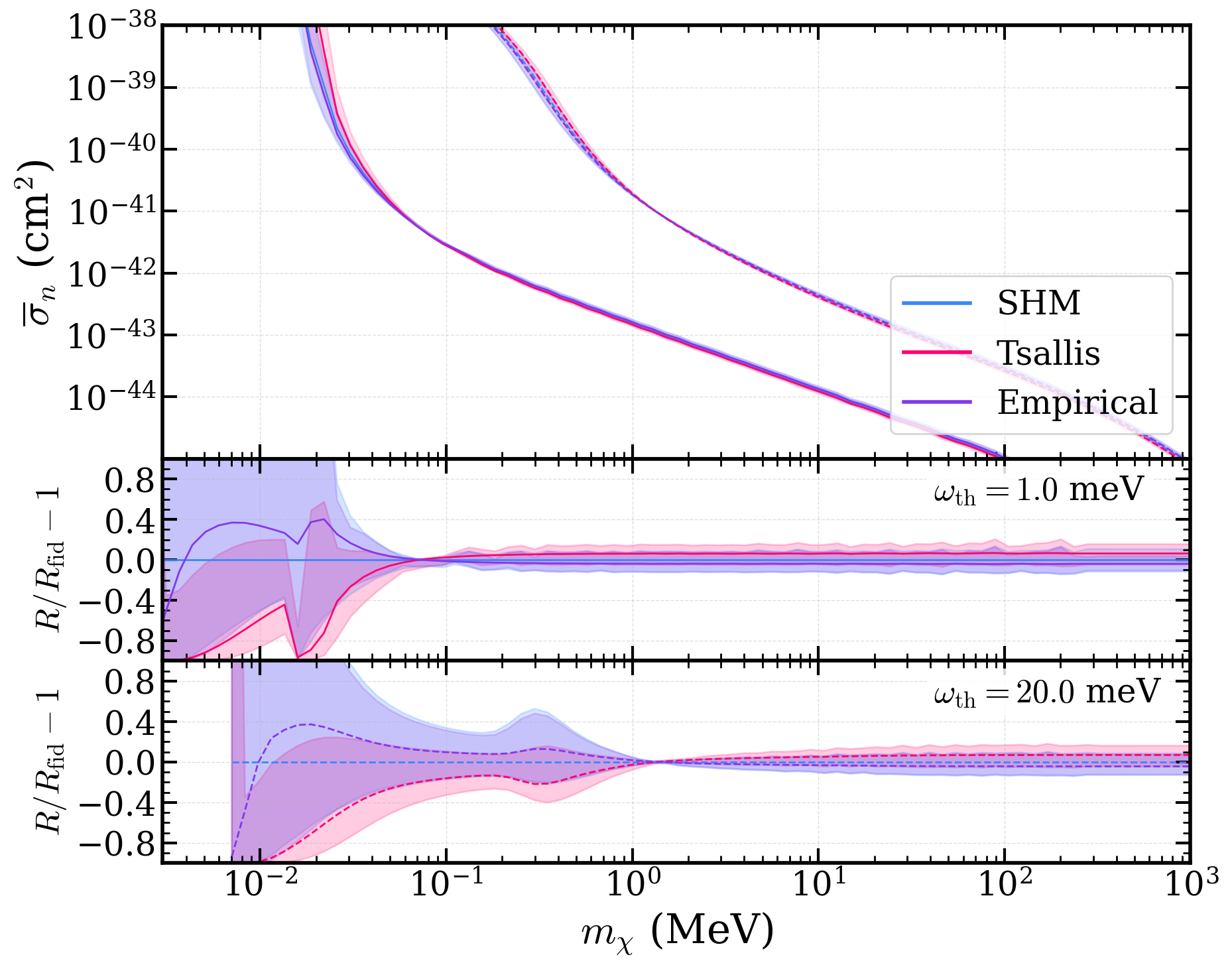}
}
\hfill
\subfloat[$\mathrm{SiO}_2$]{%
    \includegraphics[width=0.49\textwidth]{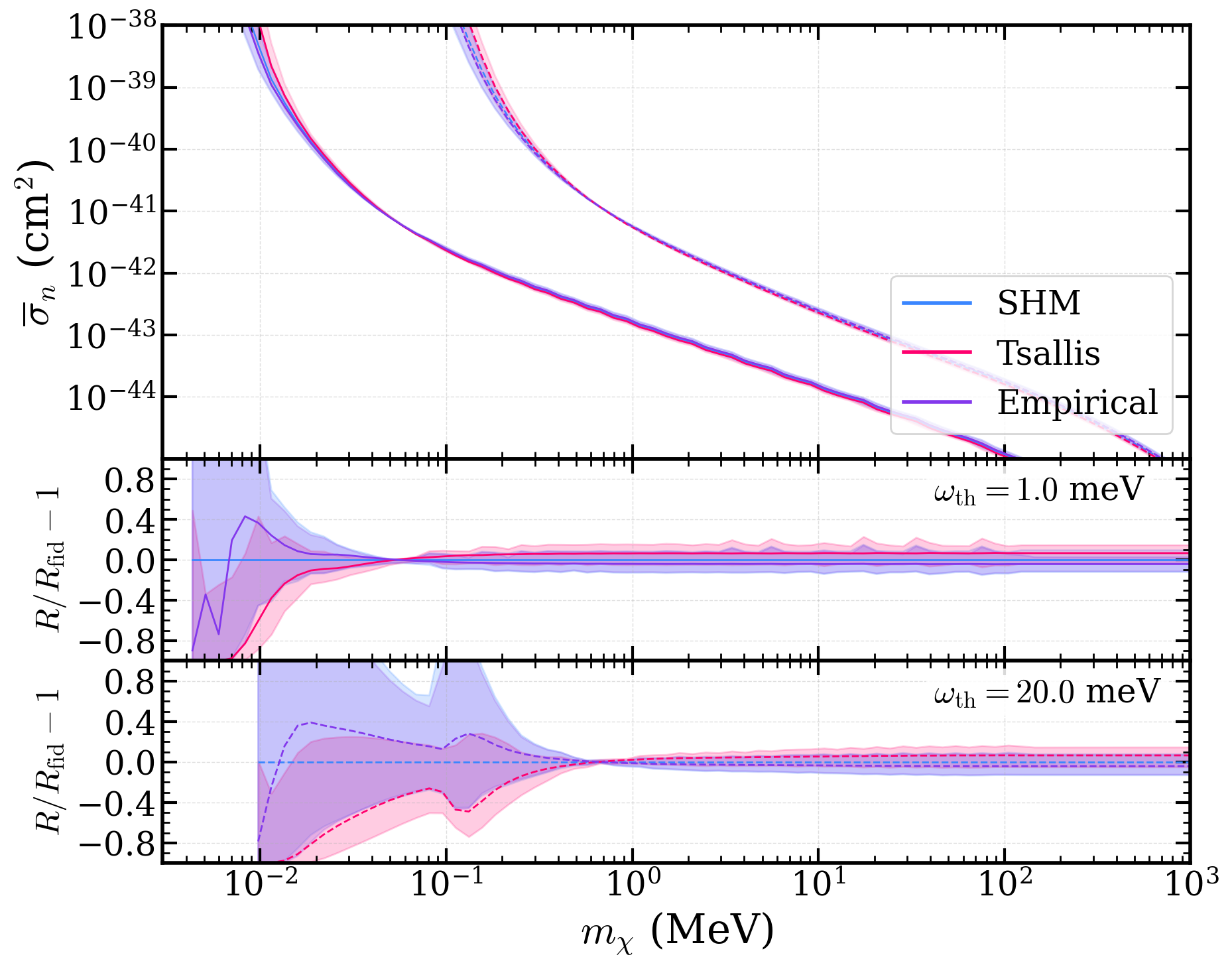}
}

\subfloat[$\mathrm{Al}_2\mathrm{O}_3$]{%
    \includegraphics[width=0.49\textwidth]{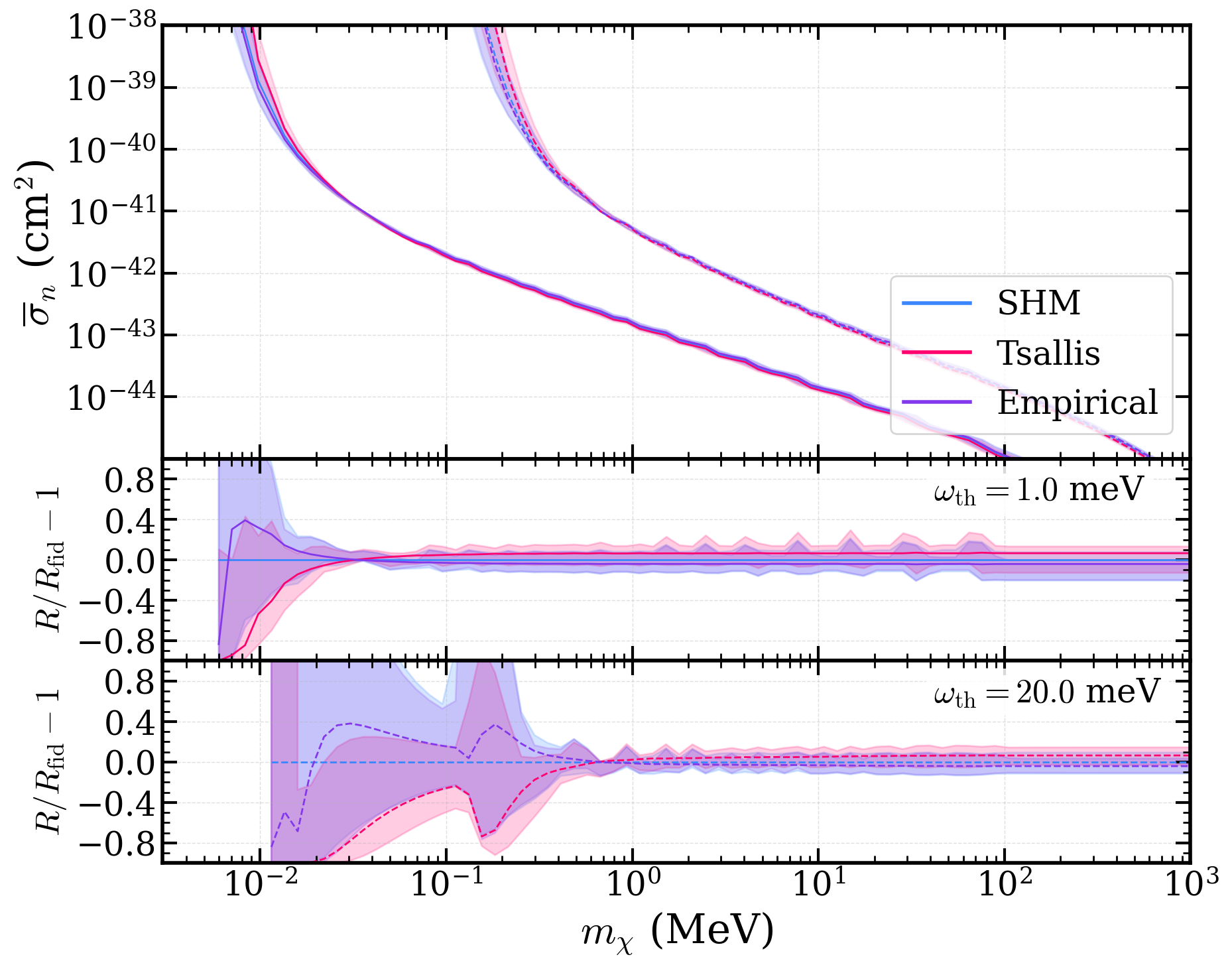}
}
\hfill
\subfloat[$\mathrm{GaAs}$]{%
    \includegraphics[width=0.49\textwidth]{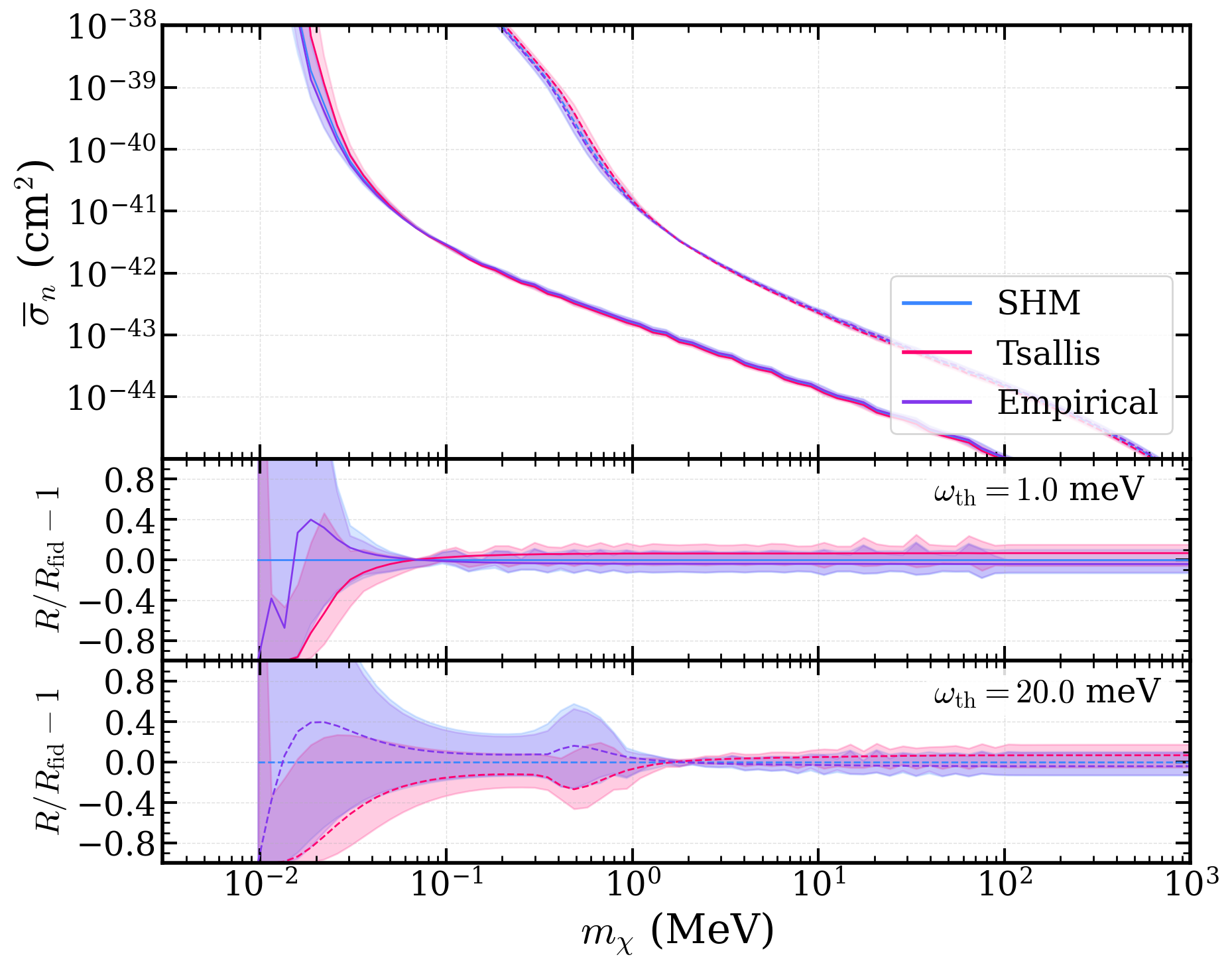}
}

\caption{Same as~\cref{fig:LDP_reach_QQQ} for light hadrophilic scalar mediated scattering.
\label{fig:LM_reach_QQQ}}
\end{figure}

\begin{figure}[t]
\centering

\subfloat[$\mathrm{CaWO}_4$]{%
    \includegraphics[width=0.49\textwidth]{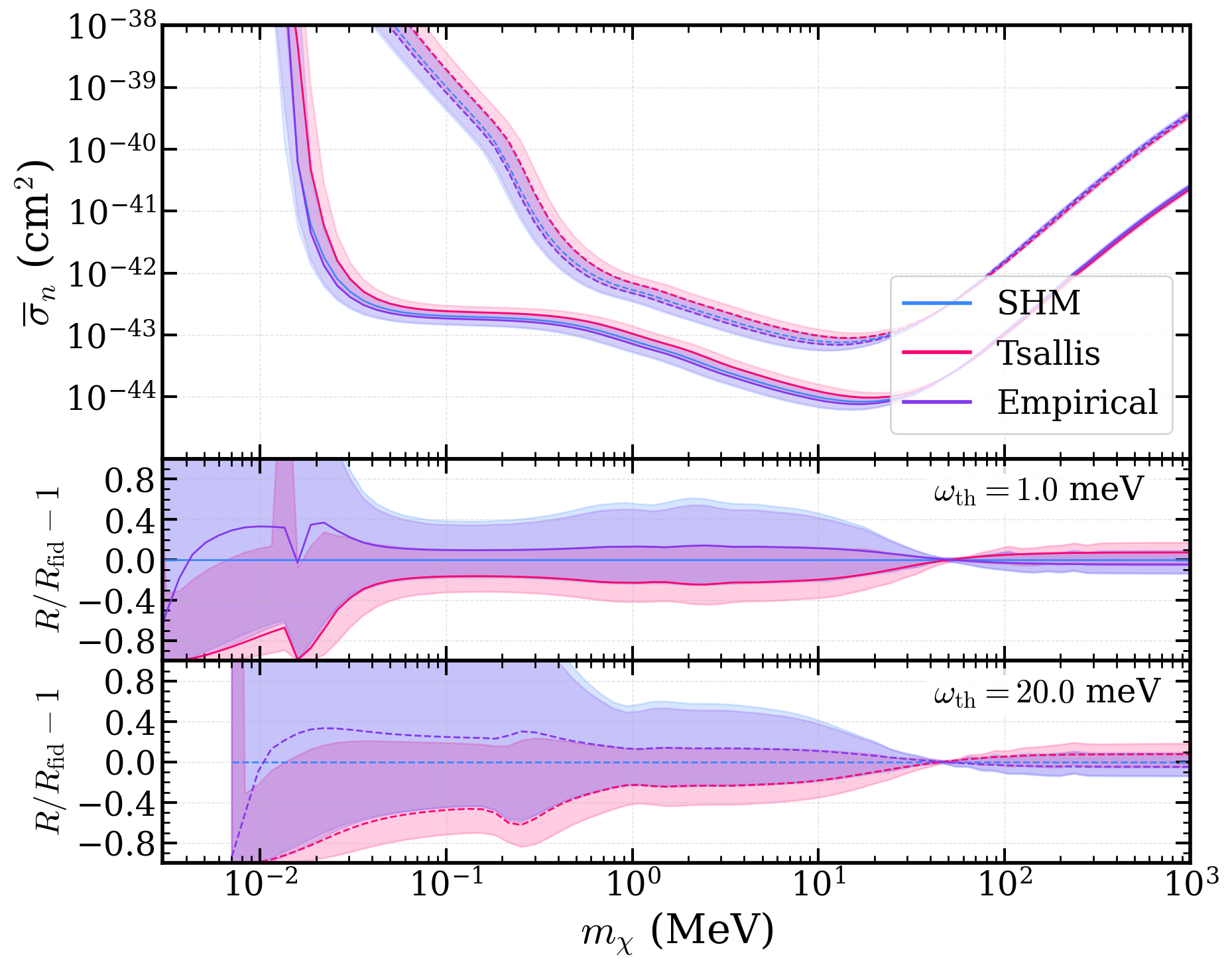}
}
\hfill
\subfloat[$\mathrm{SiO}_2$]{%
    \includegraphics[width=0.49\textwidth]{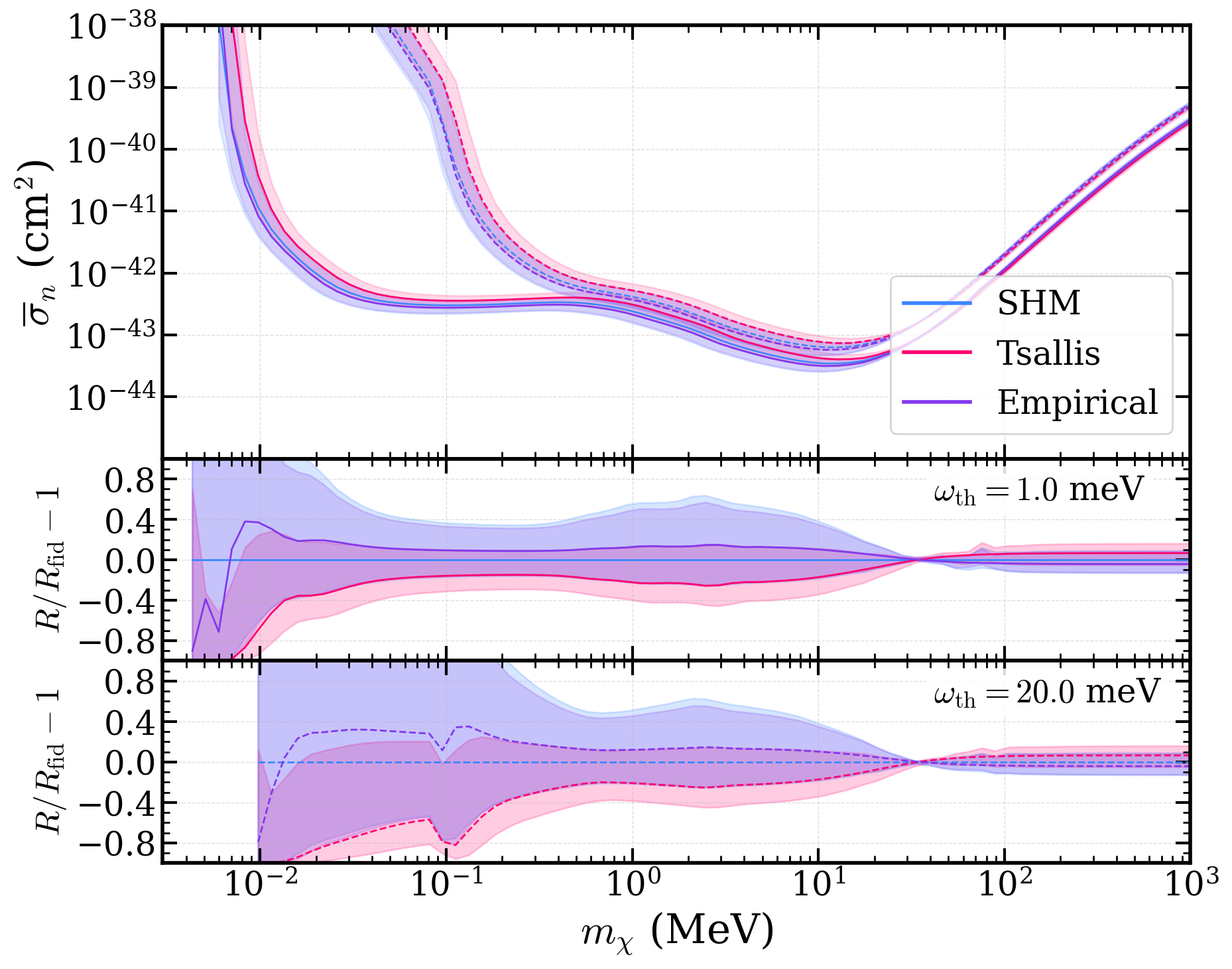}
}

\subfloat[$\mathrm{Al}_2\mathrm{O}_3$]{%
    \includegraphics[width=0.49\textwidth]{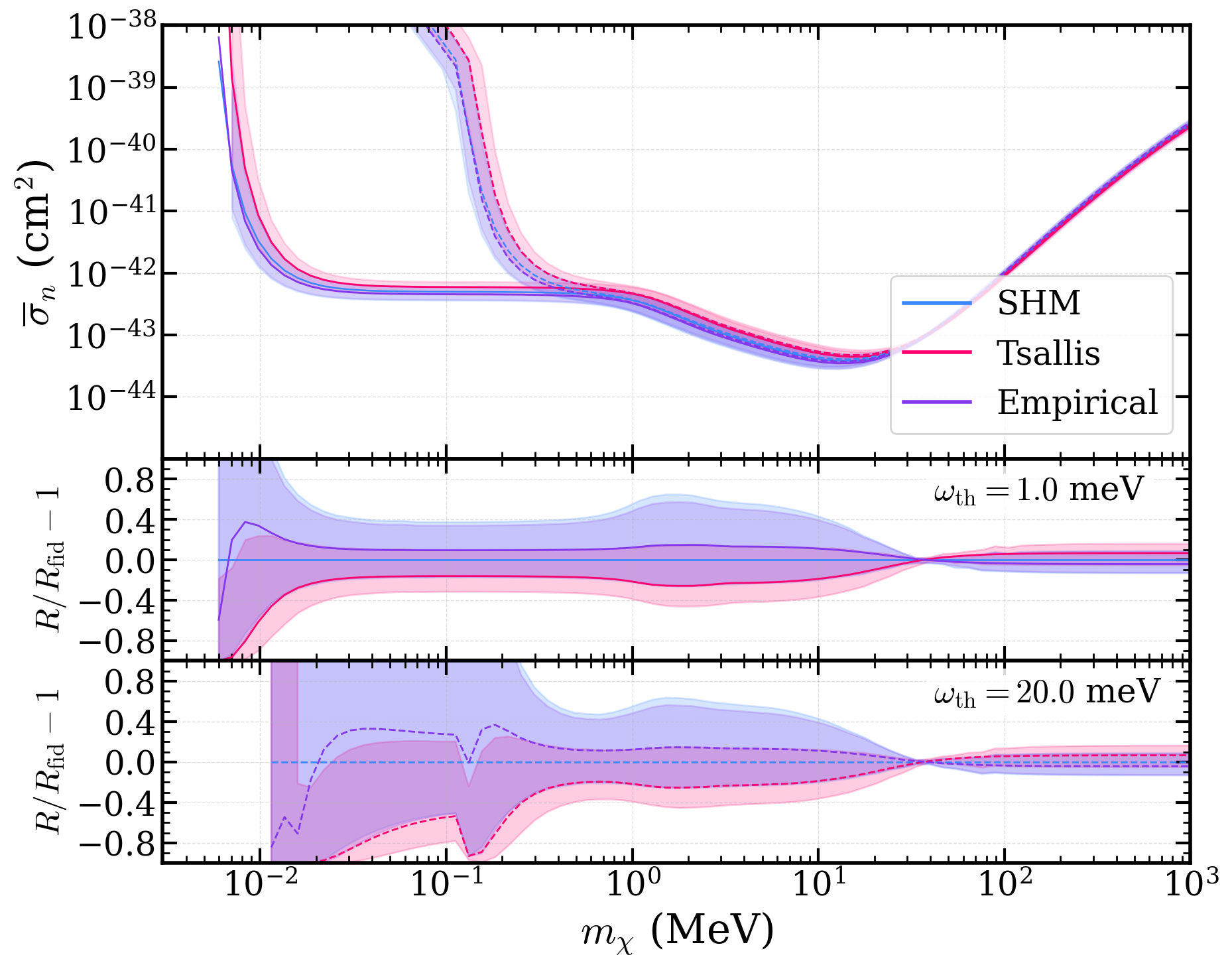}
}
\hfill
\subfloat[$\mathrm{GaAs}$]{%
    \includegraphics[width=0.49\textwidth]{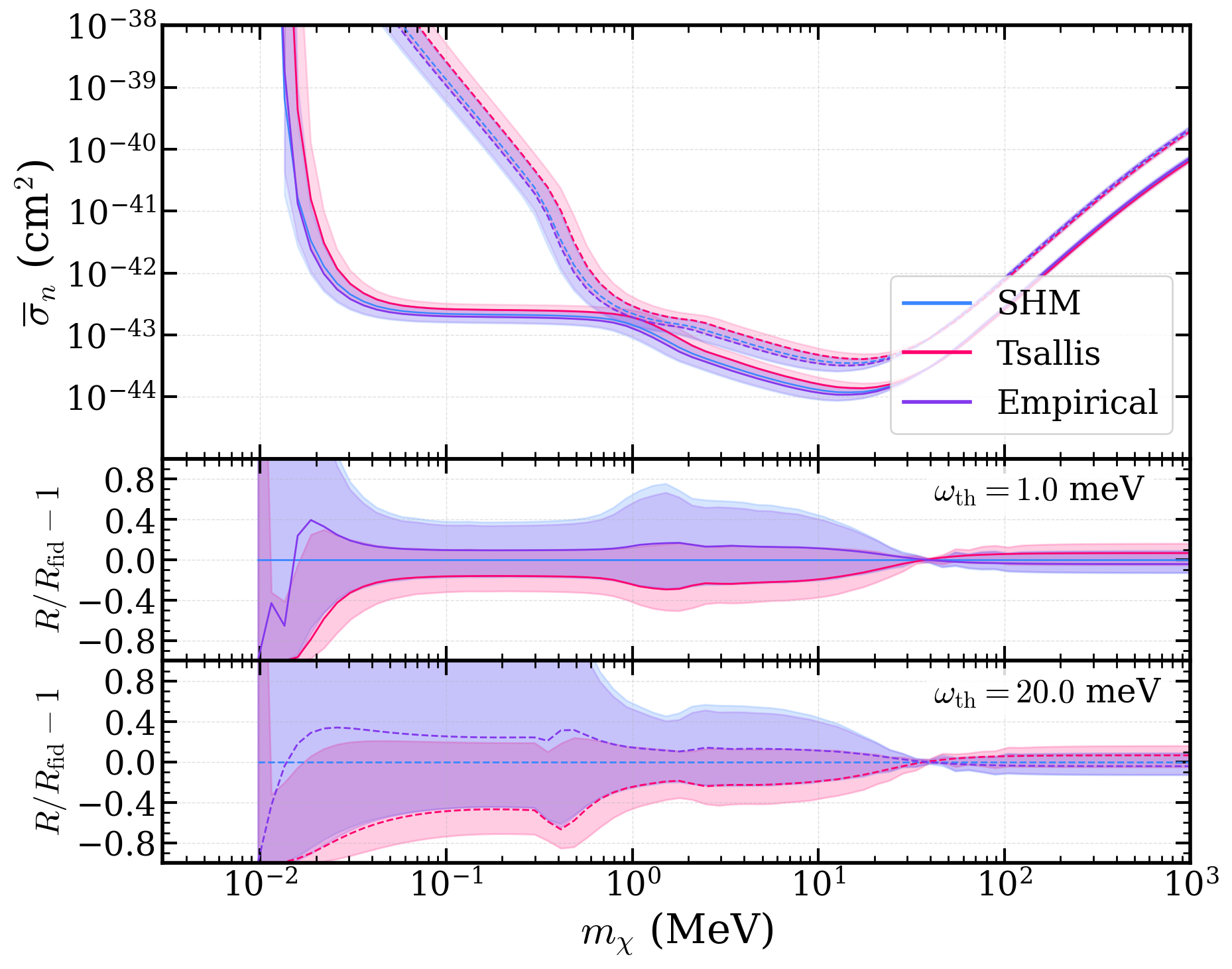}
}

\caption{Same as~\cref{fig:LDP_reach_QQQ} for heavy hadrophilic scalar mediated scattering.
\label{fig:HM_reach_QQQ}}
\end{figure}

In this appendix, we present additional results obtained using the standard matching prescription discussed in~\cref{sec:prescriptions}. As seen in~\cref{fig:HM_reach_QQQ,fig:LM_reach_QQQ,fig:LDP_reach_QQQ}, the standard prescription exhibits the same qualitative behavior as the rms-matching case: the astrophysical dependence is strongest at low dark matter masses, where the rate is controlled by the high-velocity tail. In this low-mass regime, clear differences between halo models emerge. Across all target materials and mediator scenarios, we find that the SHM and the empirical velocity distribution yield nearly identical reach over most of the mass range, while the Tsallis distribution leads to a weaker reach at low masses. This behavior can be understood from the structure of the velocity distributions shown in~\cref{fig:VDF_conservative}: under identical functional parameters $(v_\mathrm{0}, v_\mathrm{esc})$, the SHM and empirical models exhibit similar support at high velocities, whereas the Tsallis distribution has a more rapidly suppressed high-velocity tail. Consequently, fewer particles are available to satisfy the kinematic threshold in the Tsallis case, leading to reduced sensitivity. This demonstrates that, under the standard prescription, halo model dependence at low masses is primarily driven by differences in the velocity tail, in contrast to the rms-matching prescription where such differences are largely absorbed into a rescaling of the velocity scale. At higher dark matter masses, the dependence on the velocity distribution becomes weaker as a broader range of velocities contributes to the rate, and the projected reach converges across halo models. Additional dependencies on mediator type and phonon threshold follow the same qualitative trends discussed in the main text.

The contrast with the rms-matching results of the main text illustrates the practical consequence of the prescription choice: the apparent suppression of the Tsallis reach at low masses under the standard prescription is partly an artifact of assigning it a lower effective energy scale relative to the SHM, rather than a genuine reflection of its distributional shape. Under rms-matching, this difference is largely absorbed into a rescaling of $v_0$, and the three models converge to similar reach curves, isolating the residual effect of functional shape.

\section{Results using aggressive velocity choices}
\label{App_C}

\begin{figure}[t]
\centering

\subfloat[$\mathrm{CaWO}_4$]{%
    \includegraphics[width=0.49\textwidth]{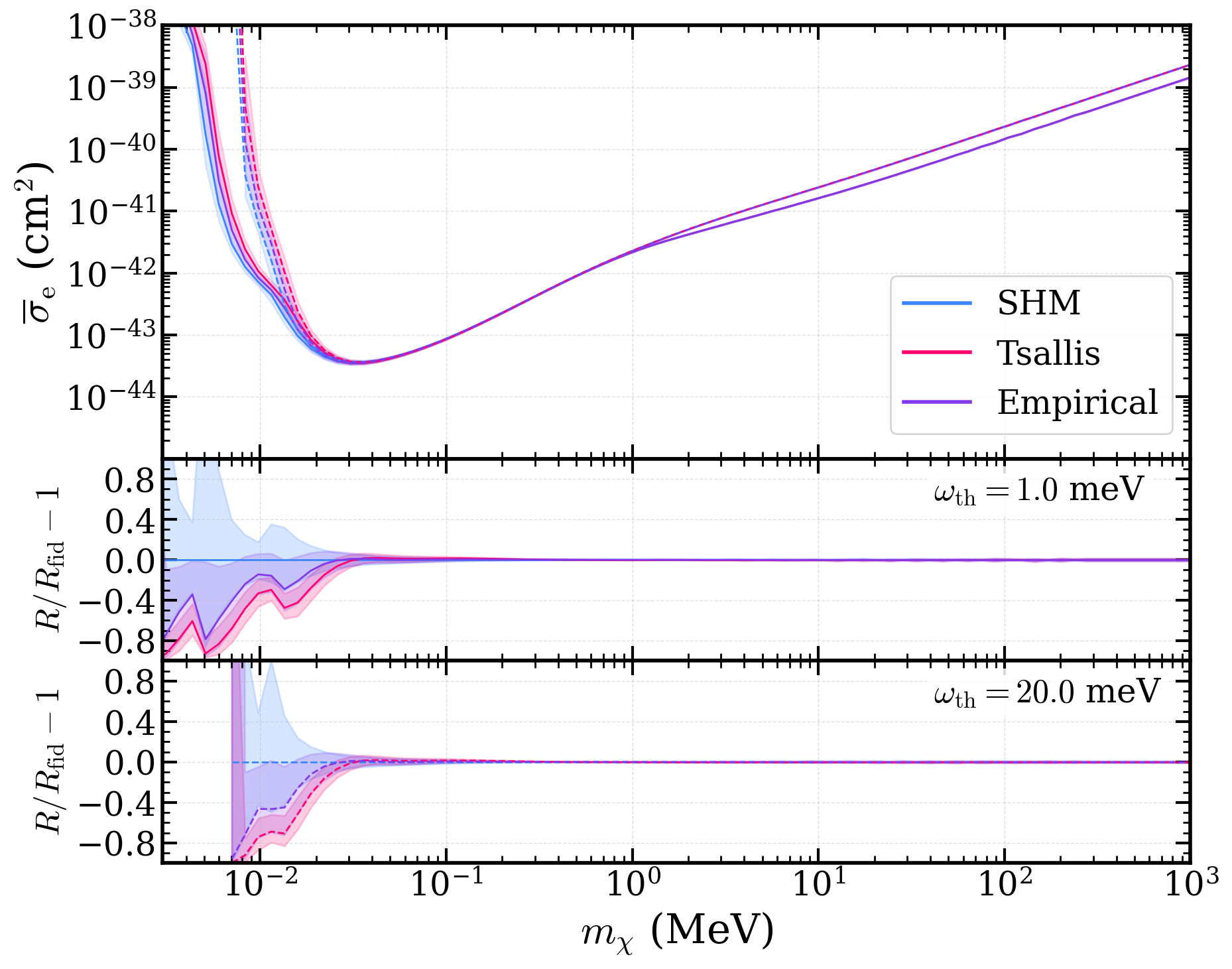}
}
\hfill
\subfloat[$\mathrm{SiO}_2$]{%
    \includegraphics[width=0.49\textwidth]{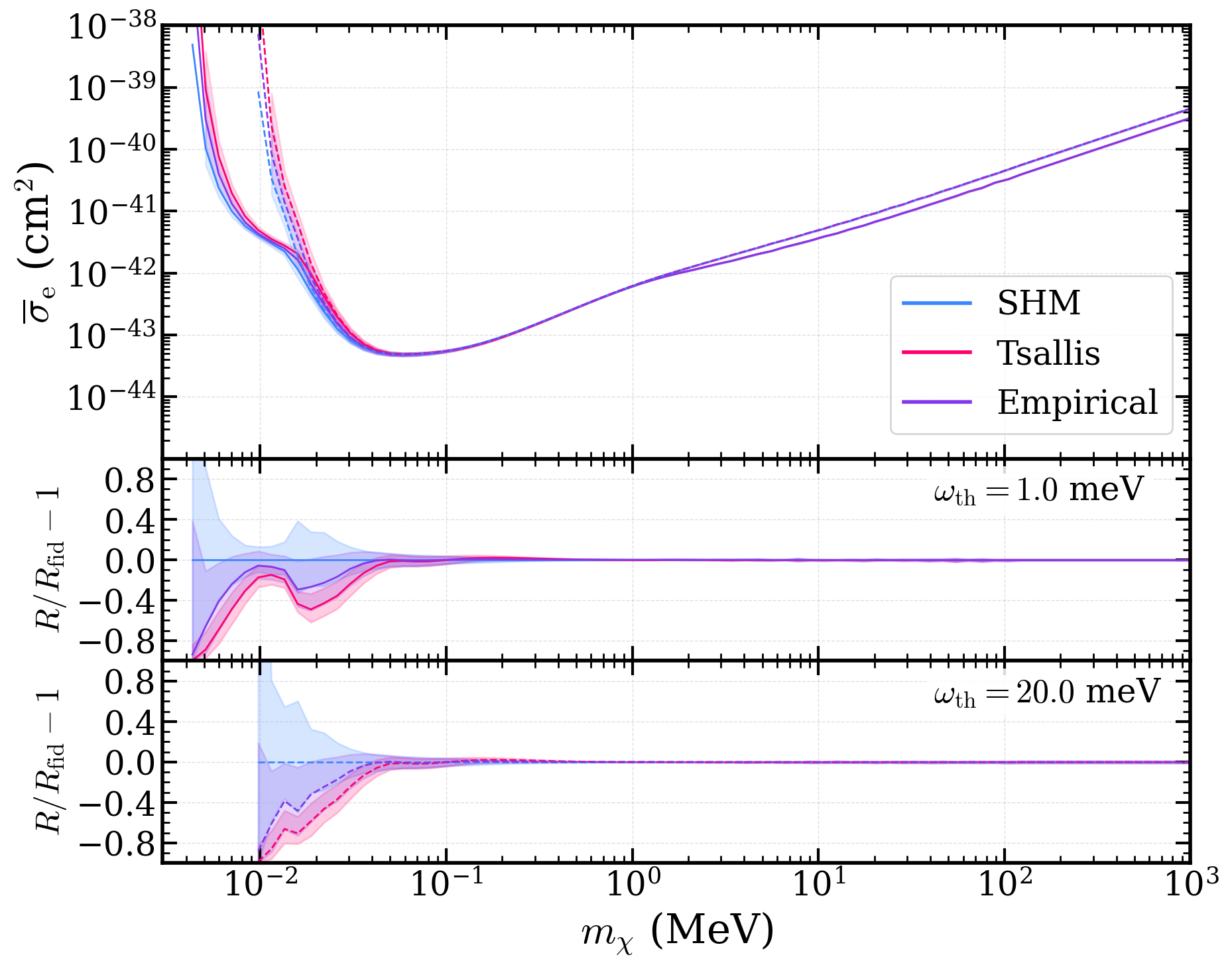}
}

\subfloat[$\mathrm{Al}_2\mathrm{O}_3$]{%
    \includegraphics[width=0.49\textwidth]{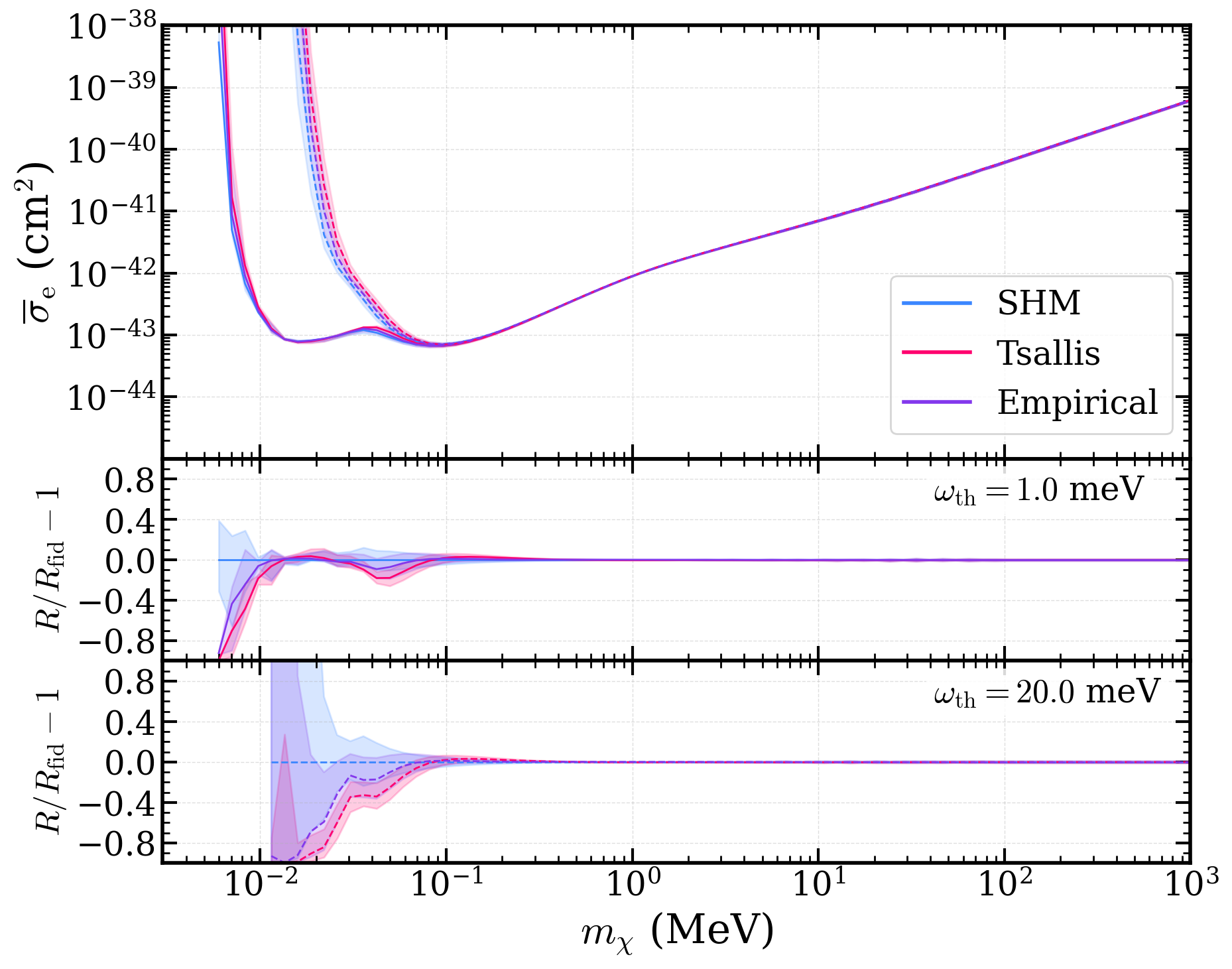}
}
\hfill
\subfloat[$\mathrm{GaAs}$]{%
    \includegraphics[width=0.49\textwidth]{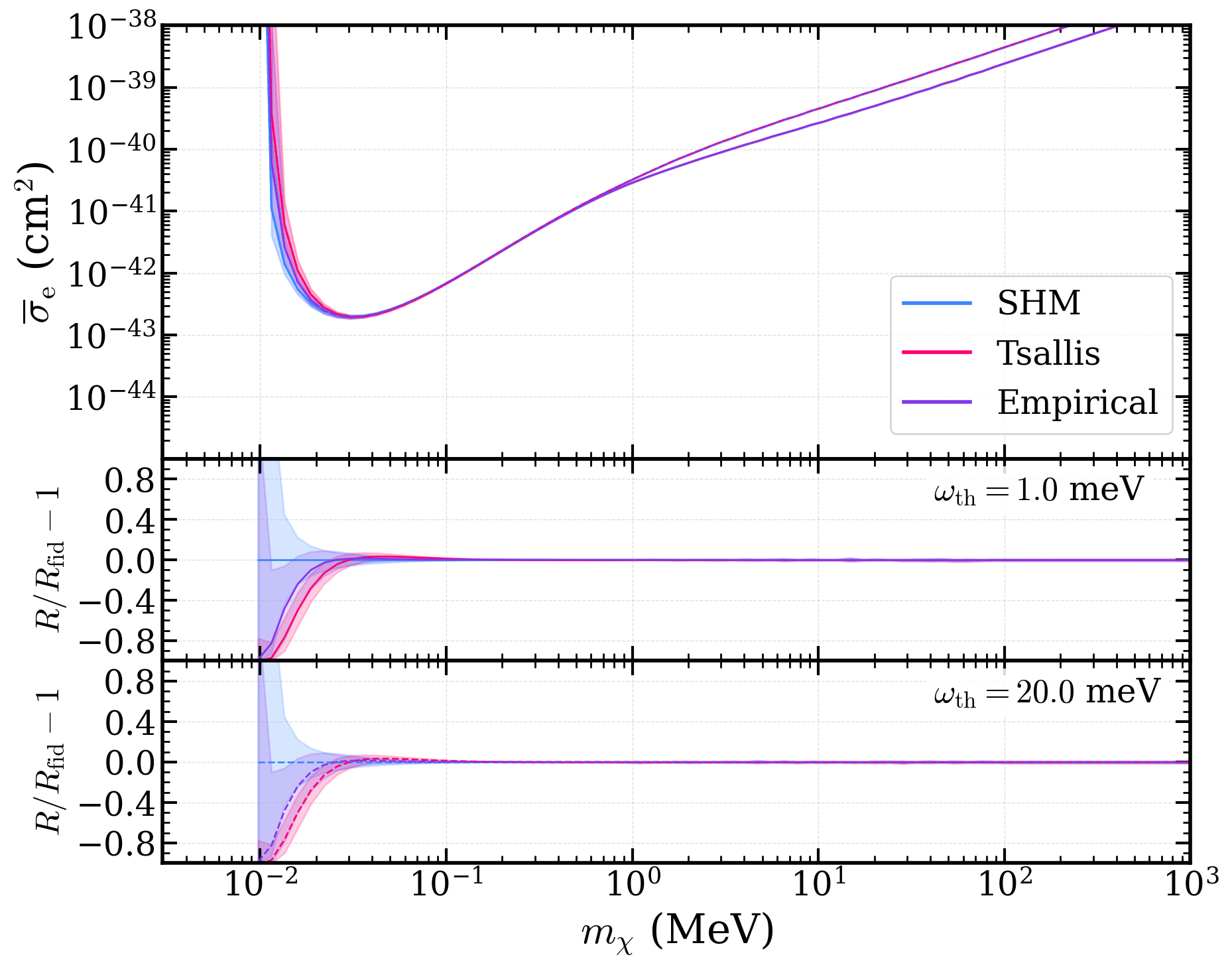}
}

\caption{Projected reach with aggressive uncertainty bands under different halo model assumptions (SHM, Tsallis, empirical) for light dark photon mediated scattering. Solid (dashed) curves correspond to central values of the velocity parameters and $\omega_\mathrm{min}=1~\mathrm{meV}$ ($20~\mathrm{meV}$). Shaded bands represent the uncertainties from varying the velocity parameters within the conservative ranges listed in~\cref{tab:Halo_Parameters}. Lower panels show the fractional deviation of the rate with respect to the SHM prediction at the central values of the velocity parameters.
\label{fig:LDP_reach_AAA}}
\end{figure}

\begin{figure}[t]
\centering

\subfloat[$\mathrm{CaWO}_4$]{%
    \includegraphics[width=0.49\textwidth]{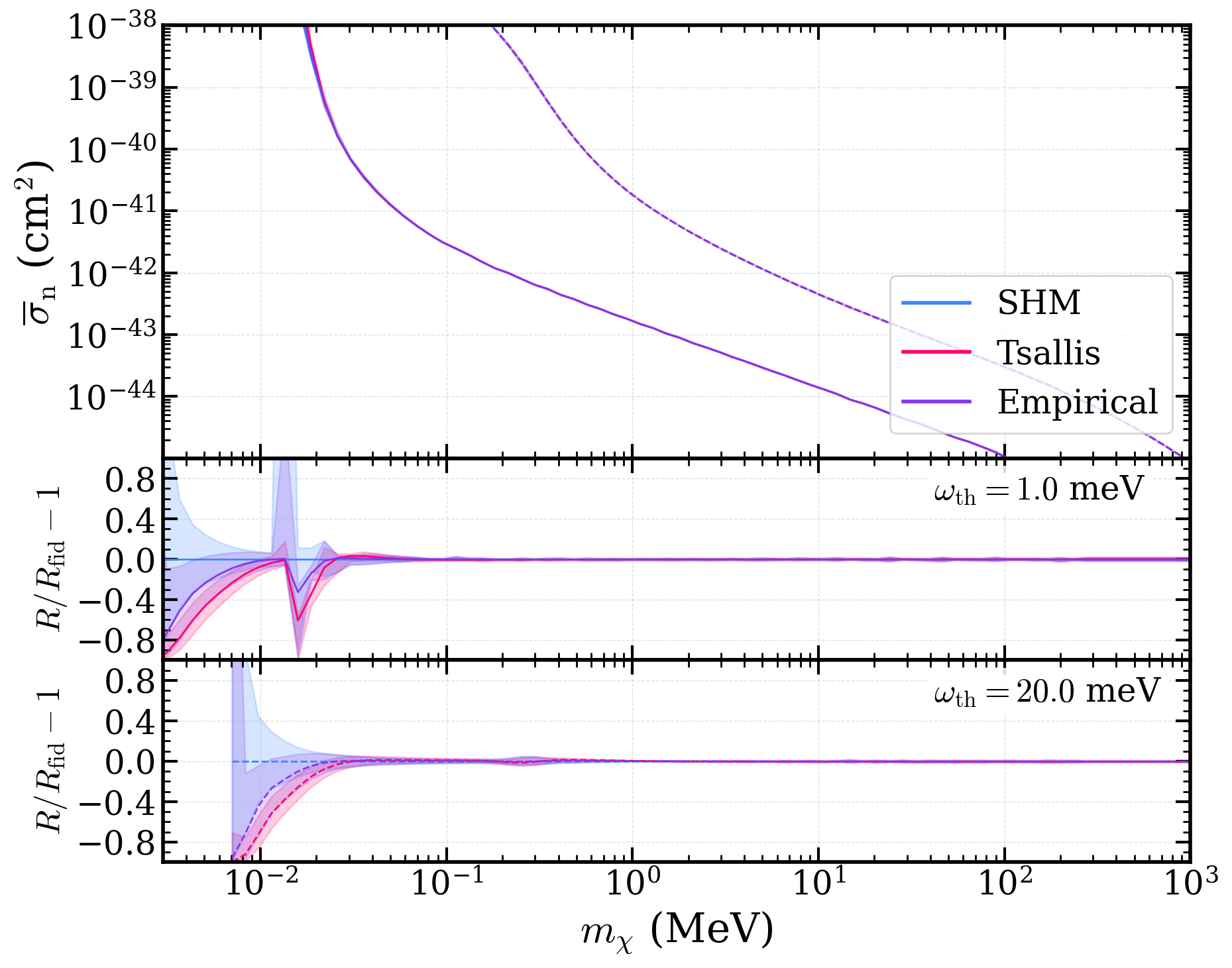}
}
\hfill
\subfloat[$\mathrm{SiO}_2$]{%
    \includegraphics[width=0.49\textwidth]{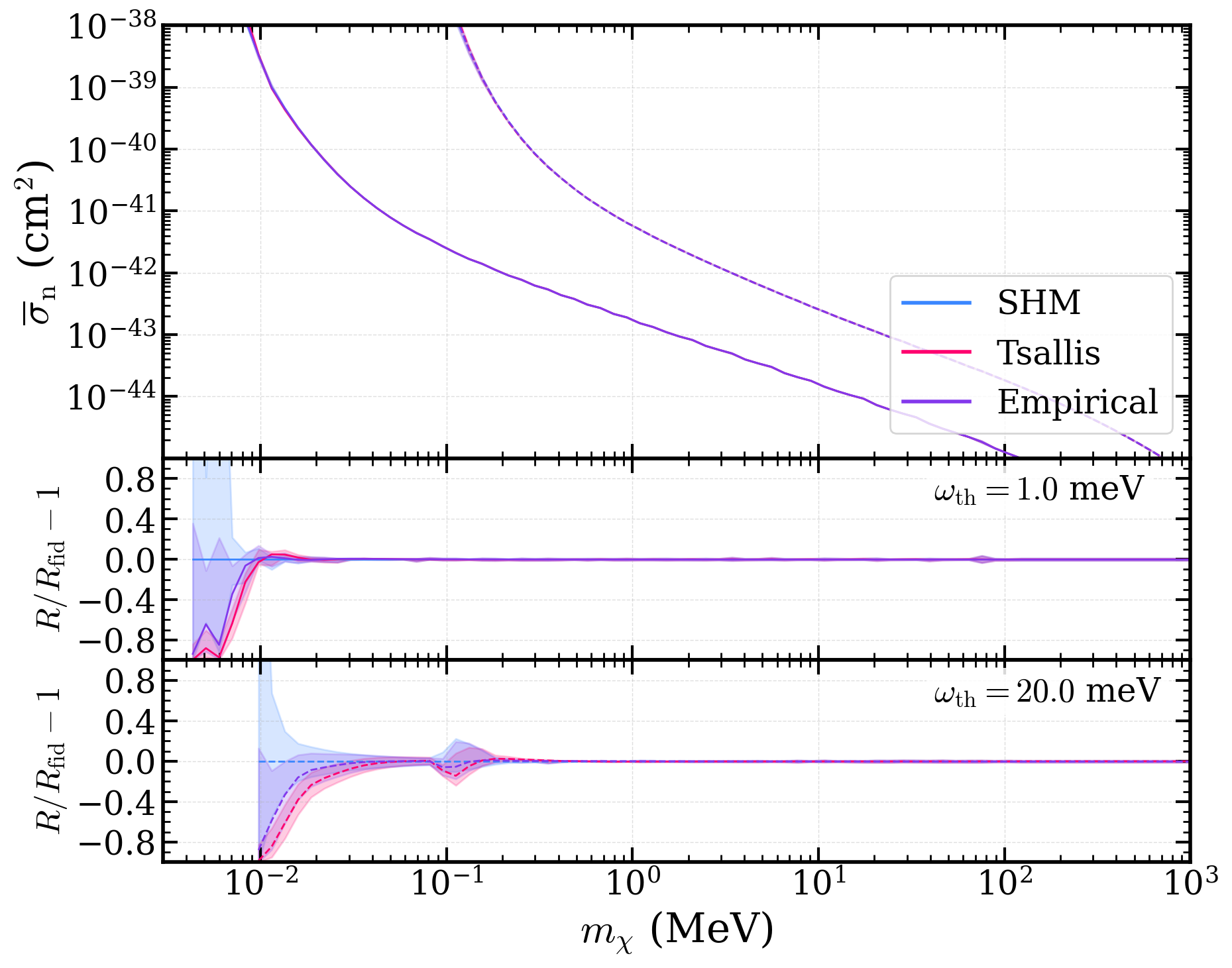}
}

\subfloat[$\mathrm{Al}_2\mathrm{O}_3$]{%
    \includegraphics[width=0.49\textwidth]{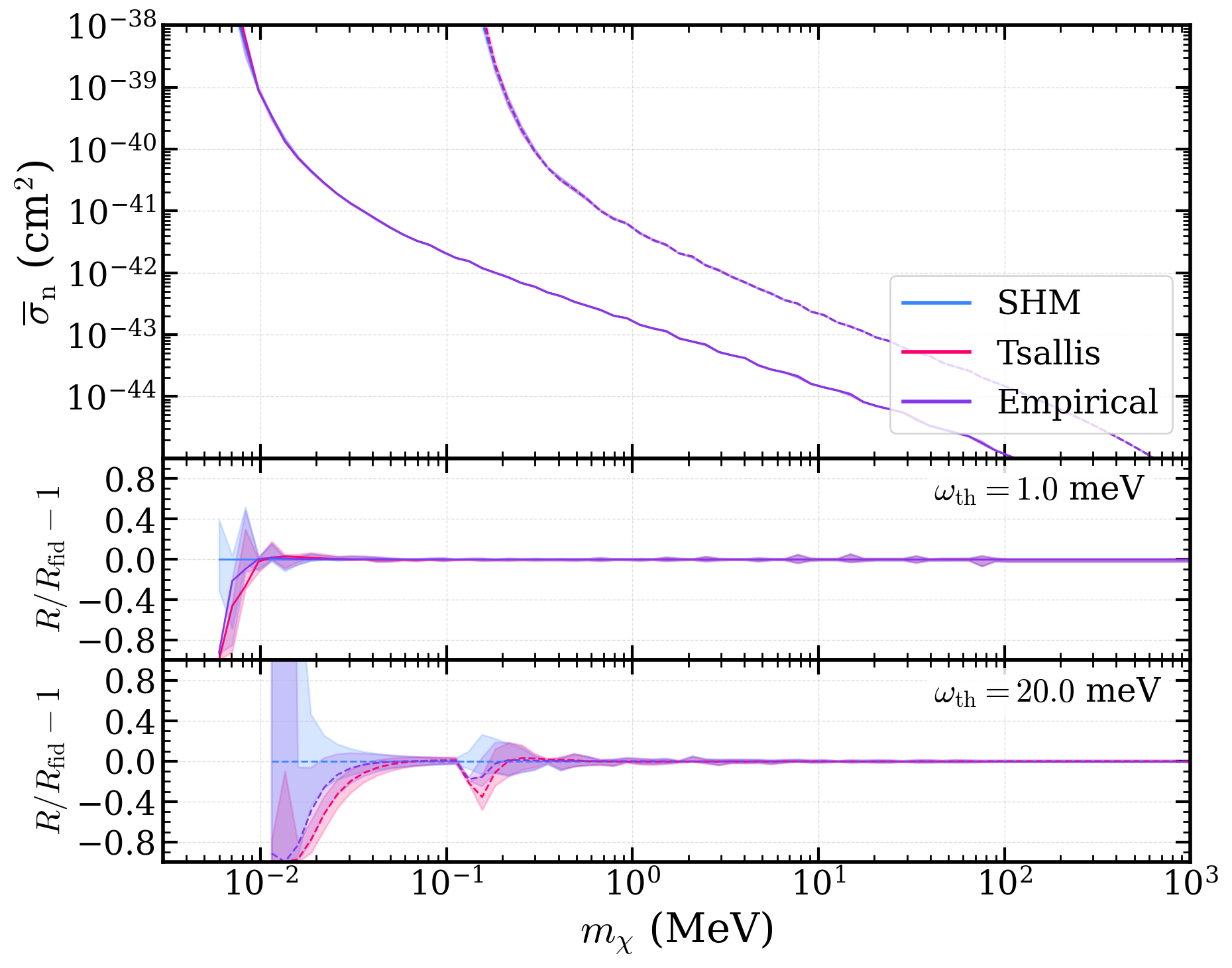}
}
\hfill
\subfloat[$\mathrm{GaAs}$]{%
    \includegraphics[width=0.49\textwidth]{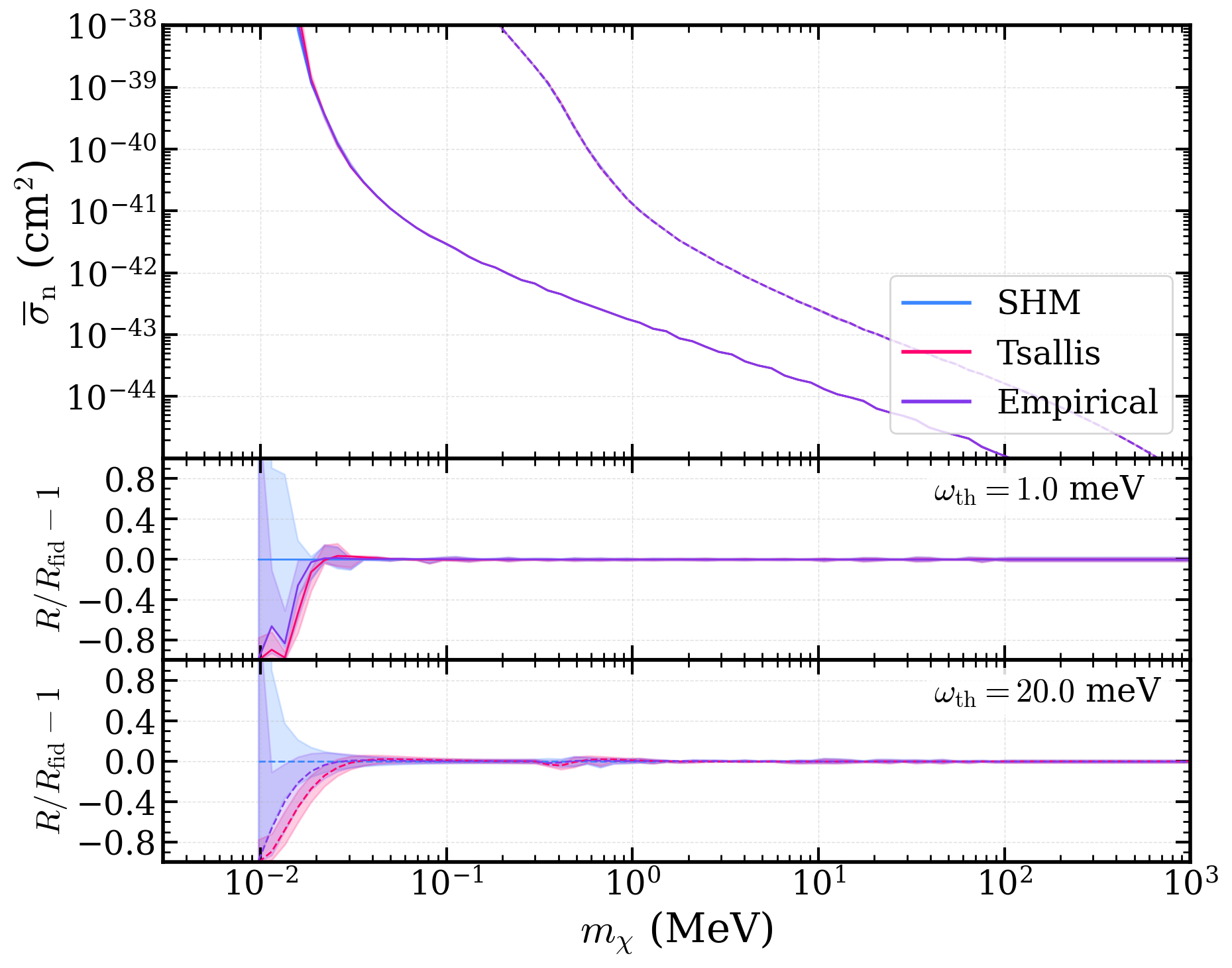}
}

\caption{Same as~\cref{fig:LDP_reach_AAA} for light hadrophilic scalar mediated scattering.
\label{fig:LM_reach_AAA}}
\end{figure}

\begin{figure}[t]
\centering

\subfloat[$\mathrm{CaWO}_4$]{%
    \includegraphics[width=0.49\textwidth]{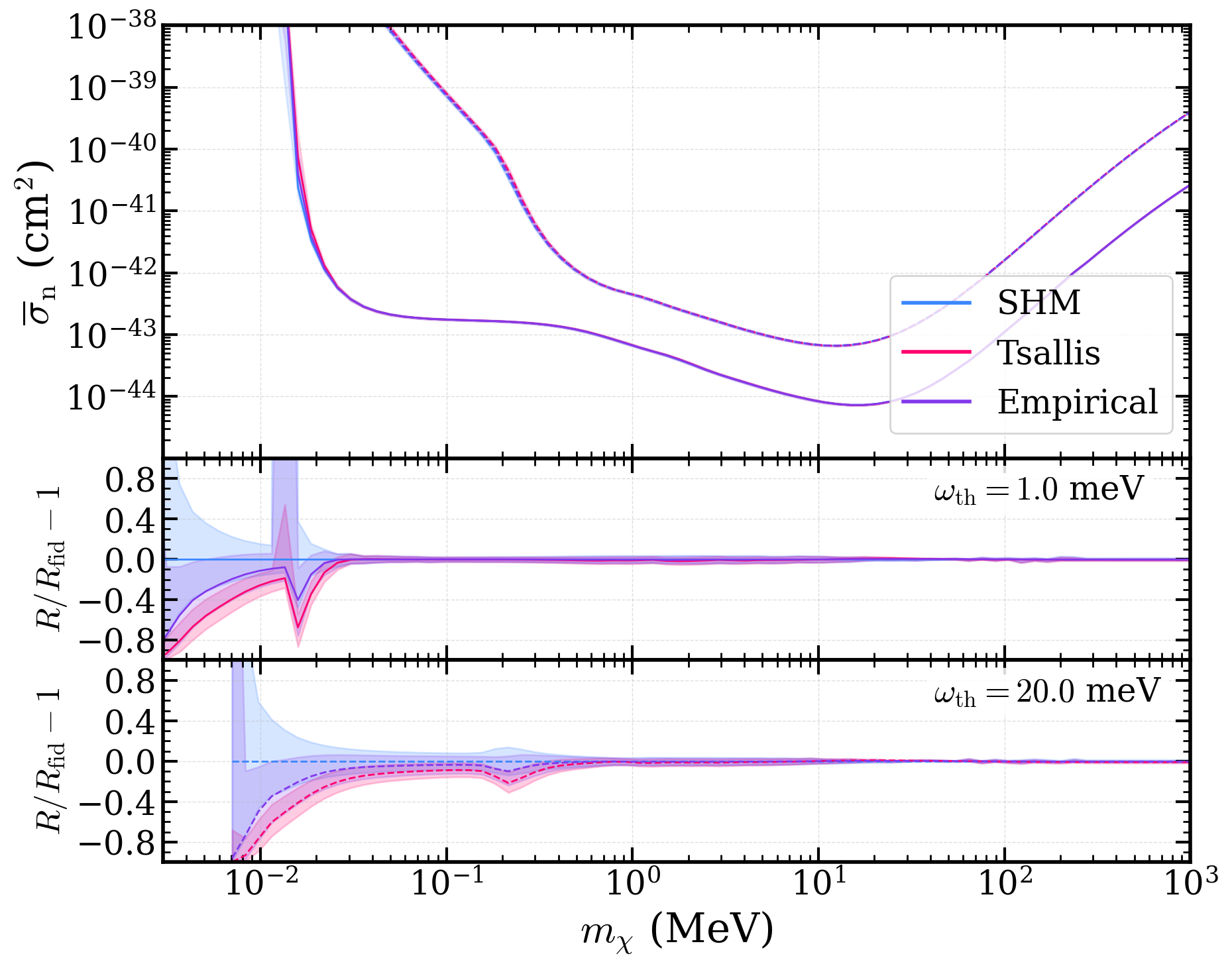}
}
\hfill
\subfloat[$\mathrm{SiO}_2$]{%
    \includegraphics[width=0.49\textwidth]{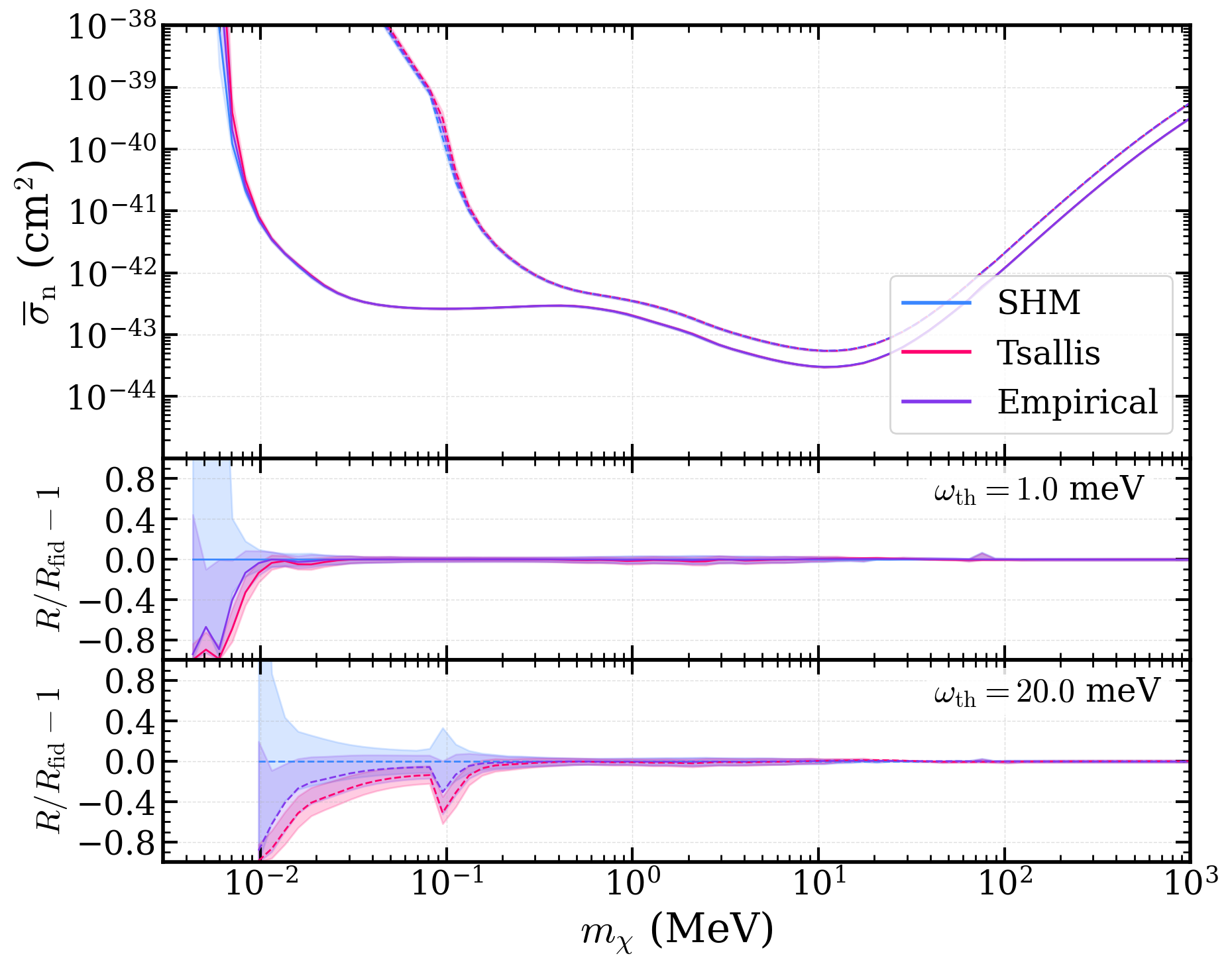}
}

\subfloat[$\mathrm{Al}_2\mathrm{O}_3$]{%
    \includegraphics[width=0.49\textwidth]{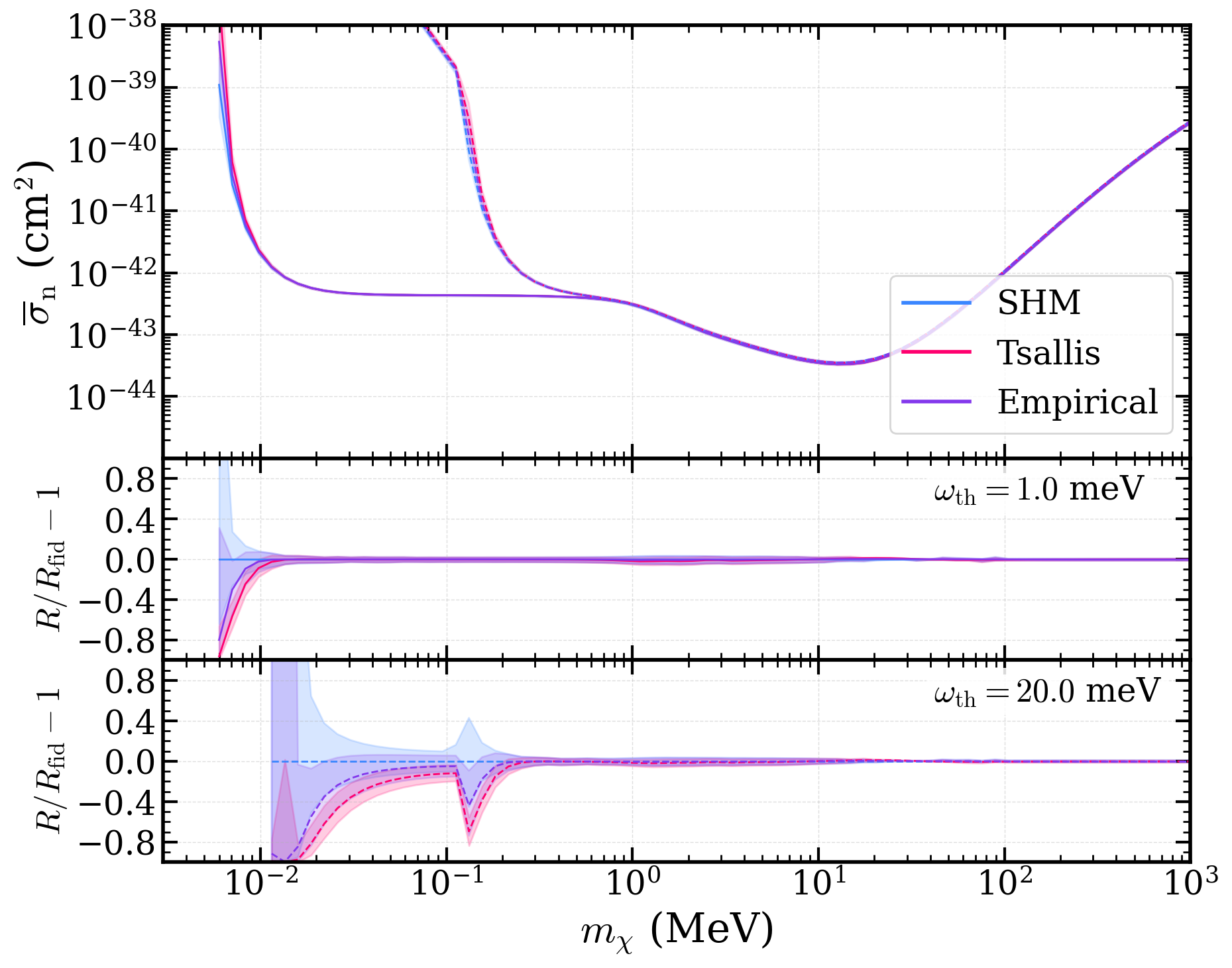}
}
\hfill
\subfloat[$\mathrm{GaAs}$]{%
    \includegraphics[width=0.49\textwidth]{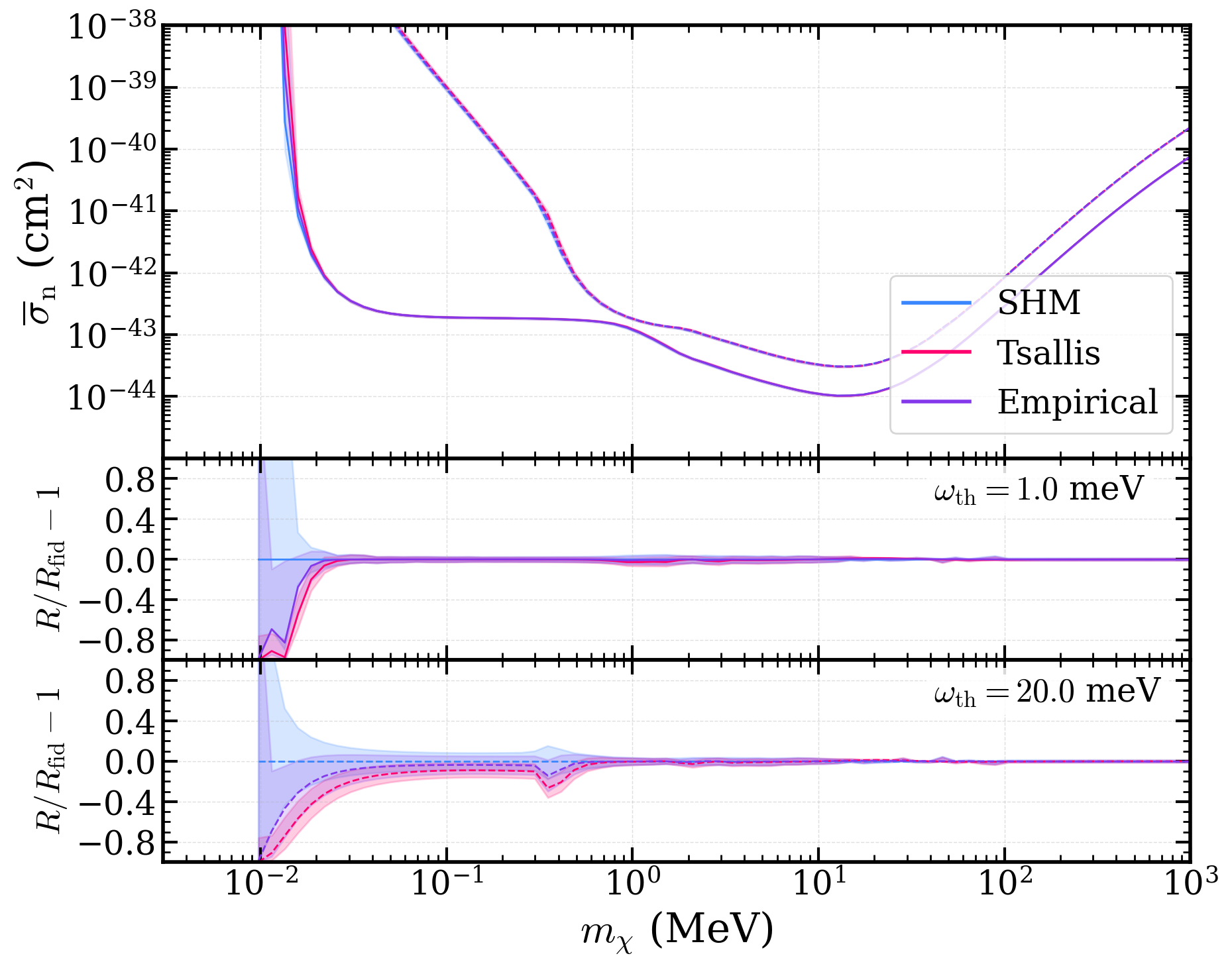}
}

\caption{Same as~\cref{fig:LDP_reach_AAA} for heavy hadrophilic scalar mediated scattering.
\label{fig:HM_reach_AAA}}
\end{figure}

In this appendix, we present results obtained using the aggressive halo parameter ranges defined in \cref{tab:Halo_Parameters}, evaluated under the rms-matching prescription introduced in~\cref{sec:prescriptions}. Compared to the conservative choice adopted in the main analysis, the aggressive ranges correspond to tighter constraints on $(v_\mathrm{c}, v_\mathrm{e}, v_\mathrm{esc})$. As illustrated in~\cref{fig:VDF_bands_aggr}, the allowed spread of the velocity distributions narrows considerably under these assumptions: with the rms-matching prescription ensuring a common mean kinetic energy across halo models, the tighter parameter ranges further compress the envelope of viable speed distributions, leaving little room for model-to-model or parameter-driven variation in the bulk of the distribution.

Consequently, the corresponding uncertainty bands in the projected reach are substantially reduced relative to the conservative case (see~\cref{fig:HM_reach_AAA,fig:LM_reach_AAA,fig:LDP_reach_AAA}). Across most of the dark matter mass range, the three halo models---SHM, Tsallis, and empirical---yield nearly indistinguishable reach curves under the rms-matching prescription with aggressive parameters, reinforcing the conclusion that differences in the functional form of the velocity distribution are subdominant once the overall energy scale is fixed. Near the kinematic threshold, however, residual differences between models persist, reflecting the distinct tail behaviors of the three distributions in the high-velocity regime.

\clearpage
\bibliographystyle{JHEP}
\bibliography{biblio.bib}

\end{document}